\documentclass[a4paper,11pt]{article}

\usepackage{jheppub}

\usepackage{bbm}
\usepackage{graphicx}
\usepackage{slashed}
\usepackage{subfigure}
\usepackage[utf8]{inputenc}
\graphicspath{ {./images/} }

\newcommand{\Tr}{\mathrm{Tr}}
\newcommand{\Pexp}{\, \mathrm{P} \, \exp}
\newcommand{\td}{\mathrm{d}}

\newcommand{\SU}[1]{\mathrm{SU}(#1)}

\newcommand{\TT}[1]{\mathrm{#1}}

\newcommand{\tdf}[2]{\dfrac{\td #1}{\td #2}}

\renewcommand{\v}[1]{\mathbf{#1}}

\renewcommand{\>}{\right\rangle}
\newcommand{\<}{\left\langle}
\newcommand{\lkl}{\left|}
\newcommand{\rkl}{\right|}

\newcommand{\qqquad}{\qquad\qquad}
\newcommand{\qqqquad}{\qquad\qquad\qquad}

\hypersetup{urlcolor=black}

\title{\boldmath Parton branching at amplitude level}

\preprint{\begin{flushright}MAN/HEP/2019/004\\ UWTHPH-2019-11
    \\   MCnet-19-10\end{flushright}}    

\author[a]{Jeffrey R. Forshaw,}
\author[a]{Jack Holguin,} 
\author[b]{and Simon Pl\"{a}tzer}

\affiliation[a]{Consortium for Fundamental Physics, School of Physics \& Astronomy, \\
	University of Manchester, Manchester M13 9PL, United Kingdom}
\affiliation[b]{Particle Physics, Faculty of Physics, \\
	University of Vienna, 1090 Wien, Austria}
\emailAdd{jeffrey.forshaw@manchester.ac.uk}
\emailAdd{jack.holguin@manchester.ac.uk}
\emailAdd{simon.plaetzer@univie.ac.at}

\abstract{We present an algorithm that evolves hard processes at the
  amplitude level by dressing them iteratively with (massless) quarks
  and gluons. The algorithm interleaves collinear emissions with soft
  emissions and includes Coulomb/Glauber exchanges. It includes all
  orders in $N_{\TT{c}}$, is spin dependent and is able to accommodate
  kinematic recoils. Although it is specified at leading logarithmic
  accuracy, the framework should be sufficient to go beyond. Coulomb
  exchanges make the factorisation of collinear and soft emissions
  highly non-trivial. In the absence of Coulomb exchanges, we show how
  factorisation works out and how a partial factorisation is manifest
  in the presence of Coulomb exchanges. Finally, we illustrate the use
  of the algorithm by deriving DGLAP evolution and computing the
  resummed thrust, hemisphere jet mass and gaps-between-jets
  distributions in $e^+ e^-$.}

\begin{document} 
\maketitle
\flushbottom

\section{Introduction}
\label{sec:intro}

Modern day experimental particle physics is often performed at hadron
colliders. As an unavoidable consequence, QCD corrections play a large
role. Contributions from coloured radiation, when evaluated Feynman
diagrammatically, diverge at multiple points in the phase space. When
regularised and cancelled, the divergences may leave behind large
logarithms. The accurate inclusion of logarithmically enhanced corrections is
of importance to both the theoretical and experimental communities.
Historically there have been two main approaches to dealing with QCD radiative
corrections: resummations and parton showers.

Resummations look to re-organise the perturbative expansion by classifying the
large logarithms and then summing the perturbation series such that the most
dominant logarithmically enhanced terms are included. Towers of logarithms may
be further simplified by making the leading colour (LC) approximation. The
re-organised expansions are referred to by their logarithmic accuracy; leading
log (LL), next-to-leading log (NLL), etc. This procedure has recently been
further formalised by work in soft-collinear effective field theories
\cite{Bauer:2000ew,Bauer:2000yr,Bauer:2001yt,SCET}. From this perspective,
resummations are renormalisation group flows that evolve `safe' perturbative
predictions into regions of phase space where perturbative expansions would be
otherwise `unsafe'.

In contrast, parton showers may be thought of as providing an all-purpose
approximation to the resummation procedure. Modern parton showers generate an
evolving, classical system of partons whilst cleverly encoding quantum
interference effects (made possible by working in the LC approximation). The
majority of currently available parton showers claim LL accuracy using the LC
approximation
\cite{colour_dipole_model,Herwig_dipole_shower,Herwig_shower,DIRE,Banfi:2004yd,Pythia,Pythia8}. The
quest to better understand the data from the LHC is a major driver for
increasingly precise parton showers. At present, there is a growing list of
phenomena that parton showers do not encapsulate. This includes effects
sub-leading in colour, some non-global logarithms \cite{Dasgupta:2001sh},
Coulomb/Glauber exchanges, super-leading logarithms
\cite{Forshaw:2006fk,SuperleadingLogs,Banfi:2010xy} and the violation of QCD
coherence (or collinear factorisation)
\cite{Catani:2011st,factorisationBreaking}. Moreover, recent fixed-order
studies have cast further doubt on the accuracy of modern parton showers. It
has been shown in \cite{Dasgupta:2018nvj} that the \textsc{Pythia}
\cite{Pythia,Pythia8} and \textsc{Dire} \cite{DIRE} showers suffer from both
incorrect next-to-leading logarithms at leading colour and incorrect
contributions from sub-leading colour (NLC) at LL. Although these showers
never claim NLL or NLC accuracy, the findings of Dasgupta et al questions the
fruitfulness of attempts to extend conventional parton showers beyond LL and
LC in general. In recent years, there has been movement towards finding new
constructions for partons showers; constructions more suited to including NLC
or NLL corrections
\cite{Platzer:2018pmd,Platzer:2012np,Nagy:2017ggp,Nagy:2008eq,Nagy:2012bt,Nagy:2015hwa,SoftEvolutionAlgorithm,Nagy:2019pjp,Banfi:2004yd}. However,
as of yet, success has been limited.

The algorithm we present here aims to provide a framework for the development
of future parton showers, enabling them to be systematically improved. We hope
it will also help make more rigorous the link between resummations and parton
showers. Our starting point is the soft-gluon evolution algorithm explored in
\cite{SoftEvolutionAlgorithm}, which we refer to as the FKS algorithm. The
evolution generated by the FKS algorithm is systematic to all orders in colour
and it accounts for the leading soft logarithms. The FKS algorithm was
originally used to derive the super-leading logarithms that may occur in
hadron-hadron collisions \cite{Forshaw:2006fk,SuperleadingLogs}. It has been
analytically verified for a general hard process dressed with up to two soft
real emissions and one loop
\cite{ColoumbGluonsOrdering,ColoumbGluonsOrderingLetter}. It has also been
shown to generate the BMS equation \cite{BMSEquation} (it presumably also
includes the NLC corrections to it) and it correctly accounts for the leading
non-global logarithms for various observables
\cite{SoftEvolutionAlgorithm}. The main goal of this paper is to improve the
FKS algorithm by including collinear emissions, spin dependence and kinematic
recoil. The algorithm we present is Markovian and can be solved iteratively,
making it well suited for use as a parton shower.

The remainder of the paper is organised as follows. In the next section, we
introduce the algorithm in a form we refer to as variant A, in which we
interleave soft and collinear emissions. Variant A has the virtue of being a
simple extension of the FKS algorithm, though it suffers from unnecessarily
complex colour evolution in the soft-collinear sector. It also suffers from
the fact that we cannot uniquely identify a parent parton in the case of
soft-gluon emission, which complicates the issue of longitudinal momentum
conservation. We are thus motivated to re-cast the algorithm in a more
convenient form, which we refer to as variant B. Specifically, in variant B we
manipulate the colour structures of variant A to isolate the full collinear
splitting functions, after which we are able to implement longitudinal
momentum conservation in a simple way. We also spend some time illustrating
how recoils may be included in both variants, though this will only be
relevant beyond the LL approximation. As it stands, either variant A or B
could be used to create a fully functioning parton shower, though B will be
computationally more efficient. In Section \ref{sec:algorithmIR} we present a
manifestly infra-red finite version of the algorithm. This reformulation is
particularly useful for the resummation of specific observables, though it is
not so well suited for use as a general purpose parton shower. This is because
the infra-red singularities are regularised by the explicit inclusion (and
exponentiation) of a measurement function.

The second half of the paper is devoted to issues of collinear factorisation
and to providing examples to illustrate how the algorithm is used. In Section
\ref{sec:Coll} we discuss the factorisation of collinear physics from soft
physics. We start by considering the case when Coulomb/Glauber exchanges are
turned off (such as would be the case in $e^+ e^-$ collisions). After this we
discuss how Coulomb exchanges can be introduced one-by-one. We will see that
collinear factorisation occurs below the scale of the last Coulomb
exchange. This discussion shows consistency between our approach and the
proofs of collinear factorisation by Collins, Soper and Sterman
\cite{Collins:1987pm,Collins:1988ig}. After this, we show how DGLAP evolution
for the parton distribution functions emerges
\cite{Dokshitzer:1977sg,Gribov:1972ri,APSplitting}. We finish the paper by
illustrating the use of the algorithm; by calculating the thrust, hemisphere
jet mass, and gaps-between-jets distributions in $e^+ e^-$. We leave an
extensive discussion of spin correlations to an appendix.

\section{The algorithm}
\label{sec:algorithm}

\begin{figure}[h]
	\centering
	\includegraphics[width=0.33\textwidth]{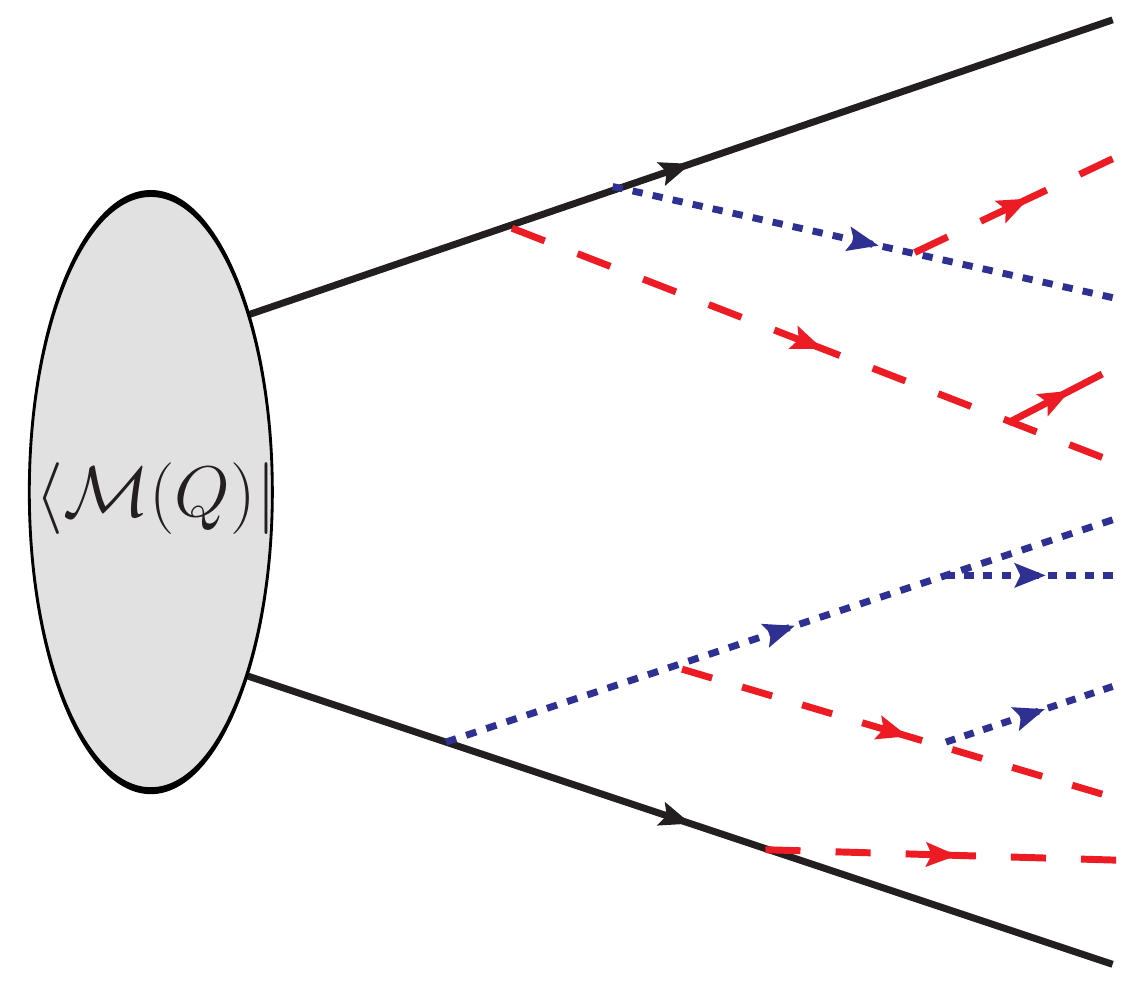}
	\caption{A term contributing to evolution of the conjugate amplitude
          (it contributes to $\v{A}_{9}$). Red dashed lines represent the
          emission of soft gluons and collinear partons are represented by
          blue dotted lines. Loops (Sudakov factors) have been neglected to
          avoid clutter. We draw all particles heading to the right, away from
          the hard process, including incoming particles. In contrast,
          evolution of the amplitude will have all particles drawn heading to
          the left and away from the hard process.}
	\label{fig:SC1_A}
\end{figure}

In this section we present the algorithm. It is Markovian and interleaves soft
emissions and virtual corrections with collinear emissions and virtual
corrections, see Figure \ref{fig:SC1_A}. Successive real emissions are
strongly ordered in an appropriately defined transverse momentum. We will
present two variants of the algorithm, which we refer to as A and B. The two
differ only in where we put the soft-collinear emissions: in A they are in the
soft sector and in B they are in the collinear sector. The second approach
allows us to exploit the colour-diagonal nature of collinear emissions and it
makes kinematic recoil more straightforward to implement. Variant A has the
virtue that it is an almost trivial extension of the purely soft evolution
presented in \cite{SoftEvolutionAlgorithm}. We present both A and B with the
momentum mappings after each real emission parametrised into two initially
unspecified functions. This is so the algorithm is able to accommodate
partonic recoil. Later, in Section \ref{sec:SC2_spectator_recoil}, we discuss
specific examples of recoil in action. For processes with coloured incoming
partons, the algorithm should be convoluted with parton distributions
functions. We leave a full description of how to do this to Section
\ref{sec:pheno}.

Before plunging in, we should explain the theoretical basis for what
follows. Our algorithm is based on Feynman diagram calculations
\cite{SuperleadingLogs,ColoumbGluonsOrdering,SoftEvolutionAlgorithm,Dokshitzer:1991wu,Dokshitzer:1977sg,DOKSHITZER1980269,Bassetto:1984ik}
and, in its present form, captures all of the logarithms associated with the
leading amplitude-level singularities. Therefore the algorithm captures
leading logarithms from wide-angle soft emissions, hard-collinear emissions
and simultaneously soft and collinear emissions. This means the algorithm is
guaranteed to capture only the most leading logarithms at cross-section
level. That said, it is also able to capture the leading single non-global
logarithms, even if the global part is double-logarithmic, as is the case with
the hemisphere jet mass for example (see Section \ref{sec:pheno}). For any
process involving incoming hadrons or measured outgoing hadrons, the single
logarithms from DGLAP evolution are recovered as well (i.e. parton
distribution function and by a simple extension fragmentation function
evolution). We believe our framework to be sufficiently flexible that we can,
in the future, extend it beyond the LL approximation.

\subsection{Parton branching with interleaved soft and collinear evolution (A)}
\label{sec:SC1}

The algorithm evolves a hard-scattering matrix, $\v{H}(Q; \{p\})$, which is
defined at some hard scale $Q$ and is a function of the hard-particle
four-momenta, $\{ p \}$. It does so by dressing with successive soft and/or
collinear real emissions and virtual corrections. $\v{H}(Q; \{p\})$ is a
tensor in the product space of colour and helicity\footnote{This paper only
  concerns itself with massless partons and so all particles have a definite
  helicity.}, defined as $\v{H}(Q; \{p\}) \equiv (\lkl \TT{colour} \> \otimes
\lkl \TT{spin} \>) \otimes (\< \TT{colour} \rkl \otimes \< \TT{spin}
\rkl)$. The hard-scattering matrix is defined so that $\Tr \, \v{H}(Q; \{p\})$
is the hard matrix element squared, summed over colour and spin\footnote{We
  may also choose to include averaging factors, a flux factor and the hard
  process phase-space, so that it is then the hard-process differential cross
  section.}. Successive real emissions are added via `rectangular' operators,
which act as a map increasing the dimension of the representation of $\SU{3}
\times \TT{E}(2)$ in which $\v{H}(Q; \{p\})$ resides. The virtual evolution
operators are `square' and preserve the representation of $\v{H}(Q;
\{p\})$. Specifically,
\begin{eqnarray}
\label{eqn:FKS}
\begin{split}
	\td\sigma_{0} &= \Tr\left( \v{V}_{\mu,Q} \v{H}(Q; \{p\}) \v{V}^{\dagger}_{\mu,Q} \right) = \Tr \, \v{A}_{0}(\mu; \{p\}), \\
	\td \sigma_{1} &= \int \prod_{i=1}^{n_\text{H}+1} \td^{4} p_{i} \, \Tr\left( \v{V}_{\mu,q_{1 \, \bot}} \v{D}_{1} \v{V}_{q_{1 \, \bot},Q} \v{H}(Q; \{p\}) \v{V}^{\dagger}_{q_{1 \, \bot},Q} \v{D}^{\dagger}_{1}  \v{V}^{\dagger}_{\mu,q_{1 \, \bot}} \right) \td \Pi_{1} \\
	& \;  = \Tr \, \v{A}_{1}(\mu; \{\tilde{p}\}\cup q_{1}) \, \td \Pi_{1}, \\
	\td \sigma_{n} & \; = \Tr \, \v{A}_{n}(\mu; \{p\}_{n}) \prod^{n}_{i =1} \td \Pi_{i},
\end{split}
\end{eqnarray}
where
\begin{equation}
	\v{A}_{n}(q_{\bot}; \{\tilde{p}\}_{n-1} \cup q_{n}) = \int \prod_{i=1}^{n_\text{H}+n} \td^{4} p_{i} \v{V}_{q_{\bot},q_{n \, \bot}} \v{D}_{n} \v{A}_{n-1}(q_{n \, \bot}; \{p\}_{n-1}) \v{D}^{\dagger}_{n} \v{V}^{\dagger}_{q_{\bot},q_{n \, \bot}} \Theta(q_{\bot} \leq q_{n \, \bot}). \label{eq:Aevo}
\end{equation}
At each step, the emission operators ($\mathbf{D}_n$) add one new particle, of
four-momentum $q_n$, to the set $\{ p \}_{n-1}$, to produce the set $\{ p
\}_n$. We use $p_j \in \{ p \}_n= \{ P_1, P_2, \cdots
P_{n_\text{H}},q_1,\cdots q_n \}$ to denote the momentum of the $j^\text{th}$
parton and $1 < j < n_\text{H}+n$, where $n_\text{H}$ is the number of partons
associated with the original hard process and $n$ is the number of emitted
partons.  Hidden in the emission operators is a map from $\{ p \}_{n-1}$ to a
new set, $\{ \tilde{p}\}_{n-1} $. The difference between these two sets is
determined by the way we implement energy-momentum conservation (i.e. the
recoil prescription) and it is why there is an extra integral over $p_i$ (it
is not a phase-space integral). The virtual evolution operators
$\mathbf{V}_{a,b}$ encode the loop corrections.  To avoid cumbersome notation
we write $\{p\}_n = \{\tilde{p}\}_{n-1} \cup \{q_n\}$ is the set of $n$
momenta including the last emission, $q_n$. We have not yet defined the
ordering variable, $q_{i \perp}$; we will do that shortly.  A generalised
observable $\Sigma$, with measurement function $u_{n}(q_{1},...,q_{n})$, is
then given by\footnote{For fixed Born-level kinematics. Generally the
  measurement function will depend upon the hard process momenta $P_j$, which
  we do not show explicitly.}
\begin{eqnarray}
\begin{split}
\Sigma(\mu) & = \int \sum_{n} \td \sigma_{n} \, u_{n} (q_{1},...,q_{n}), \\
& = \int \sum_{n} \left(\prod^{n}_{i = 1} \td \Pi_{i} \right) \Tr \, \v{A}_{n}(\mu; \{p\}_{n}) \, u_{n}(q_{1},...,q_{n}),
\end{split}
\end{eqnarray}
where $\td \Pi_i$ is the phase-space for the $i^{\text{th}}$ emission (see
below). $\mu$ should be taken either to $0$ or to the scale below which the
observable is inclusive over all radiation. The virtual (Sudakov) evolution
operator is\footnote{The path ordering ensures that the operators are ordered
  in $k^{(ij)}_{\bot}$ with the largest to the right.}
\begin{align}
\label{eqn:V_defs}
	\v{V}_{a,b} =& \Pexp \left[-\frac{\alpha_{s}}{\pi} \sum_{i<j} \int^{b}_{a} \frac{\td k^{(ij)}_{\bot}}{k^{(ij)}_{\bot}} (-\mathbb{T}^{g}_{i} \cdot \mathbb{T}^{g}_{j}) \left\{ \int \frac{\td y \, \td \phi}{4\pi}(k^{(ij)}_{\bot})^{2}\frac{\tilde{p}_{i}\cdot \tilde{p}_{j}}{(\tilde{p}_{i}\cdot k)(\tilde{p}_{j} \cdot k)}\theta_{ij}(k) - i\pi \,\tilde{\delta}_{ij}\right\} \right. \nonumber \\
	& \; \left. \times \mathcal{R}^{\TT{soft}}_{ij}(k, \{\tilde{p}\}) - \frac{\alpha_{s}}{\pi} \sum_{i} \int^{b}_{a} \frac{\td k^{(i\vec{n})}_{\bot}}{k^{(i\vec{n})}_{\bot}}   \sum_{\upsilon \in \{q,g\}} \mathbb{T}^{\bar{\upsilon} \, 2}_{i} \int \frac{\td z \, \td \phi}{8\pi} \, \overline{\mathcal{P}}^{\,\circ}_{\upsilon \upsilon_{i}}(z) \, \theta_{i}(k) \, \mathcal{R}^{\TT{coll}}_{i}(k, \{\tilde{p}\}) \right],
\end{align}
where $i$ and $j$ run over all external legs (those from the initial hard
process and also previous emissions in the evolution). $\tilde{\delta}_{ij} =
1$ if both partons $i$, $j$ are incoming or both outgoing and
$\tilde{\delta}_{ij} = 0$ otherwise. $\theta_{ij}(k) = \Theta(p_{i}\cdot
p_{j}-k \cdot (p_{j} + p_{i}))$ and ensures that the phase space of the
integration corresponds to that of a real gluon. Likewise, the $z$ integral is
over the range\footnote{We specify the range corresponding to emission off a
  final state particle, for emission off an initial state particle exchange
  $p_{i} \rightarrow \tilde{p}_{i}$.}
\begin{align}
z \in \left[\frac{\alpha}{2}  - \frac{1}{2} \sqrt{\alpha^{2} - \frac{4k_\perp^{(i \vec{n}) \, 2}}{(n . p)^{2}}} \, , \; \frac{\alpha}{2} + \frac{1}{2} \sqrt{\alpha^{2} - \frac{4k_\perp^{(i \vec{n}) \, 2}}{(n . p)^{2}}}\right], \quad \alpha = \frac{2 p \cdot p_{i} + p_{i} \cdot n \, p \cdot n}{(p \cdot n)^{2}} \, , \label{eqn:zlimits}
\end{align} 
which can be expressed via a single theta function $$\theta_{i}(k) = \Theta((n
\cdot p_{i} - n \cdot k) n \cdot p + 2 p \cdot p_{i} - 2 p \cdot k).$$ The
vectors $p$ and $n= (1, \vec{n})$ will be defined shortly: to LL accuracy $p =
p_{i}$ and $\alpha = 1$. $\upsilon_{i}, \upsilon \in \{q, g\}$ label parton
species. $\bar{\upsilon} = g$ in all cases except when $\upsilon_{i} = g$ and
$\upsilon = q$, then $\bar{\upsilon} =
q$. $\overline{\mathcal{P}}_{\upsilon\upsilon_{i}}^\circ$ is the $\upsilon_{i}
\rightarrow \upsilon$ hard-collinear splitting function and it is defined in
Appendix \ref{Appendix_A} along with the conventions we use for helicity
states and antiparticles. $\mathcal{R}^{\TT{soft}}_{ij}(k, \{p\})$ and
$\mathcal{R}^{\TT{coll}}_{i}(k, \{p\})$ are concerned with the recoil
prescription and are included to preserve unitarity, they are defined in
\eqref{eq:rsoft} and \eqref{eq:rcol} below.  $k_\perp^{(ij)}$ and $y$ are the
transverse momentum and rapidity in the $ij$ zero-momentum frame. To make the
(unitarity) link to the real emissions more explicit, we choose not to use the
substitution
\begin{equation}
	(k^{(ij)}_{\bot})^{2}\frac{\tilde{p}_{i}\cdot \tilde{p}_{j}}{(\tilde{p}_{i}\cdot k)(\tilde{p}_{j} \cdot k)} = 2.
\end{equation}
The real-emission operator is built using two operators:
\begin{eqnarray}
\label{eqn:SC_defs}
\begin{split}
&\v{S}_{i} = \sum_{j} \left( \frac{q^{(j\vec{m})}_{i \, \bot}}{2\tilde{p}_{j} \cdot q_{i}} \mathbb{T}^{g}_{j} \otimes (\tilde{p}_{j} \cdot \epsilon^{*}_{+}(q_{i})   \mathbb{S}^{1_{i}} +  \tilde{p}_{j} \cdot\epsilon^{*}_{-}(q_{i}) \mathbb{S}^{-1_{i}})\right) \, \mathfrak{R}^{\TT{soft}}_{ij}(\{p\},\{\tilde{p}\},q_{i}), \\ 
&\v{C}_{i} = \sum_{j}  \frac{q^{(j\vec{n})}_{i\,\bot}}{2\sqrt{z_{i}}}\Delta_{ij} \,\overline{\v{P}}_{ij} \, \mathfrak{R}^{\TT{coll}}_{ij}(\{p\},\{\tilde{p}\},q_{i}),
\end{split}
\end{eqnarray}
such that $\v{D}_{i}$ acts as
\begin{eqnarray}
\label{eqn:D_defs}
\begin{split}
...\v{D}_{i} \v{\mathcal{O}} \v{D}^{\dagger}_{i}... = ...\v{S}_{i} \v{\mathcal{O}} \v{S}^{\dagger}_{i}... + ...\v{C}_{i} \v{\mathcal{O}} \v{C}^{\dagger}_{i}....
\end{split}
\end{eqnarray}
$j$ again runs over all external legs and $i$ labels the emitted
parton. $\v{S}_{i}$ generates soft emissions and $\v{C}_{i}$ hard-collinear
emissions. The symbol $\Delta_{ij}$ is defined so that 
\mbox{$\Delta_{ij}\Delta_{i k} = \delta_{jk}$} and $\delta^{\TT{final}}_{j}$
($\delta^{\TT{initial}}_{j}$) is unity when parton $j$ is in the final
(initial) state and zero otherwise. $\overline{\v{P}}_{ij}$ are the
amplitude-level hard-collinear splitting functions and are defined in Appendix
\ref{Appendix_A}. The splitting functions encode DGLAP evolution
\cite{Dokshitzer:1977sg,Gribov:1972ri,APSplitting} including the
spin-dependence. $\mathbb{T}^{g}_{i}$ is a basis independent colour charge
operator. We have indexed each $\mathbb{T}^{g}_{i}$ with the leg on which it
acts, $i$, and by whether it corresponds to the emission of a gluon or not
(i.e. the index $q$ refers to a $g \to q \bar{q}$
splitting). $\mathbb{S}^{s_{i}}$ updates the helicity state by adding the
helicity of the emitted parton, $s_i$. The operators $\mathbb{S}$ and
$\mathbb{T}$ are also defined in Appendix \ref{Appendix_A}.

\begin{figure}[t]
	\centering
	\subfigure[]{\includegraphics[width=0.3\textwidth]{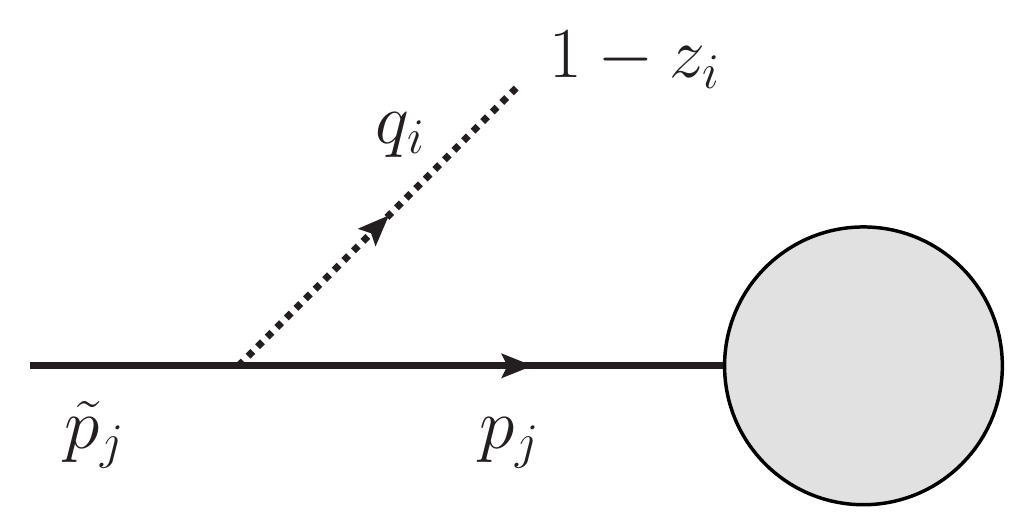}} ~~~~~~~~~~~
	\subfigure[]{\includegraphics[width=0.3\textwidth]{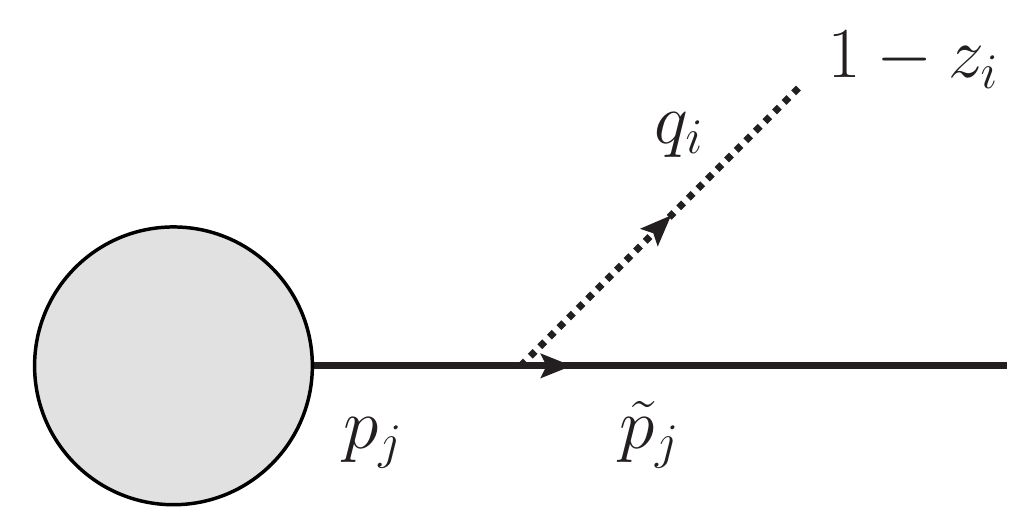}}
	\caption{Defining the kinematics: (a) $q_i$ is emitted off an incoming
          leg; (b) $q_i$ is emitted off an outgoing leg.}
	\label{fig:momenta_definitions}
\end{figure}

In the soft sector we have introduced an auxiliary vector $\vec{m}$. It is
uniquely determined, but only at cross-section level, since we require
$q^{(i\vec{m})}_{\bot} q^{(j\vec{m}')}_{\bot}|_{p_{i} \neq p_{j}} =
(q^{(ij)}_{\bot})^{2}$, which corresponds to choosing $\vec{m}$ to lie in the
direction of $j$ and $\vec{m}'$ (the corresponding vector in the conjugate
amplitude) in the direction of $i$. It is only ever this combination that
appears at cross-section level. In the collinear sector, the momentum fraction
$z_i$ is defined by (see Figure \ref{fig:momenta_definitions}):
\begin{align}
z_i &= \frac{\tilde{p}_j \cdot n}{p \cdot n} ~~\text{for final-state emissions} \nonumber \\
\text{and}~~~~z_i &= \frac{{p}_j \cdot n}{p \cdot n} ~~\text{for initial-state emissions}, \label{eqn:momentum_fraction}
\end{align}
where the light-like four-vector $n$ satisfies $n \cdot
q^{(j\vec{n})}_{i\perp}=0$. The light-like four-vector $p$ satisfies $p \cdot
q^{(j\vec{n})}_{i \perp}=0$. Neglecting terms suppressed by the transverse
momentum of the emission (which is permissible in the LL approximation) we may
take $p = p_j$ for final-state emissions and $p = \tilde{p}_j$ for
initial-state emissions, in which case $z_i$ is the light-cone momentum
fraction. The precise definition of $p$ is dependent on the recoil
prescription, as we illustrate in Section \ref{sec:SC2_spectator_recoil}.

Now we can define the ordering variable, i.e. the definition of $a$ and $b$ in
the Sudakov operator $\mathbf{V}_{a,b}$. We use transverse momentum ordering,
where the transverse momentum should be defined by the parent partons of the
emitted parton. Doing this means that we really ought not to sum over partons
in \eqref{eqn:SC_defs} and we should replace \eqref{eq:Aevo} by
\begin{equation}
	\v{A}_{n}= \sum_{j_n,j_n'}\int \prod_{i=1}^{n_\text{H}+n} \td^{4} p_{i} \v{V}_{q_{\bot},q_{n \, \bot}} \v{D}_{n}^{j_n} \v{A}_{n-1}\v{D}_{n}^{j_n'\dagger} \v{V}^{\dagger}_{q_{\bot},q_{n \, \bot}} \Theta(q_{\bot} \leq q_{n \, \bot}) \label{eq:Aevo1}
\end{equation}
where $\mathbf{D}_n^{j_n}$ is defined by
$$ \mathbf{D}_{n} = \sum_{j_n} \mathbf{D}_n^{j_n}~.$$ The ordering variable is
then $q_{n\bot} = q_n^{(j_n j_n')}$ if the emission is soft or $q_{n\bot} =
q_n^{(j_n,\vec{n}) }$ if the emission is collinear. At LL, this choice of
ordering variable is somewhat arbitrary but being a transverse momentum it is
able to generate the super-leading logarithms correctly
\cite{ColoumbGluonsOrdering}. That said, it is not equivalent to the ordering
indicated by the results in
\cite{ColoumbGluonsOrdering,ColoumbGluonsOrderingLetter}, which is based on
fixed-order Feynman diagram calculations. We have not yet figured out a way to
implement the latter ordering to all orders. In the remainder of the paper, we
will use the simpler (though potentially misleading) notation of equation
\eqref{eq:Aevo}.

The recoil functions,
$\mathfrak{R}^{\TT{soft}\,  *}_{ij}\,\mathfrak{R}^{\TT{soft}}_{ij'}$ and
$\mathfrak{R}^{\TT{coll}\, *}_{ij}\,\mathfrak{R}^{\TT{coll}}_{ij}$, encode the maps
that implement energy-momentum conservation. As the algorithm proceeds, each $\mathfrak{R}^{\TT{soft}}_{ij}$ and $\mathfrak{R}^{\TT{coll}}_{ij}$ will always collect into the pairs just given. The functions only ever appear singularly to aid book keeping\footnote{ Singular definitions would require $\mathfrak{R}$ functions to contain integrals of delta functions which independently evolve momenta in the amplitude and similarly for $\mathfrak{R}^{*}$ in the conjugate amplitude. The external momentum integrals, $\int \prod_{k} \td^{4} p_{k}$, would the be used to force the two separately evolving momenta to coincide. Giving these definitions provides an entirely unnecessary extra complexity. }. The recoil functions should be constructed out of
delta functions and algebraic pre-factors relating the momentum from the
current step in the algorithm, $\{p\}$, to the momentum that will be carried
forwards to the next step of the algorithm, $\{\tilde{p}\}$ and
$q_{i}$. $\mathcal{R}^{\TT{soft}}_{ij}$ and $\mathcal{R}^{\TT{coll}}_{i}$ are
fixed by $\mathfrak{R}^{\TT{soft}}_{ij}$ and $\mathfrak{R}^{\TT{coll}}_{ij}$,
i.e.
\begin{equation}
\int \prod_{k} \td^{4} p_{k} \, \mathfrak{R}^{\TT{soft} \, *}_{ij}(\{p\},\{\tilde{p}\},q_{i})\mathfrak{R}^{\TT{soft}}_{ij'}(\{p\},\{\tilde{p}\},q_{i})  = \mathcal{R}^{\TT{soft}}_{jj'}(q_{i}, \{\tilde{p}\}),	
\label{eq:rsoft}
\end{equation}
and
\begin{equation}
\int \prod_{k} \td^{4} p_{k} \, \mathfrak{R}^{\TT{coll} \, *}_{ij}(\{p\},\{\tilde{p}\},q_{i})\mathfrak{R}^{\TT{coll}}_{ij}(\{p\},\{\tilde{p}\},q_{i})  = \mathcal{R}^{\TT{coll}}_{j}(q_{i}, \{\tilde{p}\}). \label{eq:rcol}
\end{equation}
In the LL approximation, $\mathcal{R}^{\TT{soft}}_{jj'} =
\mathcal{R}^{\TT{coll}}_{j} = 1$.  $\v{S}_{i}$ generates soft emissions and
one might suppose that a suitable choice of recoil (to LL accuracy) is
\begin{equation}
\mathfrak{R}^{\TT{soft}}_{ij}\mathfrak{R}^{\TT{soft} \, *}_{ij'} = \prod_{k} \delta^{4}(p_{k}-\tilde{p}_{k}).
\label{eq:r1}
\end{equation}
We will shortly see that things are not quite so simple, and that this
requires modification.  $\v{C}_{i}$ generates hard-collinear emissions,
however only the longitudinal component of the recoil is hard. Therefore, in
the LL approximation, we may implement recoils in the collinear sector via
\begin{eqnarray}
\mathfrak{R}^{\TT{coll} \, *}_{ij}\mathfrak{R}^{\TT{coll}}_{ij} = \left(\delta^{4}(p_{j}-z^{-1}_{i}\tilde{p}_{j})\delta^{\TT{final}}_{j} + \delta^{4}(p_{j}-z_{i}\tilde{p}_{j})\delta^{\TT{initial}}_{j}\right)\prod_{k \neq j} \delta^{4}(p_{k}-\tilde{p}_{k}). \label{eqn:SC1_recoil}
\end{eqnarray}
Finally, the phase-space is included via 
\begin{eqnarray}
\label{eqn:pi_defs}
\begin{split}
\td\Pi_{i} &= \frac{2\alpha_{s}}{\pi^{2}q^{2}_{i \, \bot}}\frac{\td^{3} q_{i}}{2E_{i}}.
\end{split}
\end{eqnarray} 
The pre-factor has been included to simplify the definitions of $\v{S}_{i}$
and $\v{C}_{i}$, as well as to make each term in the algorithm dimensionless
and keep explicit dependence on the ordering variable in $\v{D}_{i}$. To
simplify the notation, in \eqref{eqn:pi_defs} and elsewhere, we will drop the
dipole labels on transverse momenta. It should be clear from the context which
partons are intended. In the case of \eqref{eqn:pi_defs}, it means we should
use the transverse momentum defined by the parent parton and the vector
$\vec{n}$ in the case of collinear emissions or $q_\perp^{(ij)}$ in the case
of soft emissions.  It is often useful to note that 
\begin{eqnarray}
\begin{split}
\td\Pi_{i} &= \frac{2\alpha_{s}}{\pi }\frac{\td q_{i\, \bot}}{q_{i \, \bot}}\frac{\td z_{i}}{1- z_{i}}\frac{\td \phi_{i}}{2 \pi} = \frac{2\alpha_{s}}{\pi }\frac{\td q_{i\, \bot}}{q_{i \, \bot}}\frac{\td y \, \td \phi}{2 \pi},
\end{split}
\end{eqnarray} 
where $y$ is the rapidity in the frame defining $q_{i\, \bot}$. Using these
last two relations the link between real emissions and virtual corrections is
clear, i.e. the square of the emission operators $\int
\v{D}_{i}^\dagger\v{D}_{i} \, \td\Pi_i$ is, for $e^+e^-$
collisions\footnote{This caveat is necessary to avoid complications associated
  with emissions off coloured incoming legs.}, equal to minus twice the real
part of the exponent in \eqref{eqn:V_defs}.

Using the naive recoil prescription of \eqref{eq:r1} and
\eqref{eqn:SC1_recoil}, the array of parton momenta gets modified after a
collinear emission, generated by $\overline{\v{P}}_{ij}$, but not after a soft
emission (except to add one new soft gluon of course). Specifically, this
means acting with $\overline{\v{P}}_{ij}\mathfrak{R}^{\TT{coll}}_{ij}$ maps
$p_{j} \mapsto \tilde{p}_{j} = z_{i}p_{j} + \mathcal{O}(q_{\bot})$ (for final
state partons), and a parton with momentum $q_{i} = (1-z) p_{j} +
\mathcal{O}(q_{\bot})$ is added. As usual, $p_{j}$ is the momentum of parton
$j$ prior to the action of
$\overline{\v{P}}_{ij}\mathfrak{R}^{\TT{coll}}_{ij}$. A more careful treatment
of momenta is not required to reproduce the leading logarithms for many
observables. However, any observable dependent upon parton distribution
functions or fragmentation functions will be incorrectly calculated because
this naive recoil prescription does not reproduce DGLAP evolution. This is
because the terms with soft-collinear poles are handled in the `soft side' of
the algorithm and do not conserve longitudinal momentum. This manifests as
DGLAP evolution with an incorrect plus prescription, i.e.
\begin{eqnarray}
\left(\frac{1+z^{2}}{1-z}\right)_{+} = \left(\frac{2}{1-z}\right)_{+} -(1+z)_{+} \stackrel{\TT{variant} \, \TT{A}}{\longmapsto} -(1+z)_{+},
\end{eqnarray}
as the soft poles have been removed from the hard-collinear splitting
functions defining $\overline{\v{P}}_{ij}$. On the flip side, the algorithm
works well for event shape observables in $e^+ e^-$ collisions. We will refer
to the framework in this section as variant A of the algorithm. Within variant
A, this problem could be solved by implementing a universal recoil for all
emissions, soft and collinear, i.e.
\begin{align}\int \prod_{k} \td^{4} \, p_{k} \, \mathfrak{R}^{\TT{coll}}_{ij}\mathfrak{R}^{\TT{coll} \, *}_{ij} & = \int \prod_{k} \td^{4} p_{k} \, \sum_{j'} \, \mathfrak{R}^{\TT{soft}}_{ij}\mathfrak{R}^{\TT{soft} \, *}_{ij'}. \label{eq:unirec} 
\end{align}
We will not consider universal recoils in this paper and will instead solve
this `plus prescription problem' another way; by putting the soft-collinear
emissions in the collinear sector of the algorithm. Doing this will lead us to
variant B of the algorithm. In Section \ref{sec:SC2_spectator_recoil}, we will
use the insight gained from formulating B to show how to solve the
plus-prescription problem within the framework of A.

\subsection{Parton branching using complete collinear splitting functions (B)}
\label{sec:SC2}

Soft-collinear poles can be exchanged reasonably simply between eikonal
currents and collinear splitting functions. We will now define variant B of
our algorithm, which restores the soft-collinear poles in the collinear
splitting functions and removes them from the eikonal currents. This is a good
thing to do for two reasons:
\begin{enumerate}
	\item Collinear evolution is generated by unit operators in colour
          space. Making this manifest for the soft-collinear poles simplifies
          the colour evolution of the algorithm.
	\item Putting the soft-collinear poles into the collinear `side' of
          the algorithm simplifies the recoil prescription because every
          collinear emission has a uniquely identifiable parent.
\end{enumerate}
Variant B is very similar in form to variant A:
\begin{eqnarray}
\label{eqn:FKS2_defs1}
\begin{split}
\v{S}_{i} &= \sum_{j} \left( \frac{q^{(j\vec{m})}_{i \, \bot}}{2\tilde{p}_{j} \cdot q_{i}} \mathbb{T}^{g}_{j} \otimes (\tilde{p}_{j} \cdot \epsilon^{*}_{+}(q_{i})   \mathbb{S}^{1_{i}} +  \tilde{p}_{j} \cdot\epsilon^{*}_{-}(q_{i}) \mathbb{S}^{-1_{i}}) \right) \,  \mathfrak{R}^{\TT{soft}}_{ij}(\{p\},\{\tilde{p}\},q_{i}), \\ 
\v{C}_{i} &= \sum_{j} \frac{q^{(j\vec{n})}_{i\,\bot}}{2\sqrt{z_{i}}}\Delta_{ij}  \, \v{P}_{ij} \, \mathfrak{R}^{\TT{coll}}_{ij}(\{p\},\{\tilde{p}\},q_{i}), 
\end{split}
\end{eqnarray}
with
\begin{eqnarray}
\label{eqn:FKS2_defs2}
\begin{split}
...\v{D}_{i} \v{\mathcal{O}} \v{D}^{\dagger}_{i}... &= ...\v{S}_{i} \v{\mathcal{O}} \v{S}^{\dagger}_{i}  \,\mathfrak{f}_{jj'}(q_{i},\{p\},\{\tilde{p}\})... + ...\v{C}_{i} \v{\mathcal{O}} \v{C}^{\dagger}_{i}... \, .
\end{split}
\end{eqnarray}
The form of $\v{S}_{i}$ is the same as in A and the Sudakov changes as
\begin{align}
 \label{eqn:SC2_Sudakov}
\v{V}_{a,b} = &\Pexp \left[-\frac{\alpha_{s}}{\pi} \sum_{i<j} \int^{b}_{a} \frac{\td k^{(ij)}_{\bot}}{k^{(ij)}_{\bot}} (-\mathbb{T}^{g}_{i} \cdot \mathbb{T}^{g}_{j}) \left\{ \int \frac{\td y \, \td \phi}{4\pi} (k^{(ij)}_{\bot})^{2} \frac{\tilde{p}_{i}\cdot \tilde{p}_{j}}{(\tilde{p}_{i}\cdot k)(\tilde{p}_{j} \cdot k)}\theta_{ij}(k) \, \right. \right. \\
& \; \; \left. \left. \times \mathcal{F}_{ij}(k,\{ \tilde{p} \})- i\pi \tilde{\delta}_{ij} \right\}\mathcal{R}^{\TT{soft}}_{ij}  - \frac{\alpha_{s}}{\pi} \sum_{i} \int^{b}_{a} \frac{\td k^{(i\vec{n})}_{\bot}}{k^{(i\vec{n})}_{\bot}} \sum_{\upsilon} \mathbb{T}^{\bar{\upsilon} \, 2}_{i} \int \frac{\td z  \,\td \phi}{8\pi} \mathcal{P}^{\,\circ}_{\upsilon \upsilon_{i}} \, \theta_i(k) \, \mathcal{R}^{\TT{coll}}_{i} \right]. \nonumber
\end{align}
The only changes relative to variant A are the appearance of
$\mathfrak{f}_{jj'}(q_{i},\{p\},\{\tilde{p}\})$ and $\mathcal{F}_{ij}(k,\{
\tilde{p} \})$, which specify the prescription for the subtraction of
soft-collinear poles from the eikonal currents, and the replacement of
$\overline{\v{P}}_{ij}$ with $\v{P}_{ij}$. Explicit dependence on $\v{P}_{ij}$
in $\v{C}_{i}$ means that $\v{C}_{i}\mathcal{O}\v{C}^{\dagger}_{i}$ now
contains the full spin-dependent DGLAP splitting functions
\cite{HelicitySplitting}. Unitarity requires that
\begin{align}
\int \prod_{i}\td^{4} \, p_{i} \, \mathfrak{f}_{jj'}(q_{i},\{p\},\{\tilde{p}\}) \, \mathfrak{R}^{\TT{soft} \, *}_{ij}\mathfrak{R}^{\TT{soft}}_{ij'}= \mathcal{F}_{jj'}(q_{i},\{ \tilde{p} \}) \, \mathcal{R}^{\TT{soft}}_{ij}(q_{i},\{\tilde{p} \}). \label{eqn:unitary_link}
\end{align}
The functional forms of $\mathfrak{f}_{jj'}(q_{i},\{p\},\{\tilde{p}\})$ and
$\mathcal{F}_{jj'}(q_{i},\{ \tilde{p} \})$ are uniquely fixed by the choice of
$\mathfrak{R}^{\TT{coll}}_{ij}$ and $\mathfrak{R}^{\TT{soft}}_{ij}$ once we
have fixed $\v{P}_{ij}$.  Specifically, we can derive variant B from A by
adding and subtracting a function:
\begin{align}
\v{S}^{\TT{B}}_{i}\mathcal{O}\v{S}^{\TT{B} \, \dagger}_{i} \, \td \Pi_{i} + \v{C}^{\TT{B}}_{i}\mathcal{O}\v{C}^{\TT{B} \, \dagger}_{i} \, \td \Pi_{i} = \underbrace{\v{S}^{\TT{A}}_{i}\mathcal{O}\v{S}^{\TT{A} \, \dagger}_{i} \, \td \Pi_{i}  - \v{s}_{i}\mathcal{O}\v{s}^{\dagger}_{i}}_{\equiv \v{S}^{\TT{B}}_{i}\mathcal{O}\v{S}^{\TT{B} \, \dagger}_{i} \, \td \Pi_{i}} + \underbrace{\v{C}^{\TT{A} \, \dagger}_{i}\mathcal{O}\v{C}^{\TT{A}}_{i} \, \td \Pi_{i} + \v{s}_{i}\mathcal{O}\v{s}^{\dagger}_{i}}_{\equiv \v{C}^{\TT{B}}_{i}\mathcal{O}\v{C}^{\TT{B} \, \dagger}_{i} \, \td \Pi_{i}}, \label{eqn:term_grouping}
\end{align}
where we have labelled each operator with a superscript indicating which
variant it corresponds to and where $\mathcal{O}$ is some general operator in
colour and spin. The subtraction term was constructed so that
\begin{align}
\v{s}_{i}\mathcal{O}\v{s}^{\dagger}_{i} \equiv \sum_{j} \frac{(q^{(j\vec{n})}_{i\,\bot})^{2}}{4z_{i}} (\v{P}_{ij}\mathcal{O}\v{P}^{\dagger}_{ij}-\overline{\v{P}}_{ij}\mathcal{O}\overline{\v{P}}^{\dagger}_{ij}) \, \mathfrak{R}^{\TT{coll}}_{ij}\mathfrak{R}^{\TT{coll} \, *}_{ij} \, \td \Pi_{i} \label{eqn:subtermP}
\end{align}
and after some manipulation is equal to
\begin{align}
\v{s}_{i}\mathcal{O}\v{s}^{\dagger}_{i} = \frac{2\alpha_{s}}{\pi}\sum_{j} &\mathbb{T}^{g}_{j}\otimes\left( \frac{\mathbb{S}^{1_{i}}}{\< q_{i} \tilde{p}_{j} \>} + \frac{\mathbb{S}^{-1_{i}}}{[q_{i} \tilde{p}_{j}]} \right)\mathcal{O} \,\mathbb{T}^{g \, \dagger}_{j}\otimes\left( \frac{\mathbb{S}^{1_{i}}}{\< q_{i} \tilde{p}_{j} \>} + \frac{\mathbb{S}^{-1_{i}}}{[q_{i} \tilde{p}_{j}]} \right)^{\dagger} \nonumber \\
& \times q_{i} \cdot \tilde{p}_{j} \frac{\td q^{(j\vec{n})}_{i \, \bot}}{q^{(j\vec{n})}_{i \, \bot}} \frac{\td y \, \td \phi}{2\pi} \theta_{j}(q_{i}) \, ( \delta^{\TT{final}}_{j} +  \delta^{\TT{initial}}_{j}(\tilde{p}_{j} \leftrightarrow p_{j}))\, \mathfrak{R}^{\TT{coll}}_{ij}\mathfrak{R}^{\TT{coll} \, *}_{ij}. \label{eqn:subtraction_term}
\end{align}
 $\< q_{i} \tilde{p}_{j} \>$ and $[q_{i} \tilde{p}_{j}]$ are Weyl products in
the spinor-helicity formalism \cite{HelicityTechniques}. Also note,
$\v{s}_{i}\mathcal{O}\v{s}^{\dagger}_{i}$ is equal to the collinear limit of
$\v{S}^{\TT{A}}_{i}\mathcal{O}\v{S}^{\TT{A} \, \dagger}_{i} \, \td \Pi_{i}$
with $q^{(j\vec{m})}_{i \, \bot} \approx q^{(j\vec{n})}_{i \, \bot}$. To see
the equality we express polarisation vectors using spinor products,
\begin{align}
\epsilon^{\mu}(q_{i}, \pm 1) =  \frac{1}{\sqrt{2}}\frac{\<q_{i} \mp \rkl \sigma^{\mu}_{\mp} \lkl n \mp \>}{\<q_{i} \pm | n \mp\>},
\end{align}
where $\sigma^{\mu}_{\mp} =
(\mathbbm{1},\mp\sigma_{1},\mp\sigma_{2},\mp\sigma_{3})^{\TT{T}}$ are vectors
of Pauli matrices and $n$ is an auxillary light-like vector (best chosen to be
either $p_{j}$ or $p_{j'}$).

To complete the definition of variant B we must compute
$\mathfrak{f}_{jj'}(q_{i},\{p\},\{\tilde{p}\})$ and
$\mathcal{F}_{jj'}(q_{i},\{\tilde{p}\})$. Note that
$\v{s}^{\dagger}_{i}\v{s}_{i}$ is proportional to the unit operator in colour
and helicity. After taking the trace over helicity space,
\eqref{eqn:term_grouping} leads to
\begin{align}
& \sum_{j,j'}\mathbb{T}^{g}_{j}\mathcal{O}\mathbb{T}^{g \, \dagger}_{j'} \, \frac{\td q^{(jj')}_{i \, \bot}}{q^{(jj')}_{i \, \bot}} \frac{\td y \, \td \phi}{4\pi} \theta_{jj'}(q_{i}) \, \mathfrak{R}^{\TT{soft}}_{ij}\mathfrak{R}^{\TT{soft} \, *}_{ij'} \mathfrak{f}_{jj'}(q_{i},\{p\},\{\tilde{p}\}) = \nonumber \\
& \sum_{j,j'}\mathbb{T}^{g}_{j}\mathcal{O}\mathbb{T}^{g \, \dagger}_{j'} \, \frac{\td q^{(jj')}_{i \, \bot}}{q^{(jj')}_{i \, \bot}} \frac{\td y \, \td \phi}{4\pi} \theta_{jj'}(q_{i}) \, \mathfrak{R}^{\TT{soft}}_{ij}\mathfrak{R}^{\TT{soft} \, *}_{ij'} + \sum_{j}\mathbb{T}^{g}_{j}\mathcal{O}\mathbb{T}^{g \, \dagger}_{j} \, \frac{\td q^{(j\vec{n})}_{i \, \bot}}{q^{(j\vec{n})}_{i \, \bot}} \frac{\td y \, \td \phi}{4\pi} \theta_{j}(q_{i}) \, \mathfrak{R}^{\TT{coll}}_{ij}\mathfrak{R}^{\TT{coll} \, *}_{ij}.
\end{align}
We can use colour conservation to factorise the colour operators and simplify
the second term on the right-hand side, i.e.
\begin{align}
\mathfrak{f}_{jj'}(q_{i},\{p\},\{\tilde{p}\}) = 1 - \tdf{q^{(j\vec{n})}_{i \, \bot}}{q^{(jj')}_{i \, \bot}} \, \frac{q^{(jj')}_{i \, \bot}\theta_{j}(q_{i})}{q^{(j\vec{n})}_{i \, \bot}\theta_{jj'}(q_{i})} \, \frac{\mathfrak{R}^{\TT{coll} \, *}_{ij}\mathfrak{R}^{\TT{coll}}_{ij}}{\mathfrak{R}^{\TT{soft} \, *}_{ij}\mathfrak{R}^{\TT{soft}}_{ij'}} \label{eqn:Ffrak}
\end{align}
and
\begin{align}
\mathcal{F}_{jj'}(q_{i}, \{ p_{j} \}) = 1 - \tdf{q^{(j\vec{n})}_{i \, \bot}}{q^{(jj')}_{i \, \bot}} \, \frac{q^{(jj')}_{i \, \bot}\theta_{j}(q_{i})}{q^{(j\vec{n})}_{i \, \bot}\theta_{jj'}(q_{i})} \, \frac{\mathcal{R}^{\TT{coll}}_{j}}{\mathcal{R}^{\TT{soft}}_{jj'}}.  \label{eq:frakjj}
\end{align}

For a universal recoil it is possible to employ colour conservation and write
\begin{align}
\sum_{j}\mathbb{T}^{g}_{j}\mathcal{O}\mathbb{T}^{g \, \dagger}_{j} \, \frac{\td q^{(j\vec{n})}_{i \, \bot}}{q^{(j\vec{n})}_{i \, \bot}} \frac{\td y \, \td \phi}{4\pi} \theta_{j}(q_{i}) \, \mathfrak{R}^{\TT{coll}}_{ij}\mathfrak{R}^{\TT{coll} \, *}_{ij} = \sum_{j,j'}\mathbb{T}^{g}_{j}\mathcal{O}\mathbb{T}^{g \, \dagger}_{j'} \, \frac{\td q^{(j\vec{n})}_{i \, \bot}}{q^{(j\vec{n})}_{i \, \bot}} \frac{\td y \, \td \phi}{4\pi} \theta_{j}(q_{i}) \, \mathfrak{R}^{\TT{soft}}_{ij}\mathfrak{R}^{\TT{soft} \, *}_{ij'}, \label{eqn:recoil_cancel}
\end{align}
which enables us to re-write \eqref{eqn:Ffrak} as
\begin{align}
\mathfrak{f}_{jj'}(q_{i},\{p\},\{\tilde{p}\}) = 1 - \tdf{q^{(j\vec{n})}_{i \, \bot}}{q^{(jj')}_{i \, \bot}} \, \frac{q^{(jj')}_{i \, \bot}\theta_{j}(q_{i})}{q^{(j\vec{n})}_{i \, \bot}\theta_{jj'}(q_{i})}. \label{eqn:LLsub}
\end{align} 
In the case of a universal recoil prescription, the effects of the recoil can
be factorised out of the emission operators and into a redefinition of the
phase space measure. Recoil schemes that may be universal include the more
`true to Feynman diagrams' global prescriptions which put momenta of partons
higher in the chain of emissions off-shell (e.g. see \cite{Bewick:2019rbu} and
references therein) and schemes which democratically share recoil across a jet
or every parton in the shower. Depending on their implementation, such schemes
can be universal since they globally redistribute momentum across the whole
event as a $n \rightarrow n+1$ parton processes. We leave the specification a
universal recoil scheme to future work. For now we re-iterate that it is only
when considering effects beyond LL that \eqref{eqn:LLsub} and
\eqref{eqn:Ffrak} differ.

Our implementation of recoil is not unique, and it remains to be seen (by
performing analytic calculations at NLL and beyond) the extent to which we
will eventually be able to capture the salient sub-leading logarithms in the
framework of our algorithm. A slightly different approach would be to start
with variant B (recall we started from variant A above) and assume
\eqref{eqn:LLsub} holds true. Variant A could then be constructed but it would
now include subtraction functions akin to $\mathfrak{f}_{jj'}$. In the case of
universal recoils, none of this matters of course.

\subsection{Collinear subtractions and the ordering variable}
\label{sec:subtractionterm}

Before moving on, we'd like to present a slightly more general approach to
subtracting the soft-collinear contribution. This calculation will shed some
light on the role played by the ordering variable.  We start by
writing\footnote{We ignore ignore hard-collinear corrections and the effects
  of recoil in this section.}
\begin{equation}
\ln {\mathbf V}_{ab} = \frac{\alpha_s}{2\pi} \sum_{i<j}{\mathbb T}_i^{g}\cdot{\mathbb T}_j^{g}\int_{a^{2}}^{b^{2}} 
\frac{{\rm d}q^2}{q^2} 
\int\frac{{\rm d}^{3}k}{2 E} \frac{K^2(p_i,p_j;k)}{\pi} \frac{\ p_i\cdot p_j}{p_i\cdot k\, p_j \cdot
	k}\delta\left(q^2-K^2(p_i,p_j;k)\right) \, \theta_{ij}(k)~,
\end{equation}
which holds for a general definition of the ordering variable, $K^2(p_i,p_j,k)$.  In order
to isolate the collinear divergence, we should first
expose, and factor, the soft divergence. To do this, it is sufficient to consider any
scaling which is linear in the emitted gluon's momentum components,
such that we can re-write
\begin{equation}
\ln {\mathbf V}_{ab} =\frac{\alpha_s}{2\pi} \sum_{i<j}{\mathbb T}_i^{g}\cdot{\mathbb T}_j^{g}\int_{a^{2}}^{b^{2}}  
\frac{{\rm d}q^2}{q^2} 
\int\frac{{\rm d}^{3}k\ }{2 E} \frac{K^2(p_i,p_j;k)}{\pi \, (S\cdot k)^2} \frac{\ n_i\cdot n_j}{n_i\cdot n\ n_j\cdot
	n}\delta\left(q^2-K^2(p_i,p_j;k)\right) \theta_{ij}(k)  ,
\end{equation}
where $n_i = q_i/(S\cdot q_i)$, $n = k/(S\cdot k)$ and $S$ is any time-like
four-vector, which we choose to satisfy $S^2=1$. The soft divergence is now
isolated from the eikonal term, which is singular only in the collinear limits
$n_{i,j}\cdot n\to 0$. The collinear divergences can be subtracted. We want
the ordering variable to become independent of the other parton's direction in
the collinear limit, such that the entire collinear divergence can be moved
into a jet factor that is trivial in colour space.

We choose to re-write the virtual evolution as $\ln {\mathbf V}_{ab} =
\ln {\mathbf W}_{ab} + \ln{\mathbf K}_{ab}$, where
\begin{align}
\ln &{\mathbf W}_{ab} =\frac{\alpha_s}{2\pi} \sum_{i<j}{\mathbb T}_i^{g}\cdot{\mathbb T}_j^{g}\int_{a^{2}}^{b^{2}}  
\frac{{\rm d}q^2}{q^2} 
\int\frac{{\rm d}^{3}k\ }{2 E } \frac{1}{\pi \, (S\cdot k)^2} \nonumber \\ 
& \Bigg(  K^2(p_i,p_j;k)\frac{n_i\cdot n_j}{n_i\cdot n\ n\cdot
	n_j}\delta(q^2-K^2(p_i,p_j;k)) \, \theta_{ij}(k) \nonumber \\ &
- \frac{K^2(p_i;k)}{n_i\cdot n}\delta(q^2-K^2(p_i;k))\theta_i(k) - \frac{K^2(p_j;k)}{n_j\cdot n}\delta(q^2-K^2(p_j;k)) \theta_j(k) \Bigg) \ ,
\end{align}
and colour conservation can now be used to obtain
\begin{equation}
\ln {\mathbf K}_{ab} =\frac{\alpha_s}{2\pi} \sum_{i}({\mathbb
  T}_i^{g})^2\int_{a^{2}}^{b^{2}}  \frac{{\rm d}q^2}{q^2} \int\frac{{\rm
    d}^{3}k\ }{2 E } \frac{2}{\pi \, (S\cdot k)^2}\frac{ K^2(p_i;k)}{n_i\cdot
  n}\delta\left(q^2-K^2(p_i;k)\right) \theta_i(k) \ .
\end{equation}
This factor contains the ordering variable in terms
of a single emitter direction, which is the limiting case of
the dipole-type definition in each collinear limit, i.e. $K^2(p_i,p_j,k)\to
K^2(p_i;k)$ as $n_i\cdot n\to 0$.
Given the Lorentz invariance of the virtual evolution
and the integration measure we can choose $S=(1,\vec{0})$.

In the case of energy ordering, we obtain the following for the subtracted soft
evolution:   
\begin{align}
\ln {\mathbf W}_{ab}\Big|_{\text{energy}} & =\frac{\alpha_s}{\pi} \sum_{i<j}{\mathbb T}_i^{g}\cdot{\mathbb T}_j^{g}\int_a^b 
\frac{{\rm d}E}{E}
\int\frac{{\rm d}\Omega}{4\pi} \left(\frac{n_i\cdot n_j - n_i\cdot n - n_j\cdot n}{n_i\cdot n\ n\cdot n_j}\right) \nonumber \\
& = \frac{\alpha_s}{\pi} \sum_{i<j}{\mathbb T}_i^{g}\cdot{\mathbb T}_j^{g}\int_a^b 
\frac{{\rm d}E}{E} \; 
\ln \frac{n_i \cdot n_j}{2}~ \label{eqn:WE}
\end{align}
where the angular integral can be performed using the same integral that gives
rise to angular ordering. And for the collinearly divergent factor:
\begin{equation}
\ln {\mathbf K}_{ab}\Big|_{\text{energy}} =\frac{\alpha_s}{\pi} \sum_{i}({\mathbb T}_i^{g})^2\int_a^b 
\frac{{\rm d}E}{E}
\int\frac{{\rm d}\Omega}{4\pi} \frac{2}{n_i\cdot n} \ .
\end{equation}
There is no need for $\theta_{ij}$ since this simply enforces that the emitted
gluon should have energy smaller than $\sqrt{\tfrac{1}{2} p_i \cdot p_j}$ in
the $ij$ zero momentum frame, which is automatically satisfied since $a < E <
b$.

Now let us consider the case of transverse momentum ordering. This
can be implemented through
\begin{equation}
K^2(p_i,p_j;k) = (k_\perp^{(ij)})^2 = \frac{2\ p_i\cdot k\ k\cdot p_j}{p_i\cdot p_j} \ ,
\end{equation}
and
\begin{equation}
  K^2(p;k) \sim 2 p \cdot k
\end{equation}
where the similarity sign refers to any function which approaches unity in the
limit $p \cdot k\to 0$. Making the minimal choice, the full evolution becomes
\begin{equation}
\ln {\mathbf V}_{ab}\Big|_{k_T} =\frac{\alpha_s}{\pi} \sum_{i<j}{\mathbb T}_i^{g}\cdot{\mathbb T}_j^{g}\int_a^b 
\frac{{\rm d}k_\perp}{k_\perp} \int \frac{{\rm d}z}{1-z} \,\frac{ {\rm d}\phi}{2\pi} \;  \theta_{ij}(k)
\end{equation}
and 
\begin{align}
\ln {\mathbf K}_{ab}\Big|_{k_T} &= \frac{\alpha_s}{2\pi} \sum_{i}({\mathbb T}_j^{g})^2\int_a^b 
\frac{{\rm d}k_\perp}{k_{\perp}} \int_\alpha^1 \frac{\text{d} z}{1-z+\alpha} \, \int \frac{\text{d} \phi}{2\pi} \nonumber \\
& = \frac{\alpha_s}{2\pi} \sum_{i}({\mathbb T}_j^{g})^2\int_a^b 
\frac{\text{d}k_\perp}{k_{\perp}} \, \int_0^{1-\alpha} \frac{\text{d} z}{1-z} \, \int \frac{\text{d} \phi}{2\pi}
\end{align}
where $\alpha = k_\perp^2/(2 S \cdot p_i)^{2}$. This is the same as the
subtraction prescription we introduced in the last section, with the only
difference being that the lower limit on the $z$ integral is (approximately)
equal to $\alpha$ in that case. Finally, using colour conservation we can
compute $\ln {\mathbf V}_{ab} - \ln {\mathbf K}_{ab}$ and get
\begin{align}
\ln {\mathbf W}_{ab}\Big|_{k_T} &= \frac{\alpha_s}{\pi} \sum_{i<j}{\mathbb T}_i^{g}\cdot{\mathbb T}_j^{g}\int_a^b 
\frac{{\rm d} k_\perp}{k_\perp} \int \frac{ \td y \, {\rm d}\phi}{2\pi} \; ( \theta_{ij}(k) - \theta_{i}(k)) \, , \nonumber \\
& = \frac{\alpha_s}{\pi} \sum_{i<j}{\mathbb T}_i^{g}\cdot{\mathbb T}_j^{g}\int_a^b 
\frac{{\rm d} k_\perp}{k_\perp} \int \frac{ \td y \,  {\rm d}\phi}{2\pi} \; \mathcal{F}_{ij}(k) \, \theta_{ij}(k) \, , \nonumber \\
&\approx \frac{\alpha_s}{\pi} \sum_{i<j}{\mathbb T}_i^{g}\cdot{\mathbb T}_j^{g}\int_a^b 
\frac{{\rm d} k_\perp}{k_\perp} \ln \frac{n_{i}\cdot n_{j}}{2}~, \label{eqn:W}
\end{align}
where the second line illustrates the equivalence with the subtraction scheme
presented in the previous section (recall we are ignoring recoil in this
section). The approximately-equal-to sign is because we neglect terms
suppressed by powers of $k_\perp^2$. As expected, this finite term is the same
as the energy ordering case in equation \eqref{eqn:WE}. The form factor
$\exp(\ln \mathbf{W})$ captures all of the truly wide-angle soft-gluon physics
and is essentially the same as the fifth form factor introduced by Dokshitzer
\& Marchesini \cite{Dokshitzer:2005ek}.

\subsection{A local recoil prescription}
\label{sec:SC2_spectator_recoil}

Next we will show how a more sophisticated recoil prescription (than
\eqref{eq:r1} and \eqref{eqn:SC1_recoil}) can be implemented. The recoil we
choose is based on the one in \cite{Schumann:2007mg,Platzer:recoil}, but
extended to work with colour off-diagonal evolution. The dipole recoil is
itself based on Catani-Seymour dipole factorisation and furthers the work in
\cite{Catani:1996vz} so that recoil can be implemented in a dipole parton
shower. As a result, this recoil scheme shares similarities with the spectator
recoil prescriptions used in modern dipole showers such as \textsc{Pythia} and
\textsc{Dire} \cite{Pythia8,DIRE}. The idea is not to present a definitive
recoil prescription but rather to illustrate how one can be implemented in our
algorithm. To that end, we calculate $\mathcal{F}_{ij}$ and
$\mathfrak{R}^{\TT{coll}}_{ij}$. We also provide a short discussion on the
successes and limitations of the prescription. We then go on to show that, at
LL, the recoil prescription can be reduced to the naive recoil prescription
when implemented in variant B (but not variant A).

To keep things as simple as possible, we will consider the dipole recoil
scheme in the case of only coloured final-state partons. The extension to
coloured initial-state partons is straightforward and can be found by
following Section 3.2 of \cite{Platzer:recoil}. First we will summarise the
dipole recoil for colour-diagonal evolution. It works by adding a spectator
particle to the standard description of a $1 \rightarrow 2$ collinear
splitting ($p_{j}\rightarrow \tilde{p}_{j}, q_{i}$). This spectator particle
absorbs the recoil from the splitting, which would otherwise put $p_{j}$
off-shell. The spectator particle has a second function: to specify the frame
in which the transverse momentum of the emission is computed. In
\cite{Platzer:recoil} it was shown that one can obtain the correct
colour-diagonal evolution by choosing the parton that is colour connected to
parton $j$ as the spectator. We will denote the momentum of the spectator
parton by $p_{j^{\TT{LR}}}$ (the reason for the LR subscript will become
clear). The Sudakov decomposition is
\begin{eqnarray}
\begin{split}
\tilde{p}_{j} &= z_{i} p_{j}-k_{\bot}+\frac{(q^{(jj^{\TT{LR}})}_{\bot})^{2}}{z_{i}}\frac{p_{j^{\TT{LR}}}}{2p_{j}\cdot p_{j^{\TT{LR}}}}, \qquad (q^{(jj^{\TT{LR}})}_{\bot})^{2} = - k_{\bot}^{2},\\
q_{i} &= (1-z_{i}) p_{j}+k_{\bot}+\frac{(q^{(jj^{\TT{LR}})}_{\bot})^{2}}{1-z_{i}}\frac{p_{j^{\TT{LR}}}}{2p_{j}\cdot p_{j^{\TT{LR}}}}, \\
\tilde{p}_{j^{\TT{LR}}} &= \left(1-\frac{(q^{(jj^{\TT{LR}})}_{\bot})^{2}}{z_{i}(1-z_{i})}\frac{1}{2p_{j}\cdot p_{j^{\TT{LR}}}}\right)p_{j^{\TT{LR}}}, \qquad k_{\bot} \cdot p_{j} = k_{\bot} \cdot p_{j^{\TT{LR}}}=0.
\end{split} \label{eqn:2_to_3_Sudakov_decomposition}
\end{eqnarray}
This now defines a $2\rightarrow 3$ splitting
($p_{j},p_{j^{\TT{LR}}}\rightarrow q_{i}, \tilde{p}_{j},
\tilde{p}_{j^{\TT{LR}}}$) in which $q_{i}$ is emitted collinear to
$p_{j}$. The prescription is momentum conserving, i.e. $$ p_{j} +
p_{j^{\TT{LR}}} = q_{i} + \tilde{p}_{j} + \tilde{p}_{j^{\TT{LR}}}$$ and it
ensures that all particles are on-shell at each stage in the evolution.  One
can check that this Sudakov decomposition does not change the functional form
of the leading-order collinear splitting functions. Comparing to
\eqref{eqn:momentum_fraction}, we see that $p$ and $n$ are now fixed:
$p=p_{j}$ and $n=p_{j^{\TT{LR}}}$. Working in the LC approximation, the effect
of this prescription amounts to a correction to the single-particle emission
phase space \cite{Platzer:recoil}, i.e.
\begin{eqnarray}
\td \sigma(q_{i}, \tilde{p}_{j}, \tilde{p}_{j^{\TT{LR}}}) = \frac{\alpha_{s}}{2\pi} \, \td \sigma(p_{j},p_{j^{\TT{LR}}})\frac{\td q^{(jj^{\TT{LR}})}_{\bot}}{q^{(jj^{\TT{LR}})}_{\bot}} \td z_{i}  \, \mathcal{P}_{\upsilon_{i}\upsilon_{j}}(z_{i}) \, \left(1-\frac{(q^{(jj^{\TT{LR}})}_{\bot})^{2}}{z_{i}(1-z_{i})}\frac{1}{2p_{j}\cdot p_{j^{\TT{LR}}}}\right). ~~~~
\end{eqnarray}
This correction contributes soft-collinear NLLs and hard-collinear NNLLs \cite{Platzer:recoil}. 

The dipole recoil prescription was developed for a leading $N_c$ shower and as
such is not completely sufficient for our purposes. That is because, beyond
the LC approximation, the left evolution (of the amplitude) and right
evolution (of the conjugate amplitude) are independent, which means they can
evolve to produce colour off-diagonal terms. These are terms for which the
parton $j$ is colour connected to different partons in the left and right
evolution. In such a case $p_{j^{\TT{LR}}}$ cannot be defined. Instead, we
must introduce parton $p_{j^{\TT{L}}}$, which is the colour connected parton
to $j$ in the left evolution, and parton $p_{j^{\TT{R}}}$, which is colour
connected to $j$ in the right evolution. We will now construct a recoil
prescription that extends the dipole recoil to include colour off-diagonal
terms but collapses back to the dipole recoil in the LC approximation.

To begin, we define a Sudakov decomposition for a $3 \rightarrow 4$ splitting
($p_{j}, p_{j^{\TT{L}}}, p_{j^{\TT{R}}} \rightarrow q_{i}, \tilde{p}_{j},
\tilde{p}_{j^{\TT{L}}}, \tilde{p}_{j^{\TT{R}}}$). We aim to construct the
decomposition so that recoil is shared equally between $p_{j^{\TT{L}}}$ and
$p_{j^{\TT{R}}}$. We also wish to leave the collinear splitting functions
unchanged. Finally, we also require all partons involved to be on-shell. These
constraints are fulfilled by the decomposition:
\begin{eqnarray}
\begin{split}
\tilde{p}_{j} &= z_{i} p_{j}-k_{\bot}+\frac{(q^{(j\vec{n})}_{\bot})^{2}}{z_{i}}\dfrac{n}{2p_{j}\cdot n}, \\
q_{i} &= (1-z_{i}) p_{j}+k_{\bot}+\frac{(q^{(j\vec{n})}_{\bot})^{2}}{1-z_{i}}\dfrac{n}{2p_{j}\cdot n}, \\
n&= p_{j^{\TT{L}}} + p_{j^{\TT{R}}}(1 - \delta_{j^{\TT{L}}, j^{\TT{R}}}) - \sqrt{\frac{2p_{j^{\TT{L}}} \cdot p_{j^{\TT{R}}}}{\hat{n}^{2}}}\hat{n}, \\
\tilde{p}_{j^{\TT{L}}} &= \left( 1 - \gamma \right)p_{j^{\TT{L}}} + \gamma \sqrt{\frac{p_{j^{\TT{L}}} \cdot p_{j^{\TT{R}}}}{2\hat{n}^{2}}}\hat{n} + \gamma l(1 - \delta_{j^{\TT{L}}, j^{\TT{R}}}), \\
\tilde{p}_{j^{\TT{R}}} &= \left( 1 -  \gamma \right)p_{j^{\TT{R}}} + \gamma \sqrt{\frac{p_{j^{\TT{L}}} \cdot p_{j^{\TT{R}}}}{2\hat{n}^{2}}}\hat{n} - \gamma l(1 - \delta_{j^{\TT{L}}, j^{\TT{R}}}), \\
\end{split} \label{eqn:3_to_4_Sudakov_decomposition}
\end{eqnarray}
where
\begin{eqnarray}
\begin{split}
&(q^{(j\vec{n})}_{\bot})^{2} = - k_{\bot}^{2}, \qquad 2q \cdot \tilde{p}_{j} =\dfrac{(q^{(j\vec{n})}_{\bot})^{2}}{z_{i}(1-z_{i})}, \qquad \hat{n} \cdot p_{j^{\TT{L}}} = \hat{n} \cdot p_{j^{\TT{R}}} = 0, \\
& \gamma = \frac{(q^{(j\vec{n})}_{\bot})^{2}}{z_{i}(1-z_{i})}\dfrac{1}{2p_{j}\cdot n}, \qquad k_{\bot} \cdot p_{j} = k_{\bot} \cdot n=0,\\
&l^{2} = \frac{p_{j^{\TT{L}}} \cdot p_{j^{\TT{R}}}}{2}, \qquad l \cdot \left( (1-\gamma)(p_{j^{\TT{L}}}+p_{j^{\TT{R}}}) +\gamma\sqrt{\frac{p_{j^{\TT{L}}} \cdot p_{j^{\TT{R}}}}{2\hat{n}^{2}}}\hat{n} \right) = 0,
\end{split} \label{eqn:3_to_4_definitions}
\end{eqnarray}
where $\delta_{j^{\TT{L}}, j^{\TT{R}}}$ is the usual Kronecker delta
symbol. Note that $p_{j} + p_{j^{\TT{L}}} +p_{j^{\TT{R}}} = q_{i} +
\tilde{p}_{j} + \tilde{p}_{j^{\TT{L}}} + \tilde{p}_{j^{\TT{R}}}$, and so
momentum is conserved in the $3 \rightarrow 4$ splitting. Also note that when
${j^{\TT{L}}} = {j^{\TT{R}}} = {j^{\TT{LR}}}$, i.e. the emission is
colour-diagonal, this reduces to the dipole $2 \rightarrow 3$ scattering with
$p_{j} + p_{j^{\TT{LR}}} = q_{i} + \tilde{p}_{j} +
\tilde{p}_{j^{\TT{LR}}}$. Finally, note that every new term relative the
dipole recoil is accompanied by a factor $\gamma$, which is two orders higher
than the leading terms in the collinear limit and one order higher in the soft
limit.  Using this decomposition, the recoil prescription for collinear
emissions is
\begin{eqnarray}
\begin{split}
\mathfrak{R}^{\TT{coll} \, *}_{ij}\mathfrak{R}^{\TT{coll}}_{ij} = \; & z_{i} (\mathcal{J}_{ij}(z_{i},q^{(j\vec{n})}_{i \, \bot},p_{j^{\TT{L}}}, p_{j^{\TT{R}}}, l, \hat{n}))^2 \, \delta^{4}\left(\tilde{p}_{j} - z_{i} p_{j}+k_{\bot}-\frac{(q^{(j\vec{n})}_{\bot})^{2}}{z_{i}}\dfrac{n}{2p_{j}\cdot n}\right) \\
&\times \delta^{4}\left(\tilde{p}_{j^{\TT{L}}} - \left( 1 - \gamma \right)p_{j^{\TT{L}}} - \gamma \sqrt{\frac{p_{j^{\TT{L}}} \cdot p_{j^{\TT{R}}}}{2\hat{n}^{2}}}\hat{n} - \gamma l(1 - \delta_{j^{\TT{L}}, j^{\TT{R}}})\right)\\
& \times \delta^{4}\left(\tilde{p}_{j^{\TT{R}}} - \left( 1 - \gamma \right)p_{j^{\TT{R}}} - \gamma \sqrt{\frac{p_{j^{\TT{L}}} \cdot p_{j^{\TT{R}}}}{2\hat{n}^{2}}}\hat{n} + \gamma l(1 - \delta_{j^{\TT{L}}, j^{\TT{R}}})\right)\prod_{k \neq j, j^{\TT{L}}, j^{\TT{R}}} \delta^{4}(p_{k}-\tilde{p}_{k}),
\end{split}
\end{eqnarray}
where the Jacobian, $\mathcal{J}_{ij}$, can (in principle) be evaluated using
\begin{align}
\mathcal{J}_{ij} &= \left[\int \td^{4} p'_{j^{\TT{L}}} \td^{4} p'_{j^{\TT{R}}}\delta^{4}\left(\tilde{p}_{j^{\TT{L}}} - \left( 1 - \gamma' \right)p'_{j^{\TT{L}}} - \gamma' \sqrt{\frac{p'_{j^{\TT{L}}} \cdot p'_{j^{\TT{R}}}}{2\hat{n}^{'2}}}\hat{n}' - \gamma' l(1 - \delta_{j^{\TT{L}}, j^{\TT{R}}}) \right)\right. \nonumber \\
& \times \left. \delta^{4}\left(\tilde{p}_{j^{\TT{R}}} - \left( 1 - \gamma' \right)p'_{j^{\TT{R}}} - \gamma' \sqrt{\frac{p'_{j^{\TT{L}}} \cdot p'_{j^{\TT{R}}}}{2\hat{n}^{'2}}}\hat{n}' + \gamma' l(1 - \delta_{j^{\TT{L}}, j^{\TT{R}}}) \right) \right]^{-1} + \mathcal{O} \left( \left( q^{(j\vec{n})}_{i \, \bot}/Q \right)^{3}  \right) , \nonumber \\
& = 1 + \mathcal{O}(\gamma)~.
\end{align}
One factor of $z_{i} \mathcal{J}_{ij}$ ensures that the integral over the
delta functions is correctly normalised whilst the additional factor of
$\mathcal{J}_{ij}$ encodes the recoil corrections. This is the factor that was
absorbed into the phase-space in \cite{Platzer:recoil}, i.e.
\begin{eqnarray}
\begin{split}
\td \sigma(q_{i}, \tilde{p}_{j}, \tilde{p}_{j^{\TT{L}}}, \tilde{p}_{j^{\TT{R}}}) =& \frac{\alpha_{s}}{2\pi} \, \td \sigma(p_{j}, p_{j^{\TT{L}}}, p_{j^{\TT{R}}})\frac{\td q^{(j\vec{n})}_{i \, \bot}}{q^{(j\vec{n})}_{i \, \bot}} \td z_{i} \, \mathcal{P}_{\upsilon_{i}\upsilon_{j}}(z_{i}) \, \mathcal{J}_{ij}. ~
\end{split}
\end{eqnarray}
We can extend this prescription to the soft sector using Catani-Seymour dipole
factorisation \cite{Catani:1996vz}. The dipole factorisation provides a unique
way to split the parent dipole of a soft emission into two halves,
identifiable by their separate collinear poles:
$$
\frac{\tilde{p}_{j'}\cdot \tilde{p}_{j}}{(q_{i} \cdot \tilde{p}_{j'})(\tilde{p}_{j} \cdot q_{i})} = \frac{\tilde{p}_{j'}\cdot \tilde{p}_{j}}{q_{i}\cdot(\tilde{p}_{j'}+\tilde{p}_{j})\tilde{p}_{j} \cdot q_{i}} + \frac{\tilde{p}_{j'}\cdot \tilde{p}_{j}}{\tilde{p}_{j'}\cdot q_{i}(\tilde{p}_{j'}+\tilde{p}_{j}) \cdot q_{i}}.
$$
This provides the means to implement a local recoil using the parton contributing to the collinear pole in each half of the dipole. Thus we can write
\begin{eqnarray}
\begin{split}
\mathfrak{R}^{\TT{soft} \, *}_{ij'}\mathfrak{R}^{\TT{soft}}_{ij} = \frac{(q^{(jj')}_{i \, \bot})^{2}\tilde{p}_{j'}\cdot \tilde{p}_{j}}{2 q_{i}\cdot(\tilde{p}_{j'}+\tilde{p}_{j})(\tilde{p}_{j} \cdot q_{i})}\mathfrak{R}^{\TT{coll} \, *}_{ij}\mathfrak{R}^{\TT{coll}}_{ij} + \frac{(q^{(jj')}_{i \, \bot})^{2}\tilde{p}_{j'}\cdot \tilde{p}_{j}}{2(\tilde{p}_{j'}\cdot q_{i})(\tilde{p}_{j'}+\tilde{p}_{j}) \cdot q_{i}}\mathfrak{R}^{\TT{coll} \, *}_{ij'}\mathfrak{R}^{\TT{coll}}_{ij'}. ~~~
\end{split} \label{eqn:soft_recoil}
\end{eqnarray}
From this, $\mathcal{R}^{\TT{soft}}_{ij}$ and $\mathcal{R}^{\TT{coll}}_{i}$ can be evaluated:
\begin{eqnarray}
\begin{split}
\mathcal{R}^{\TT{soft}}_{jj'}(q_{i}, \{p\}) = \frac{(q^{(jj')}_{i \, \bot})^{2}p_{j'}\cdot p_{j}}{2 q_{i}\cdot(p_{j'}+p_{j})(p_{j} \cdot q_{i})}\mathcal{J}_{ij} + (j \leftrightarrow j '), \qquad \mathcal{R}^{\TT{coll}}_{j}(q_{i}, \{p\}) = \mathcal{J}_{ij}.
\end{split}
\end{eqnarray}
We can now go ahead and determine the subtraction functions used to define variant B. Using \eqref{eqn:Ffrak} and \eqref{eq:frakjj} we get
\begin{align}
\mathcal{F}_{jj'}(q_{i},\{p\}) = 1 - \tdf{q^{(j\vec{n})}_{i \, \bot}}{q^{(jj')}_{i \, \bot}} \, \frac{q^{(jj')}_{i \, \bot}\theta_{j}(q_{i})}{q^{(j\vec{n})}_{i \, \bot}\theta_{jj'}(q_{i})}\frac{\mathcal{J}_{ij}}{\mathcal{R}^{\TT{soft}}_{jj'}}.
\end{align}
Before moving on we want to comment on the discussion in
\cite{Dasgupta:2018nvj}, which shows that the dipole recoil scheme, as
implemented in a dipole shower, fails at the level of the NLL even at LC due
to incorrectly assigning the longitudinal recoil after multiple emissions. It
remains to be seen whether this is also true in the scheme discussed here.

\subsubsection{A LL recoil prescription}
\label{sec:LLrecoil}
We can now consider constructing a recoil prescription where we only keep the parts contributing at LL. Firstly note that in the strictly LL soft limit $$\mathfrak{R}^{\TT{coll} \, *}_{ij}\mathfrak{R}^{\TT{coll}}_{ij} = \mathfrak{R}^{\TT{soft} \, *}_{ij'}\mathfrak{R}^{\TT{soft}}_{ij} = \prod_{k} \delta^{4}(p_{k}-\tilde{p}_{k}).$$ We can use this fact with variant B restricted to LL accuracy and find
\begin{align}
\mathfrak{f}_{jj'}(q_{i},\{p\},\{\tilde{p}\}) = \mathcal{F}_{jj'}(q_{i},p_{j},p_{j'}) = 1 - \tdf{q^{(j\vec{n})}_{i \, \bot}}{q^{(jj')}_{i \, \bot}} \, \frac{q^{(jj')}_{i \, \bot}\theta_{j}(q_{i})}{q^{(j\vec{n})}_{i \, \bot}\theta_{jj'}(q_{i})}. \label{eqn:FfrakLL}
\end{align}
Using the naive recoil prescription,
\begin{eqnarray}
&\mathfrak{R}^{\TT{coll} \, *}_{ij}\mathfrak{R}^{\TT{coll}}_{ij} = \left(\delta^{4}(p_{j}-z^{-1}_{i}\tilde{p}_{j})\delta^{\TT{final}}_{j} + \delta^{4}(p_{j}-z_{i}\tilde{p}_{j})\delta^{\TT{initial}}_{j}\right)\prod_{k \neq j} \delta^{4}(p_{k}-\tilde{p}_{k}), \nonumber\\
&\mathfrak{R}^{\TT{soft} \, *}_{ij'}\mathfrak{R}^{\TT{soft}}_{ij} = \prod_{k} \delta^{4}(p_{k}-\tilde{p}_{k}), \qquad \mathcal{R}^{\TT{soft}}_{ij} = \mathcal{R}^{\TT{coll}}_{i} = 1,
\end{eqnarray}
with \eqref{eqn:FfrakLL}, variant B fully captures the correct DGLAP evolution
as longitudinal recoil is now included in the soft-collinear region. This is
not the case for variant A. We stress that this observation is not a
reflection of any fundamental difference between A and B since, with a
complete recoil prescription, A and B are equivalent.  Indeed, we can use the
Catani-Seymour dipole factorisation \cite{Catani:1996vz}, as previously
discussed, to extend the naive recoil prescription so that it does generate
longitudinal recoil with variant A, i.e.
\begin{eqnarray}
\begin{split}
&\mathfrak{R}^{\TT{coll} \, *}_{ij}\mathfrak{R}^{\TT{coll}}_{ij} = \left(\delta^{4}(p_{j}-z^{-1}_{i}\tilde{p}_{j})\delta^{\TT{final}}_{j} + \delta^{4}(p_{j}-z_{i}\tilde{p}_{j})\delta^{\TT{initial}}_{j}\right)\prod_{k \neq j} \delta^{4}(p_{k}-\tilde{p}_{k}), \\
&
\mathfrak{R}^{\TT{soft} \, *}_{ij'}\mathfrak{R}^{\TT{soft}}_{ij} = \frac{(q^{(jj')}_{i \, \bot})^{2}\tilde{p}_{j'}\cdot \tilde{p}_{j}}{2 q_{i}\cdot(\tilde{p}_{j'}+\tilde{p}_{j})(\tilde{p}_{j} \cdot q_{i})}\mathfrak{R}^{\TT{coll} \, *}_{ij}\mathfrak{R}^{\TT{coll}}_{ij} + \frac{(q^{(jj')}_{i \, \bot})^{2}\tilde{p}_{j'}\cdot \tilde{p}_{j}}{2(\tilde{p}_{j'}\cdot q_{i})(\tilde{p}_{j'}+\tilde{p}_{j}) \cdot q_{i}}\mathfrak{R}^{\TT{coll} \, *}_{ij'}\mathfrak{R}^{\TT{coll}}_{ij'}. ~~~
\end{split}
\end{eqnarray}
With this, variant A also captures the correct DGLAP evolution. 

\subsection{A manifestly infra-red finite reformulation}
\label{sec:algorithmIR}
It is possible to re-cast both variants of our algorithm such that the IR
divergences, other than those renormalised into parton distribution and
fragmentation functions, explicitly cancel at each iteration of the
algorithm. We will demonstrate this for variant A, though the procedure is
pretty much identical for variant B. Our method closely follows that in
\cite{SoftEvolutionAlgorithm}.

We begin by expressing a generalised measurement function in the soft and collinear limits as follows
\begin{equation}
\begin{split}
u_{m}(q_{1}, ... , q_{m}) \stackrel{q_{j} \, \TT{soft}}{=} u(q_{j} , \{q_{1},... , q_{j-1}, q_{j+1}, ... ,q_{m} \}) u_{m-1}(q_{1},... , q_{j-1}, q_{j+1}, ... , q_{m}),
\end{split}
\end{equation}
and
\begin{equation}
\begin{split}
u_{m}(q_{1}, ... , q_{m}) \stackrel{q_{j} || q_{i}}{=} u(q_{j}, \{q_{1},... , q_{i} + q_{j}, ... ,q_{m} \}) u_{m-1}(q_{1},... , q_{i} + q_{j}, ... , q_{m}),
\end{split}
\end{equation}
where $u(q_{j}, \{q\} ) \rightarrow 1$ as $j$ becomes exactly soft or
collinear. For many observables, it is possible to further pull apart the
measurement function by defining an `out' region, where there is total
inclusivity over radiation. For such observables we can write
\begin{equation}
\begin{split}
u(q_{j} , \{q\}) = \Theta_{\TT{out}}(q_{j})+\Theta_{\TT{in}}(q_{j})u_{\TT{in}}(q_{j}, \{q\}),
\end{split}
\end{equation}
where $\Theta_{\TT{in} / \TT{out}}(q_{j})=1$ when $q_{j}$ is in the in/out
region and zero otherwise. For a global observable, the out region has zero
extent and so $\Theta_{ \TT{out}}(q_{j}) = 0$. First we define
\begin{equation}
\begin{split}
\v{\Gamma} &= \v{\Gamma}_{u} + \overline{\v{\Gamma}}_{u}, \\
\overline{\v{\Gamma}}_{u} &= \int \frac{\td S_{2} }{4\pi}  (1-u(k, \{q\}))\tfrac{1}{2}\v{D}^{2}_{k} + \sum_{i<j}\mathbb{T}^{g}_{i} \cdot \mathbb{T}^{g}_{j}  \, i\pi \, \tilde{\delta}_{ij}, \qquad \td S_{2} \, = \, \td y \, \td \phi = \frac{\td z \, \td \phi}{(1-z)}, \\
\tfrac{1}{2}\v{D}^{2}_{k} &= \sum_{i<j}(-\mathbb{T}^{g}_{i} \cdot \mathbb{T}^{g}_{j}) (k^{(ij)}_{\bot})^{2}\frac{p_{i}\cdot p_{j}}{(p_{i}\cdot k)(p_{j} \cdot k)}\theta_{ij}(k) \, \mathcal{R}^{\TT{soft}}_{ij} + \frac{(1-z)}{2} \sum_{i,\upsilon} \mathbb{T}^{\bar{\upsilon} \, 2}_{i} \, \overline{\mathcal{P}}^{\,\circ}_{\upsilon \upsilon_{i}} \, \theta_i(k) \, \mathcal{R}^{\TT{coll}}_{i}, \\
\overline{\v{V}}_{a,b} &= \Pexp\left(- \frac{\alpha_{s}}{\pi} \int^{b}_{a} \frac{\td k_{\bot}}{k_{\bot}} \, \overline{\v{\Gamma}}_{u} \right).
\end{split}
\end{equation}
After a simple path-ordered operator expansion,
\begin{equation}
\begin{split}
\v{V}_{a,b} =& \overline{\v{V}}_{a,b} - \int \td\Pi_{1} \, u(k_{1}, \{q\}) \, \overline{\v{V}}_{a,k_{1\, \bot}} \tfrac{1}{2}\v{D}^{2}_{1} \, \overline{\v{V}}_{k_{1 \, \bot},b}\\
+ & \int \td\Pi_{1} \td\Pi_{2} \, \Theta(k_{1 \, \bot} - k_{2 \, \bot}) \, u(k_{1}, \{q\}) u(k_{2}, \{q\}) \, \overline{\v{V}}_{a,k_{2 \, \bot}} \tfrac{1}{2}\v{D}^{2}_{2} \, \overline{\v{V}}_{k_{2 \, \bot},k_{1 \, \bot}} \tfrac{1}{2}\v{D}^{2}_{1} \,  \overline{\v{V}}_{k_{1 \, \bot},b} - ...
\end{split}
\end{equation}
the observable, $\Sigma$, can be expressed as
\begin{eqnarray}
\begin{split}
\Sigma(\mu) = \int \sum_{n} \left(\prod^{n}_{i = 1} \td \Pi_{i} \right) \Tr \, \v{B}_{n}(\mu; \{p\}_{n}) \, \Phi_{n}(q_{1},...,q_{n}),
\end{split} \label{eq:nglsum}
\end{eqnarray}
where
\begin{eqnarray}
\begin{split}
&\v{B}_{n}(q_{\bot}; \{\tilde{p}\}_{n-1} \cup q_{n})  \\
& \qquad = \overline{\v{V}}_{q_{\bot},q_{n \, \bot}}\left[\int  \prod_{i} \td^{4} p_{i} \, \delta^{R}_{n} \, \v{D}_{n}\v{B}_{n-1}(q_{n \, \bot}; \{p\}_{n-1}) \v{D}^{\dagger}_{n} \right. \\
& \qquad\quad \left. - \delta^{V}_{n}\left\{ \v{B}_{n-1}(q_{n \, \bot}; \{\tilde{p}\}_{n-1}), \frac{1}{2}\v{D}^{2}_{n} \right\} u(q_{n},  \{\tilde{p}\}_{n-1}) \ \right]\overline{\v{V}}^{\dagger}_{q_{\bot},q_{n \, \bot}}\Theta(q_{n \, \bot} - q_{\bot}),
\end{split}
\end{eqnarray}
with $\v{B}_{0}(q_{\bot}) =
\overline{\v{V}}_{q_{\bot},Q}\v{H}(Q)\overline{\v{V}}^{\dagger}_{q_{\bot},Q}$. We
define $\delta^{R}_{n}=1$ when parton $n$ is real and $\delta^{R}_{n}=0$ when
parton $n$ is virtual, and similarly $\delta^{V}_{n}=1 -
\delta^{R}_{n}$. $\Phi_{n}(q_{1},...,q_{n})$ is a measurement function on the
phase-space of real particles. We refer to \cite{SoftEvolutionAlgorithm} for
its precise definition and here just present an illustrative example:
\begin{eqnarray}
\begin{split}
&( \delta^{R}_{3}\delta^{V}_{2}\delta^{V}_{1}+\delta^{V}_{3}\delta^{R}_{2}\delta^{R}_{1}+\delta^{R}_{3}\delta^{V}_{2}\delta^{R}_{1} + \delta^{V}_{3}\delta^{V}_{2}\delta^{V}_{1})\Phi_{3}(q_{1},q_{2},q_{3}) \\ &\qqquad =  \delta^{R}_{3}\delta^{V}_{2}\delta^{V}_{1}u_{1}(q_{3})+\delta^{V}_{3}\delta^{R}_{2}\delta^{R}_{1}u_{2}(q_{1},q_{2})+\delta^{R}_{3}\delta^{V}_{2}\delta^{R}_{1}u_{2}(q_{1},q_{3}) + \delta^{V}_{3}\delta^{V}_{2}\delta^{V}_{1}.
\end{split}
\end{eqnarray}

Written in this form, each $\v{B}_{n}$ is explicitly infra-red finite provided
the measurement function is infra-red-collinear safe, and that the evolution
is not convoluted with parton distribution or fragmentation functions. In this
case $\mu$ can be safely taken to zero. In the case that the evolution is
convoluted with parton distribution or fragmentation functions, collinear
poles remain (one for each hadron). These poles are removed by renormalisation
of the parton distribution functions or fragmentation functions, generating
their $\mu$ dependence.  Finally, note that for recursively-infra-red-safe continuously-global observables \cite{Banfi:2004yd} in $e^+ e^-$
collisions $\v{B}_{n}=0$ for $n \ge 1$ at single-log accuracy. If the observable depends on fragmentation functions the $n \ge 1$ contributions give rise to the DGLAP evolution of the fragmentation
functions (see Section \ref{sec:Coll}).

\section{Collinear factorisation}
\label{sec:Coll}
Up to this point we have been interleaving the emission of soft and collinear partons to build up the complete amplitude. As is well known, it is possible to factorise the collinear emissions into the evolution of parton distribution and fragmentation functions. In this section, our aim is to explore collinear factorisation within the context of our algorithm. 

The plan is as follows. First, we will derive the factorisation of collinear physics into jet functions; one for each parton in the initial hard process. At first we do this ignoring the presence of Coulomb exchanges. This result is sufficient to derive DGLAP evolution  (which we do in Section \ref{sec:pheno}). After this, we go on to construct the complete factorisation of collinear physics into jet functions on every hard or soft leg. Finally, we use a path-ordered expansion of our Sudakov operators, $\v{V}_{a,b}$, to re-insert Coulomb exchanges one-by-one into the previous results. The result is that collinear evolution below the scale of the last Coulomb exchange can be factorised. The outcome of which is the general factorisation of collinear poles into parton distribution functions, as anticipated after the work of Collins, Soper and Sterman \cite{Collins:1988ig}. 

We provide diagrammatic proofs where possible and only sketch in the text the algebra that is going on behind the scenes. Throughout this section we leave aside the recoil functions $\mathfrak{R}^{\TT{soft}}$, $\mathfrak{R}^{\TT{coll}}$, and the integrals corresponding to the momentum maps between each iteration of the algorithm. This is to reduce the length of the algebra that remains, and it is certainly valid to LL accuracy since tracking longitudinal recoil is sufficiently simple. We also drop the inclusion of measurement functions since they too have no affect on the discussion. That said, in Sections \ref{sec:SC1f} and \ref{sec:SC2f+}, we present a summary of the results with all of these functions re-instated (for both of variants A and B).
 
\subsection{Factorisation on hard legs without Coulomb interactions}
\label{sec:hard_factorisation}

\begin{figure}[h]
	\centering
	\includegraphics[width=0.9\textwidth]{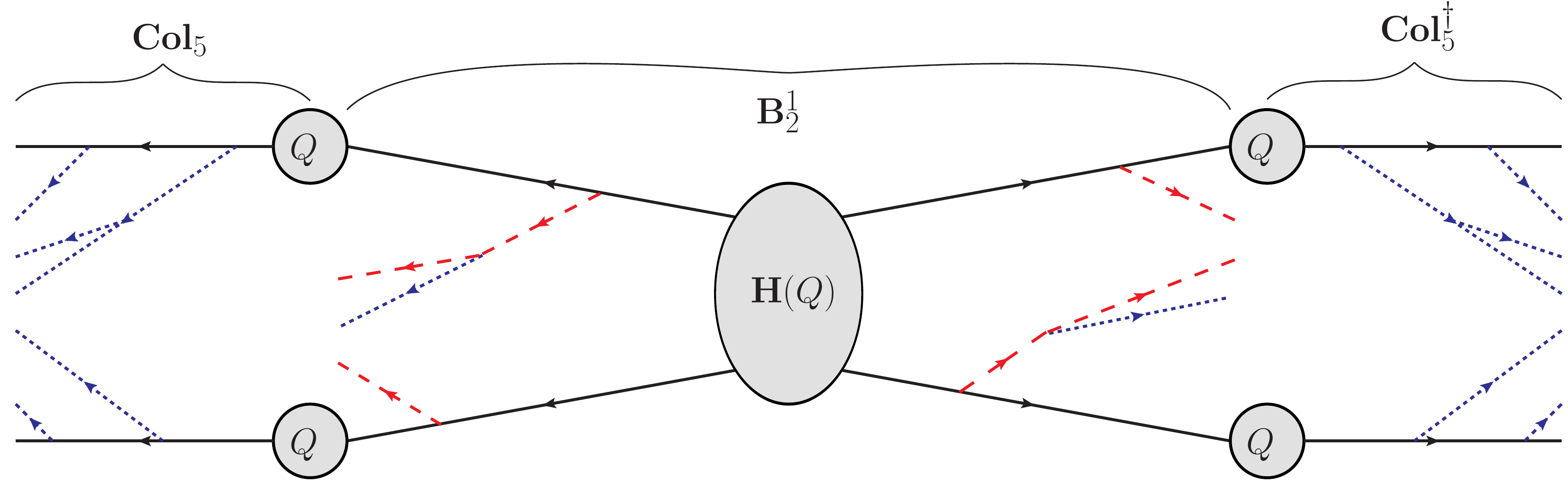}
	\caption{Illustrating hard-leg factorisation. Red dashed lines represent the emission of soft gluons and collinear emissions are represented by blue dotted lines. Circles indicate the hard scale from which subsequent evolution proceeds. Loops (Sudakov factors) have been neglected to avoid clutter.}
	\label{fig:SC1fsubfig}
\end{figure}

\begin{figure}[h]
	\centering
	\subfigure[A term contributing to the right evolution ($\v{B}^{3}_{6}$).]{\label{fig:A_hat}\includegraphics[width=0.33\textwidth]{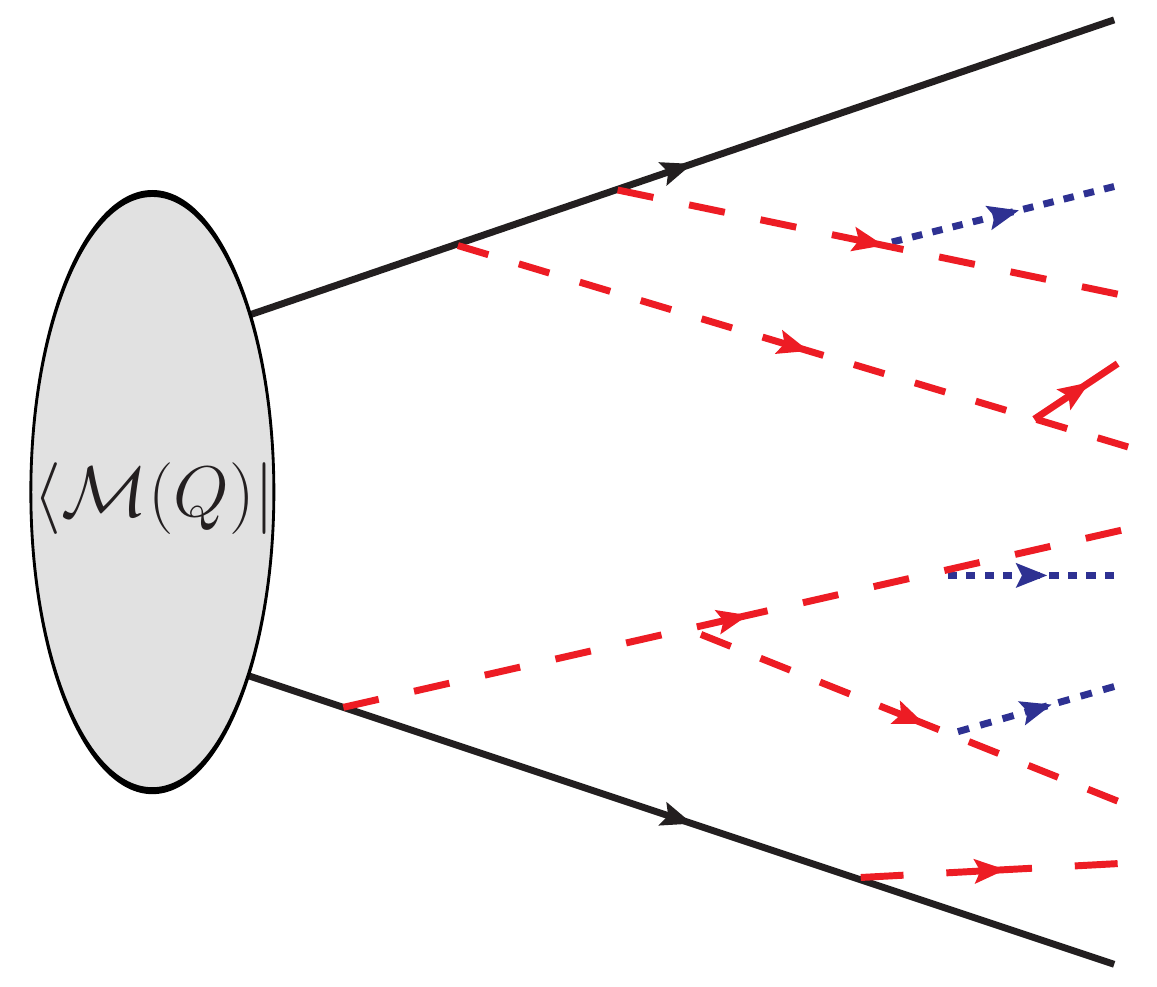}} ~~~~~~~
	\subfigure[A term contributing to the right evolution ($\v{A}^{\TT{soft}}_{9}$).]{\label{fig:A_soft}\includegraphics[width=0.33\textwidth]{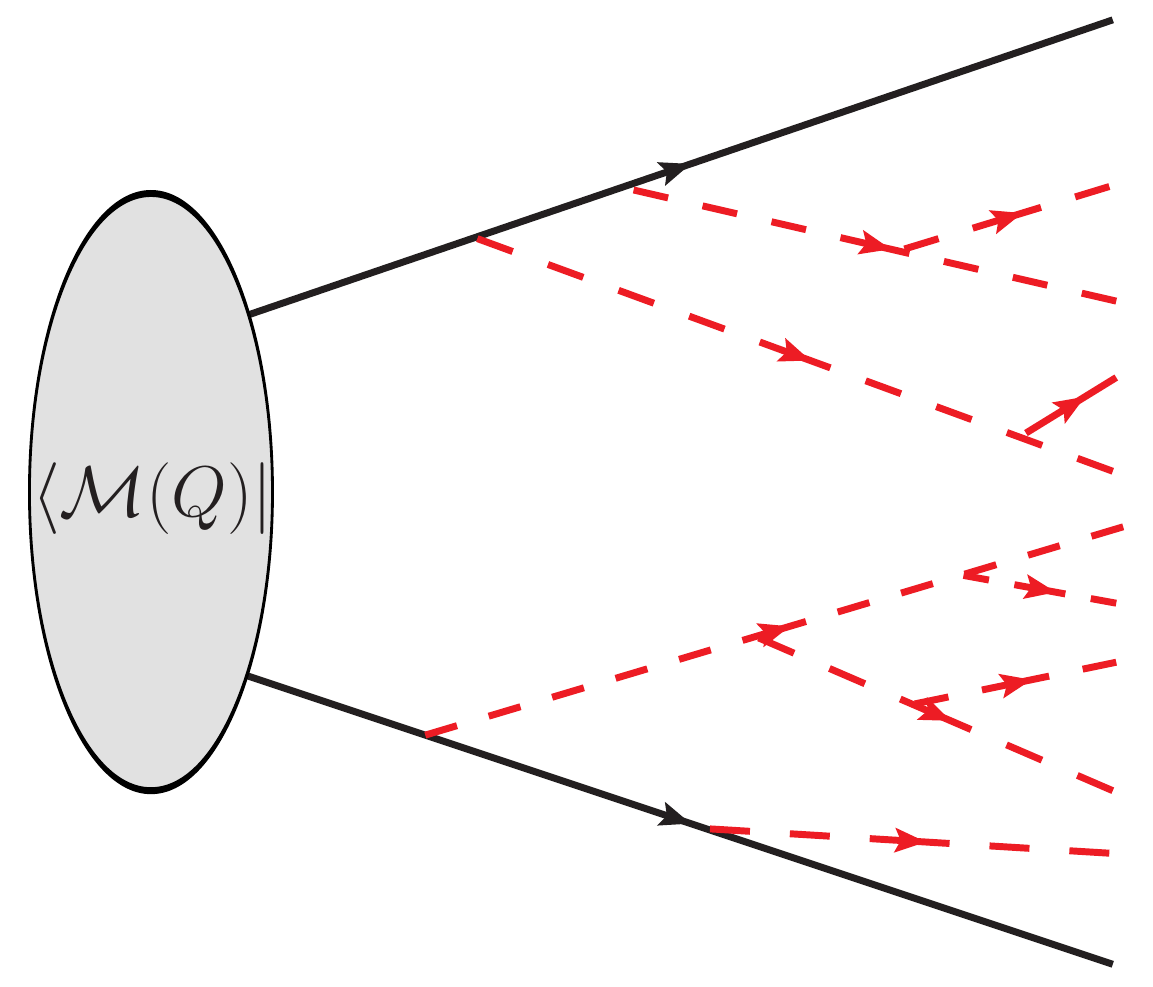}}
	\caption{The right evolution (the evolution of the conjugate amplitude) of a hard process after 9 emissions. Red dashed lines represent the emission of soft gluons and collinear emissions are represented by blue dotted lines. Loops (Sudakov factors) have been neglected to avoid clutter.}
	\label{fig:SC1f}
\end{figure}

The main result of this subsection is the factorisation of collinear physics into jet functions; one for each leg emerging from the hard scatter. We do this with Coulomb gluons removed ($\tilde{\delta}_{ij}=0$) and will discuss their re-introduction in Section \ref{sec:factorise_with_coloumb}. The following manipulations can equally well be performed using either variant A or B of the algorithm. For concreteness we will use variant B whenever an operator needs to be given an explicit definition\footnote{In the case of variant A, for the most part, all that must be done is exchange $\v{P}_{ij}$ and $\mathcal{P}_{\upsilon_{i}\upsilon_{j}}$ with the overlined versions $\overline{\v{P}}_{ij}$ and $\overline{\mathcal{P}}_{\upsilon_{i}\upsilon_{j}}$.}. Let us begin by simply stating the result:
\begin{eqnarray}
	\label{eqn:hard_col_factorisation}
	\Sigma(\mu) = \int \sum_{n} \left(\prod^{n}_{i = 1} \td \Pi_{i} \right) \sum^{n}_{m=0} \sum^{n-m}_{p=0} \Tr \,\left(\v{Col}^{\dagger}_{m}(\mu) \circ  \v{Col}_{m}(\mu) \, \v{B}^{p}_{n-m-p}(\mu) \right).
\end{eqnarray}
Figure \ref{fig:SC1fsubfig} illustrates what is going on diagrammatically (it shows a contribution with $n=8$, $m=5$ and $p=1$).
The collinear evolution operators for hard legs, which provide an operator description of a jet function, are constructed iteratively according to
\begin{eqnarray}
\begin{split}
\v{Col}_{0}(q_{\bot}) &= \v{V}^{\TT{col}}_{q_{\bot},Q}, \\
\v{Col}_{m}(q_{\bot}) &= \v{V}^{\TT{col}}_{q_{\bot},q_{m \, \bot}} \overline{\v{C}}_{m} \v{Col}_{m-1}(q_{m \, \bot}) \, \Theta(q_{\bot} \leq q_{m \, \bot}),
\end{split}
\end{eqnarray}
where
\begin{eqnarray}
\begin{split}
\v{V}^{\TT{col}}_{a,b} &= \exp \left[- \frac{\alpha_{s}}{\pi} \sum_{j} \int^{b}_{a} \frac{\td k^{(j\vec{n})}_{\bot}}{k^{(j\vec{n})}_{\bot}} \sum_{\upsilon} \mathbb{T}^{\bar{\upsilon} \, 2}_{j} \int \frac{\td z \, \td \phi}{8\pi} \mathcal{P}^{\,\circ}_{\upsilon \upsilon_{j}} \right], \\
\overline{\v{C}}_{i} &= \sum_{j} \frac{q^{(j\vec{n})}_{i\,\bot}}{2\sqrt{z_{i}}}\Delta_{ij} \, \v{P}_{ij}.
\end{split}
\end{eqnarray}
In both operators $j$ sums only over hard partons. The circle operation, $\circ$,  indicates the sharing of partons between $\v{Col}_{m}(\mu)$ and $\v{Col}^{\dagger}_{m}(\mu)$, i.e. 
\begin{eqnarray}
\begin{split}
\v{Col}_{m}(q_{\bot}) \circ \v{Col}^{\dagger}_{m}(q_{\bot}) &= \v{V}^{\TT{col}}_{q_{\bot},q_{m \, \bot}} \overline{\v{C}}_{m} \, \v{Col}_{m-1}(q_{m \, \bot}) \circ \v{Col}^{\dagger}_{m-1}(q_{m \, \bot}) \, \overline{\v{C}}^{\dagger}_{m} \v{V}^{\TT{col} \, \dagger}_{q_{\bot},q_{m \, \bot}}.
\end{split}
\end{eqnarray}
$\v{B}^{p}_{n-m-p}(\mu)$ are the scattering matrices evolved using the algorithm, modified so that all collinear emissions from hard legs have been removed. Specifically,
\begin{align}
\tilde{\v{A}}_{n-m}(\mu) = \sum_{p=0}^{n-1}\v{B}^{p}_{n-p-m}(\mu),
\end{align}
where $\tilde{\v{A}}_{n-m}(\mu)$ is computed using \eqref{eq:Aevo} with the collinear evolution off hard legs removed, i.e. with the replacements $\v{D}_{i} \mapsto \v{D}_{i} - \overline{\v{C}}_{i}$ and $\v{V}_{a,b} \mapsto \v{V}_{a,b}(\v{V}^{\TT{col}}_{a,b})^{-1}$. The number of collinear emissions not off hard legs is indexed by $p$ and $n-p-m$ is the number of soft emissions (in equation \eqref{eqn:hard_col_factorisation}, $m$ is the number of collinear emissions off hard legs). An example term, contributing to $\v{B}^{3}_{6}(\mu)$, is presented in Figure \ref{fig:A_hat}. 

It may be helpful to contrast $\v{B}^{p}_{n}(\mu)$ with scattering matrices evolved using the FKS algorithm \cite{SoftEvolutionAlgorithm}. We denote the FKS matrices as $\v{A}^{\TT{soft}}_{n}(\mu)$ and they can be evaluated using variant A with $\overline{\mathcal{P}}_{\upsilon_{i}\upsilon_{j}}=0$; an example is shown in Figure \ref{fig:A_soft}.  Note that $\v{B}^{0}_{n}(\mu) \neq \v{A}^{\TT{soft}}_{n}(\mu)$ for $n \geq 1$ since $\v{B}^{0}_{n}(\mu)$ still contains the collinear Sudakov factors `attached' to soft partonic legs. Also note that $\v{B}^{0}_{0}(\mu) = \v{A}^{\TT{soft}}_{0}(\mu)$ and $\v{B}^{i}_{0}(\mu) = 0$ for all $i > 0$. In Section \ref{sec:fullyfactorise} we will generalise the arguments presented here so that we can factorise collinear physics into jet functions that multiply $\v{A}^{\TT{soft}}_{n}(\mu)$. However, in this section we will not make any further use of $\v{A}^{\TT{soft}}_{n}(\mu)$.

Equation \eqref{eqn:hard_col_factorisation} can be written more simply by combining the collinear evolution operators (which are proportional to unit operators in colour space) into a single cross-section level jet function, $\v{Col}^{\dagger}_{m}(\mu) \circ  \v{Col}_{m}(\mu) = \mathbbm{1}\otimes\TT{Col}_{m}(\mu)$. Doing this enables \eqref{eqn:hard_col_factorisation} to be written as
\begin{eqnarray}
\Sigma(\mu)|_{u_{n}=1} = \int \sum_{n} \left(\prod^{n}_{i = 1} \td \Pi_{i} \right) \sum^{n}_{m=0} \sum^{n-m}_{p=0} \Tr_{s} \left( \TT{Col}_{m}(\mu) \, \Tr_{c}\, \v{B}^{p}_{n-m-p}(\mu) \right),
\end{eqnarray}
where the traces are over colour, $c$, and helicity, $s$. However, we avoid working with collinear factorisation in this form because it does not apply when Coulomb exchanges are present.

Having stated the result, let us now proceed to show how it is derived. The following commutation relations are important:
\begin{eqnarray}
\begin{split}
	\left[\v{D}_{i} - \overline{\v{C}}_{i}, \overline{\v{C}}_{j} \right] \simeq 0, \qquad \left[\v{V}_{a,b} (\v{V}^{\TT{col}}_{a,b})^{-1} , \overline{\v{C}}_{j} \right] \simeq 0, \qquad \\ 
	\left[\v{V}_{a,b}(\v{V}^{\TT{col}}_{a,b})^{-1}, \v{V}^{\TT{col}}_{c,d} \right] \simeq 0, \qquad \left[\v{D}_{i} - \overline{\v{C}}_{i}, \v{V}^{\TT{col}}_{a,b} \right] \simeq 0.
\end{split}
\label{eqn:commutators}
\end{eqnarray}
Here $\simeq$ denotes equality when the operator acts on a matrix element, which is all we ever encounter.

$[\v{V}_{a,b}(\v{V}^{\TT{col}}_{a,b})^{-1}, \v{V}^{\TT{col}}_{c,d} ] \simeq 0$ and $[\v{D}_{i} - \overline{\v{C}}_{i}, \v{V}^{\TT{col}}_{a,b} ] \simeq 0$ are trivial to show as $\v{V}^{\TT{col}}_{a,b}$ is proportional to the identity in colour-helicity space. Diagrammatically, proving $$[\v{V}_{a,b}(\v{V}^{\TT{col}}_{a,b})^{-1}, \v{V}^{\TT{col}}_{c,d} ] \simeq 0$$ reduces to showing
\begin{eqnarray}
\begin{split}
\begin{array}{c}
\includegraphics[width=0.25\textwidth]{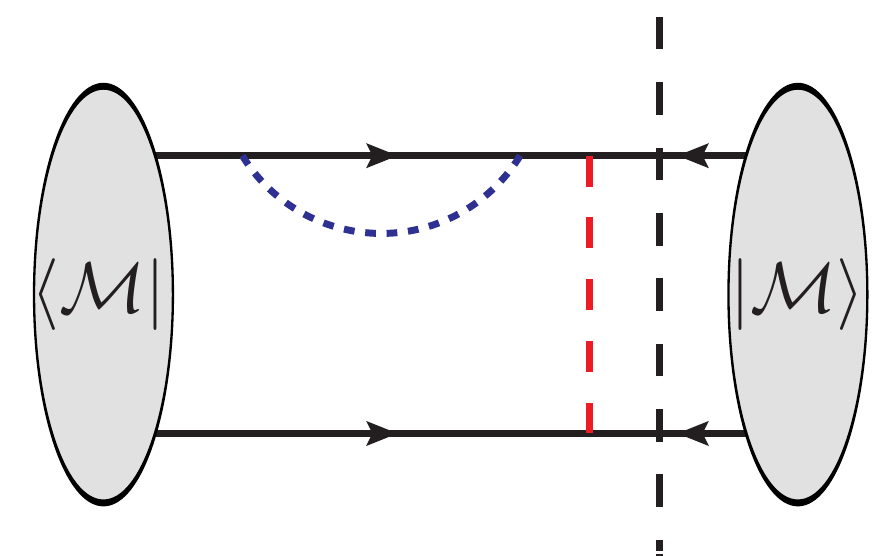} 
\end{array} = 
\begin{array}{c}
\includegraphics[width=0.25\textwidth]{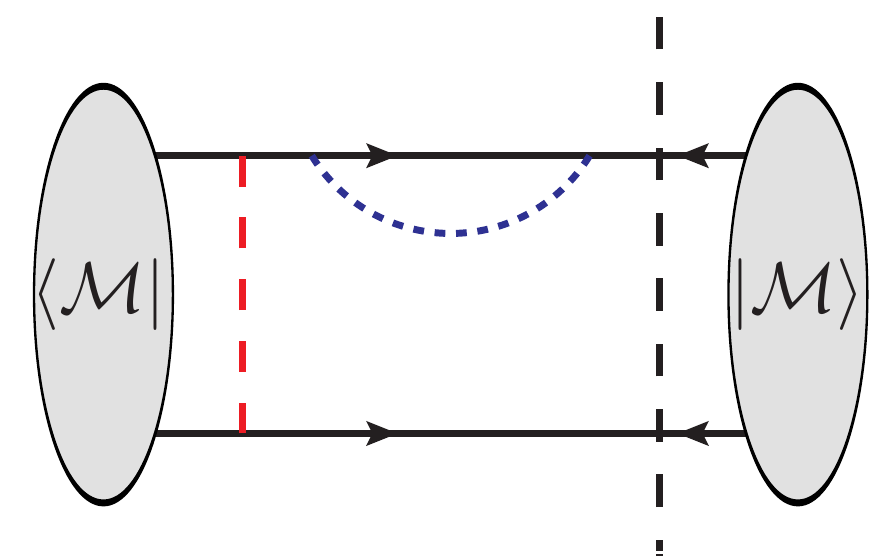} 
\end{array},
\end{split}
\end{eqnarray}
and $[\v{D}_{i} - \overline{\v{C}}_{i}, \v{V}^{\TT{col}}_{a,b} ] \simeq 0$ to showing
\begin{eqnarray}
\begin{split}
\begin{array}{c}
\includegraphics[width=0.25\textwidth]{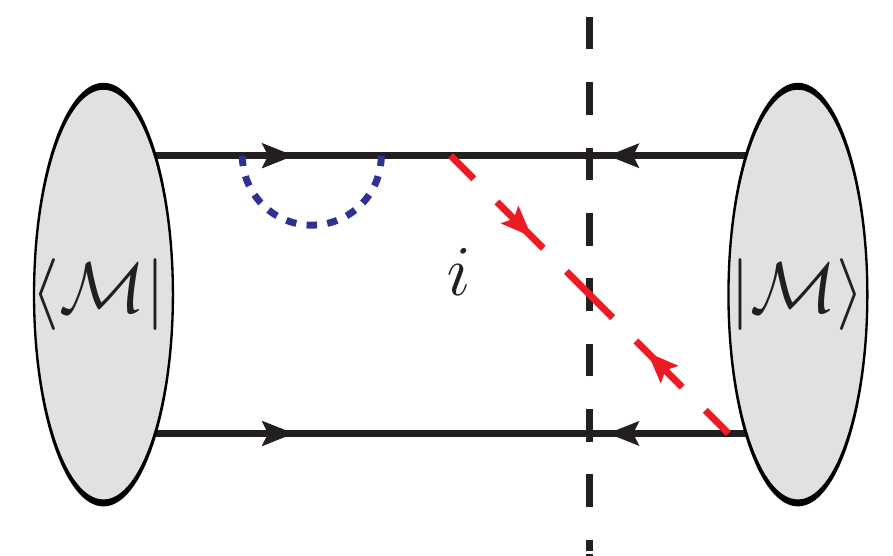} 
\end{array} = 
\begin{array}{c}
\includegraphics[width=0.25\textwidth]{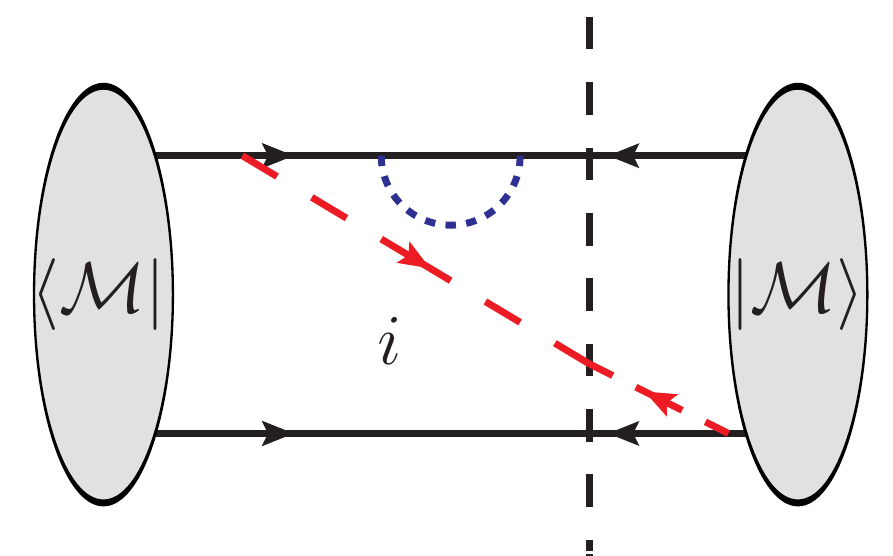}
\end{array}.
\end{split}
\end{eqnarray}
As ever, a red dashed line is used to represent a soft parton and a blue dotted line represents a collinear parton. The black dashed line indicates a cut (cut lines are on shell).

$[\v{V}_{a,b} (\v{V}^{\TT{col}}_{a,b})^{-1}, \overline{\v{C}}_{j}] \simeq 0$ and $[\v{D}_{i} - \overline{\v{C}}_{i}, \overline{\v{C}}_{j}] \simeq 0$ can be shown by factorising kinematic factors from the colour and helicity operators, then carefully tracking the action of the colour operators so that colour conservation can be applied. Proving the commutators also requires noting that both soft real emissions and soft Sudakov factors are identity operators in helicity space, and that helicity states are orthogonal. $[\v{V}_{a,b} (\v{V}^{\TT{col}}_{a,b})^{-1}, \overline{\v{C}}_{j}] \simeq 0$ presents the biggest challenge. The derivation follows closely the discussion in \cite{SuperleadingLogs}. It is most easily illustrated by expressing the operators diagrammatically. Doing so reduces the problem to showing
\begin{eqnarray}
\begin{split}
\begin{array}{c}
\includegraphics[width=0.25\textwidth]{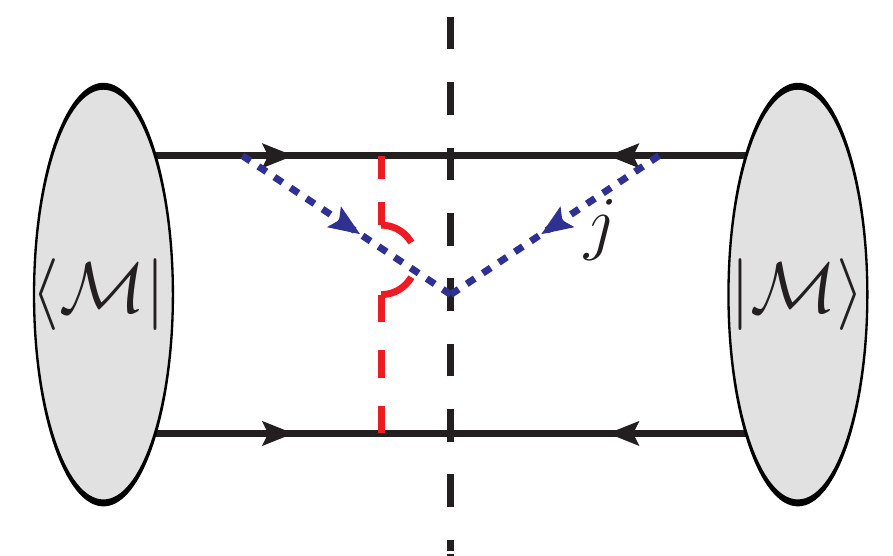} 
\end{array} &+ 
\begin{array}{c}
\includegraphics[width=0.25\textwidth]{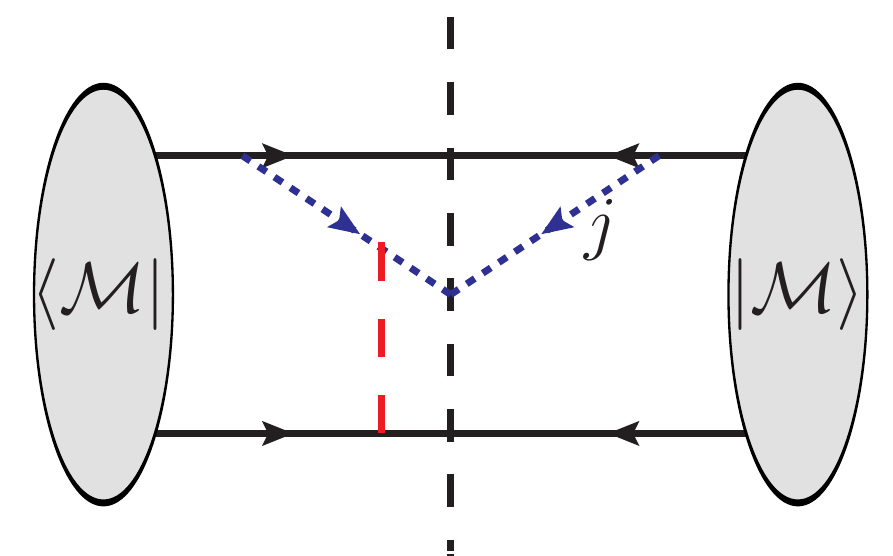} 
\end{array} +
\begin{array}{c}
\includegraphics[width=0.25\textwidth]{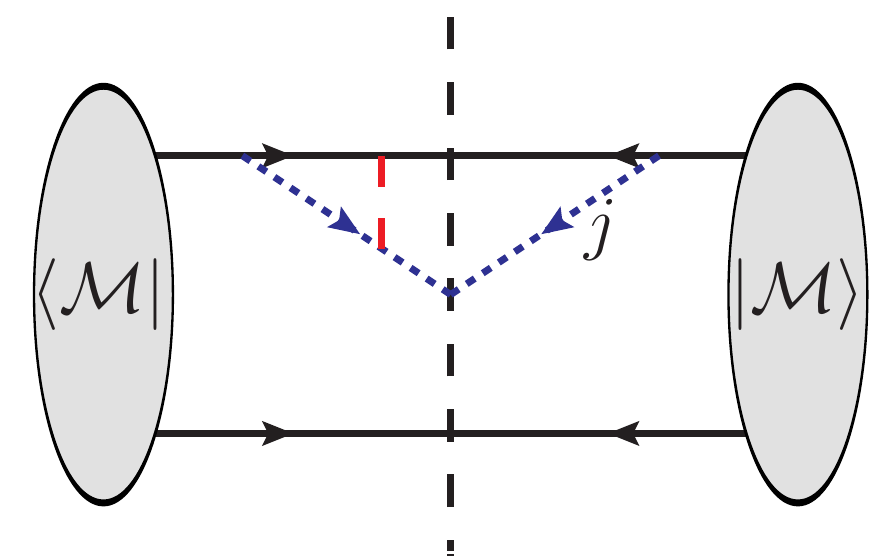}
\end{array} \\
&= \begin{array}{c}
\includegraphics[width=0.25\textwidth]{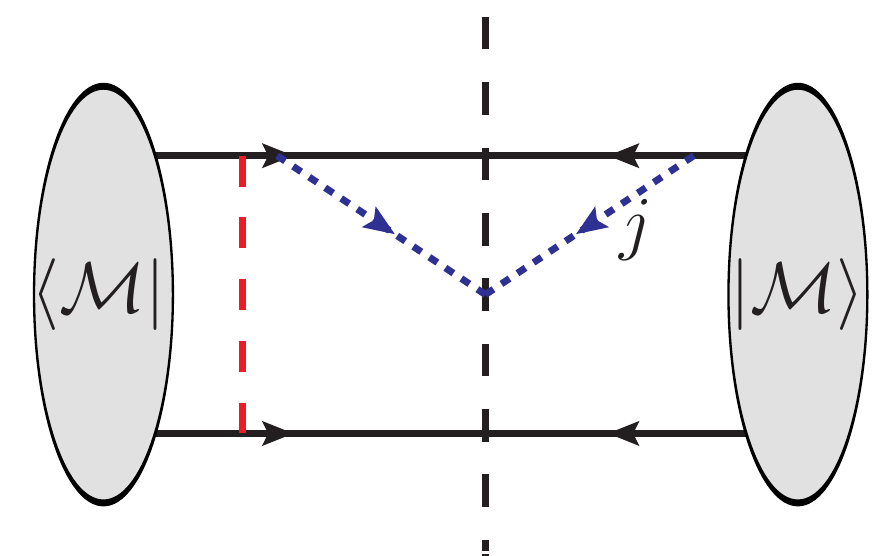}
\end{array}.
\end{split}
\label{eqn:VC_Commutator}
\end{eqnarray}
Also note that $$[\v{V}_{a,b} (\v{V}^{\TT{col}}_{a,b})^{-1}, \overline{\v{C}}_{j}] \simeq 0$$ implies $[\v{D}_{i} - \overline{\v{C}}_{i}, \overline{\v{C}}_{j}] \simeq 0$ since
\begin{eqnarray}
\begin{split}
\begin{array}{c}
\includegraphics[width=0.25\textwidth]{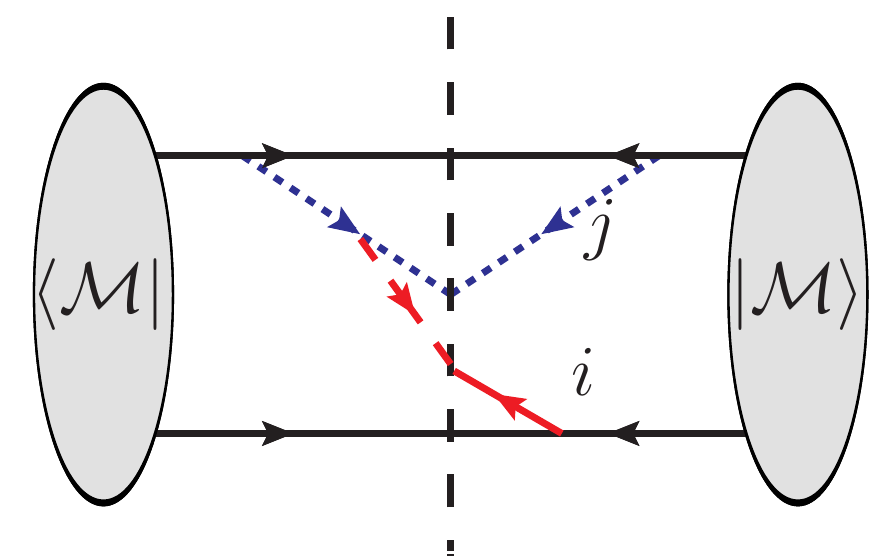} 
\end{array} + 
\begin{array}{c}
\includegraphics[width=0.25\textwidth]{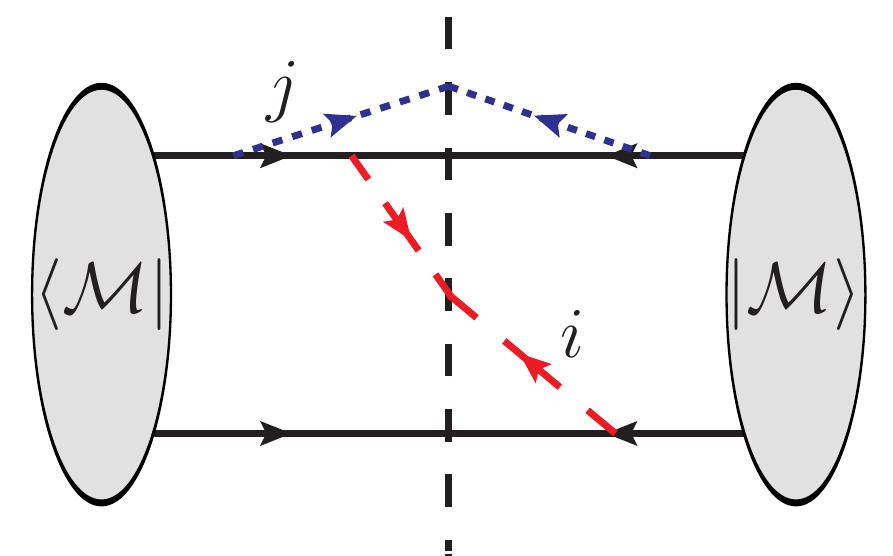} 
\end{array} =
\begin{array}{c}
\includegraphics[width=0.25\textwidth]{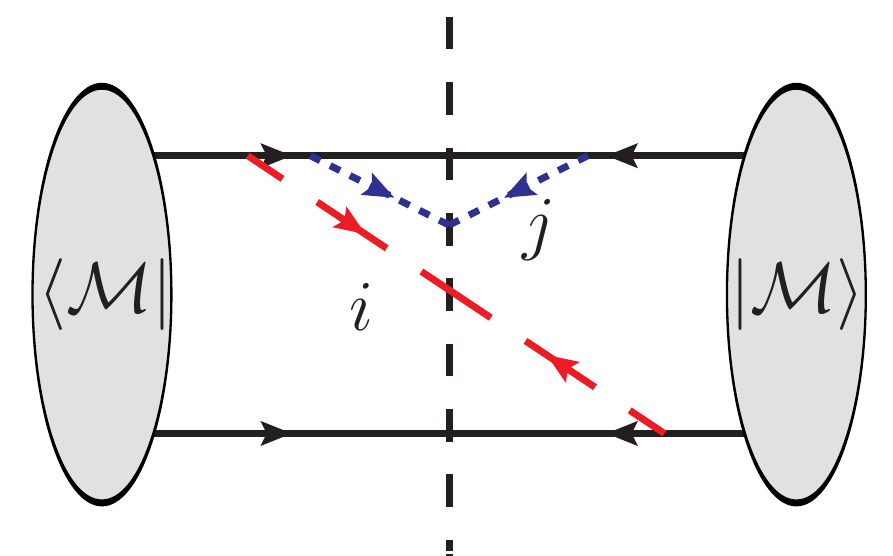} 
\end{array},
\end{split}
\end{eqnarray}
which trivially follows from \eqref{eqn:VC_Commutator}. Using these commutation relations, reconstructing the separate strong orderings of collinear and soft physics in \eqref{eqn:hard_col_factorisation} is simply a case of careful combinatorics and relabelling of momenta. For instance
\begin{eqnarray}
\begin{split}
& \left[
\begin{array}{c}
\includegraphics[width=0.2\textwidth]{DC_commutator_1} 
\end{array} + 
\begin{array}{c}
\includegraphics[width=0.2\textwidth]{DC_commutator_2} 
\end{array} \right] \Theta(q_{j \, \bot} - q_{i \, \bot}) \Theta(q_{i \, \bot} - \mu) \Theta(Q - q_{j \, \bot})\\
&+ 
\begin{array}{c}
\includegraphics[width=0.2\textwidth]{DC_commutator_3} 
\end{array} \Theta(q_{i \, \bot} - q_{j \, \bot}) \Theta(q_{j \, \bot} - \mu) \Theta(Q - q_{i \, \bot})\\ 
&= 
\begin{array}{c}
\includegraphics[width=0.2\textwidth]{DC_commutator_3} 
\end{array} \Theta(q_{i \, \bot} - \mu) \Theta(Q - q_{i \, \bot}) \Theta(q_{j \, \bot} - \mu) \Theta(Q - q_{j \, \bot}).
\end{split}
\end{eqnarray}

For the sake of completeness, in the next section we will go ahead and put back $\mathfrak{R}^{\TT{soft}}$, $\mathfrak{R}^{\TT{coll}}$, and the measurement functions.  However, we have only proven correctness at LL accuracy.  As such, Sections \ref{sec:SC1f} and \ref{sec:SC2f+} are conjectures. It might be the case that only certain classes of recoil prescription factorise in this way. We will focus on variant A in Section \ref{sec:SC1f} and turn to variant B in Section \ref{sec:SC2f+}, where we show how to re-instate the plus prescription in the collinear splitting functions.  We caution that both these versions of the algorithm will not produce super-leading logarithms because Coulomb interactions have been neglected. Therefore they are only suitable for processes with fewer than two coloured particles in either the initial or final state of the hard process, i.e. $e^{+}e^{-}$, deep-inelastic scattering and Drell-Yan.

\subsubsection{The details}
\label{sec:SC1f}
Now we summarise the results of the previous section and make a conjecture regarding the inclusion of recoils (recall we left these out of the discussions in the previous section). For concreteness we use variant A. The evolution has two phases. In the first phase soft gluons are emitted:
\begin{eqnarray}
\begin{split}
\boldsymbol{\sigma}(0, 0) &= \tilde{\v{V}}_{\mu,Q} \v{H}(Q; \{p\}) \tilde{\v{V}}^{\dagger}_{\mu,Q} = \tilde{\v{A}}_{0}(\mu; \{p\}), \\
\v{d}\boldsymbol{\sigma}(1, 0) &= \int \prod_{i} \td^{4} p_{i} \tilde{\v{V}}_{\mu,q_{1 \, \bot}} \tilde{\v{D}}_{1} \tilde{\v{A}}_{0}(q_{1 \, \bot}; \{p\}) \tilde{\v{D}}^{\dagger}_{1}  \tilde{\v{V}}^{\dagger}_{\mu,q_{1 \, \bot}} \td \Pi_{1} \\
& \; = \tilde{\v{A}}_{1}(\mu; \{p\}_{1}) \td \Pi_{1} \equiv \v{B}^{0}_{1}(\mu; \{p\}_{1}) \td \Pi_{1}, \\
\v{d}\boldsymbol{\sigma}(2, 0) &= \int \prod_{i} \td^{4} p_{i} \tilde{\v{V}}_{\mu,q_{2 \, \bot}} \tilde{\v{D}}_{2} \tilde{\v{A}}_{1}(q_{2 \, \bot}; \{p\}) \tilde{\v{D}}^{\dagger}_{2}  \tilde{\v{V}}^{\dagger}_{\mu,q_{2 \, \bot}} \td \Pi_{1} \td \Pi_{2} \\
& \; = \tilde{\v{A}}_{2}(\mu; \{p\}_{2}) \td \Pi_{1} \td \Pi_{2} \equiv \left(\v{B}^{0}_{2}(\mu)  + \v{B}^{1}_{1}(\mu) \right)\td \Pi_{1} \td \Pi_{2}, \\
\v{d} \boldsymbol{\sigma}(n, 0) & \; = \tilde{\v{A}}_{n}(\mu; \{p\}_{n}) \prod^{n}_{i =1} \td \Pi_{i} \equiv \sum_{p=0}^{n} \v{B}^{p}_{n-p}(\mu) \prod^{n}_{i =1} \td \Pi_{i} ,
\end{split} 
\end{eqnarray}
where $\tilde{\v{D}}_{i} = \v{D}_{i} - \overline{\v{C}}_{i}$ and
\begin{eqnarray}
\begin{split}
\overline{\v{C}}_{i} = \sum_{l} \frac{q^{(l\vec{n})}_{i\,\bot}}{2\sqrt{z_{i}}}\Delta_{il} \, \overline{\v{P}}_{il}\, \mathfrak{R}^{\TT{coll}}_{ij}(\{p\},\{\tilde{p}\},q_{i}),
\end{split}
\end{eqnarray}
where $l$ sums only over hard partons. And $\tilde{\mathbf{V}}_{a,b} = \v{V}_{a,b}(\overline{\v{V}}_{a,b})^{-1}$ and 
\begin{eqnarray}
\begin{split}
\overline{\v{V}}_{a,b} &= \exp \left[-\frac{\alpha_{s}}{\pi} \sum_{l} \int^{b}_{a} \frac{\td k^{(l\vec{n})}_{\bot}}{k^{(l\vec{n})}_{\bot}}  \sum_{\upsilon} \mathbb{T}^{\bar{\upsilon} \, 2}_{\bar{l}} \int \frac{\td z \, \td \phi}{8\pi} \, \overline{\mathcal{P}}^{\,\circ}_{\upsilon \upsilon_{l}} \, \mathcal{R}^{\TT{coll}}_{l}\right].	
\end{split}
\end{eqnarray}
Again, the sum over $l$ only includes hard partons.
In $\v{d} \boldsymbol{\sigma}(n, m)$, $n$ indicates the number of soft emissions, which occur during the first phase of the evolution, and $m$ indicates the number of collinear emissions, which occur during the second phase of the evolution. 

The second phase of the evolution dresses the hard legs with collinear emissions:
\begin{eqnarray}
\begin{split}
\td \sigma(n, 0) &= \Tr \,\left( \overline{\v{V}}_{\mu,Q} \, \v{d} \boldsymbol{\sigma}(n, 0) \, \overline{\v{V}}^{\dagger}_{\mu,Q} \right) = \Tr \, (\tilde{\v{A}}^{n}_{n+0}(\mu; \{p\}_{n})) \\ 
& \equiv \sum_{p=0}^{n} \Tr(\v{Col}^{\dagger}_{0}(\mu) \circ  \v{Col}_{0}(\mu) \, \v{B}^{p}_{n-p}(\mu)), \\
\td {\sigma}(n, 1) &= \int \prod_{i} \td^{4} p_{i} \, \Tr \,\left( \overline{\v{V}}_{\mu,q_{n+1 \, \bot}} \overline{\v{C}}_{n+1} \overline{\v{V}}_{q_{n+1 \, \bot},Q} \, \v{d} \boldsymbol{\sigma}(n, 0) \right. \\ 
&\qquad \left. \times \overline{\v{V}}^{\dagger}_{q_{n+1 \, \bot},Q} \overline{\v{C}}^{\dagger}_{n+1}  \overline{\v{V}}^{\dagger}_{\mu,q_{n+1 \, \bot}} \right) \td \Pi_{n+1} = \Tr \, (\tilde{\v{A}}^{n}_{n+1}(\mu; \{p\}_{n+1})) \td \Pi_{n+1} \\ 
& \equiv \sum_{p=0}^{n} \Tr(\v{Col}^{\dagger}_{1}(\mu) \circ  \v{Col}_{1}(\mu) \, \v{B}^{p}_{n-p}(\mu)) \td \Pi_{n+1}, \\
\td \sigma(n, m) & \; = \; \Tr \, (\tilde{\v{A}}^{n}_{n+m}(\mu; \{p\}_{n+m})) \prod^{m}_{i =1} \td \Pi_{n+i}\\ 
& \equiv \sum_{p=0}^{n} \Tr(\v{Col}^{\dagger}_{m}(\mu) \circ  \v{Col}_{m}(\mu) \, \v{B}^{p}_{n-p}(\mu)) \prod^{m}_{i =1} \td \overline{\Pi}_{n+i},
\end{split} 
\end{eqnarray}
where $\tilde{\v{A}}^{n}_{n+m}$ obeys the recurrence relation
\begin{eqnarray}
\begin{split}
\tilde{\v{A}}^{n}_{n+m}(q_{\bot}; \{\tilde{p}\}_{n+m-1} \cup q_{n+m})&= \int \prod_{i} \td^{4} p_{i} \overline{\v{V}}_{q_{\bot},q_{n+m \, \bot}} \overline{\v{C}}_{n+m} \tilde{\v{A}}^{n}_{n+m-1}(q_{n+m \, \bot}; \{p\}_{n+m-1}) \\
&\qquad \times\overline{\v{C}}^{\dagger}_{n+m} \overline{\v{V}}^{\dagger}_{q_{\bot},q_{n+m \, \bot}} \Theta(q_{\bot} \leq q_{n+m \, \bot}). \nonumber
\end{split} 
\end{eqnarray}
An observable can be calculated using
\begin{eqnarray}
\Sigma(\mu) = \sum_{n} \int \td \sigma_{n} \,u_{n}(q_{1},...,q_{n}), \label{eq:cobs}
\end{eqnarray}
where $\td\sigma_{n} = \sum_{m=0}^{n}\td \sigma(n-m, m)$.

\subsubsection{Recovering the `plus prescription'}
\label{sec:SC2f+}
Now let us turn to variant B. Recall that, in this variant of the algorithm collinear evolution proceeds using the full DGLAP splitting functions. Things are precisely as in the last subsection except that we now use the splitting operators without overlines ($\v{P}_{il}$) and the functions $\mathfrak{f}$ and $\mathcal{F}$ are to be included in the soft terms. We can go a little further and expand out the Sudakov factors in order to recover the familiar DGLAP plus prescription. To that end, we expand the collinear Sudakov factors ($\overline{\v{V}}$) that appear in the second phase of the evolution:
\begin{equation}
\begin{split}
\overline{\v{V}}_{a,b} =& \mathbbm{1} - \int^{b}_{a} \frac{ \td k_{1 \, \bot}}{k_{1 \, \bot}}\Gamma_{1} + \int^{b}_{a} \frac{ \td k_{2 \, \bot}}{k_{2 \, \bot}}\Gamma_{1}\int^{b}_{k_{1}} \frac{ \td k_{2 \, \bot}}{k_{2 \, \bot}}\Gamma_{2} - ...
\end{split}
\end{equation}
where
\begin{equation}
\begin{split}
\Gamma_{i} = \frac{\alpha_{s}}{\pi} \sum_{l} \sum_{\upsilon} \mathbb{T}^{\bar{\upsilon} \, 2}_{\bar{l}} \int \frac{\td z_{i} \, \td \phi_{i}}{8\pi} \overline{\mathcal{P}}^{\,\circ}_{\upsilon \upsilon_{l}} \, \mathcal{R}^{\TT{coll}}_{l}.
\end{split}
\end{equation}
Once again, the sum over $l$ is only over hard partons. Using this, we can regroup terms in the same way as Section \ref{sec:algorithmIR} to generate the plus prescription: 
\begin{eqnarray}
\begin{split}
\td {\sigma}(n, 0) &= \Tr \,\left( \v{d} \boldsymbol{\sigma}(n, 0) \right) = \Tr \, \tilde{\v{A}}^{n}_{n+0}, \\
\td {\sigma}(n, 1) &= \int \prod_{i} \td^{4} p_{i} \, \Tr \,\left( \overline{\v{D}}_{n+1}  \v{d} \boldsymbol{\sigma}(n, 0)  \, \overline{\v{D}}^{\dagger}_{n+1} \right) \td \overline{\Pi}_{n+1} \\& \; = \Tr \, \tilde{\v{A}}^{n}_{n+1}(\mu; \{p\}_{n+1}) \td \overline{\Pi}_{n+1}, \\
\td \sigma(n, m) & \; = \Tr \, \tilde{\v{A}}^{n}_{n+m}(\mu; \{p\}_{n+m}) \prod^{m}_{i =1} \td \overline{\Pi}_{n+i}\\ 
& \equiv \sum_{p=0}^{n} \Tr\left(\left[\v{Col}^{\dagger}_{m}(\mu) \circ  \v{Col}_{m}(\mu) + \mathcal{O}(\alpha^{m+1}_{s}) \right] \, \v{B}^{p}_{n-p}(\mu)\right) \prod^{m}_{i =1} \td \overline{\Pi}_{n+i} ,
\end{split} 
\end{eqnarray}
where
\begin{eqnarray}
\begin{split}
\tilde{\v{A}}^{n}_{n+m}(q_{\bot}; \{\tilde{p}\}_{n+m-1} \cup q_{n+m})&=\overline{\v{D}}_{n+m} \tilde{\v{A}}^{n}_{n+m-1}(q_{n+m \, \bot}; \{p\}_{n+m-1} ) \overline{\v{D}}^{\dagger}_{n+m} \Theta(q_{\bot} \leq q_{n+m \, \bot}), \nonumber
\end{split} 
\end{eqnarray}
using the boundary condition that
\begin{eqnarray}
\tilde{\v{A}}^{n}_{n+1}(q_{\bot})=\overline{\v{D}}_{n+1} \tilde{\v{A}}^{n}_{n+0} \overline{\v{D}}^{\dagger}_{n+1} \Theta(q_{\bot} \leq q_{n+1 \, \bot})\Theta(q_{n+1 \, \bot} \leq Q),
\end{eqnarray}
and
\begin{eqnarray} \label{eq:plusP}
\overline{\v{D}}_{i} = \sum_{\bar{l}} \frac{q^{(\bar{l}\vec{n})}_{i\,\bot}}{2}\Delta_{i\bar{l}}\, \v{P}^{+}_{i\bar{l}} \,\mathfrak{R}^{\TT{coll}}_{ij}(\{p\},\{\tilde{p}\},q_{i}).	
\end{eqnarray}
The sum over $\bar{l}$ only includes hard partons. $\v{P}_{il}$ has been redefined to include the plus prescription and labelled $\v{P}^{+}_{i\bar{l}}$. The plus prescription is defined in Appendix \ref{Appendix_A}. Observables are computed using \eqref{eq:cobs}.

\subsection{Complete collinear factorisation without Coulomb interactions}
\label{sec:fullyfactorise}

\begin{figure}[h]
	\centering
	\includegraphics[width=0.98\textwidth]{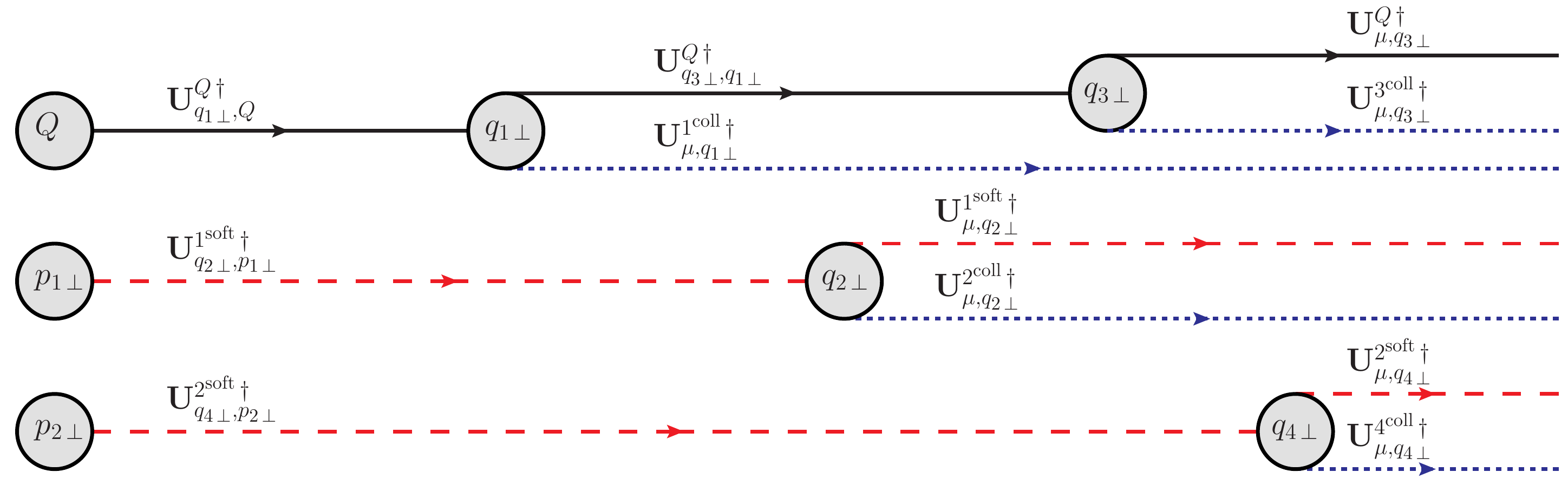}
	\caption{One of the possible contributions to $\v{tCol}_4^{\dagger}$. Red dashed lines represent soft gluons and blue dotted lines represent collinear partons. Each line is associated with a Sudakov factor and circles indicate the scale from which the subsequent evolution proceeds. Circles from which two lines leave indicate the action of the operator $\tilde{\v{C}}$. Circles from which one line leaves indicate the scales inherited from the soft evolution phase (not shown). Collinear scales $\{q_{i \, \bot}\}$ are ordered with respect to each other, as are soft scales $\{p_{i \, \bot}\}$. Scales connected along lines are also ordered, with the largest to the left and smallest to the right.}
	\label{fig:SC1F}
\end{figure}

\begin{figure}[h]
	\centering
	\includegraphics[width=0.6\textwidth]{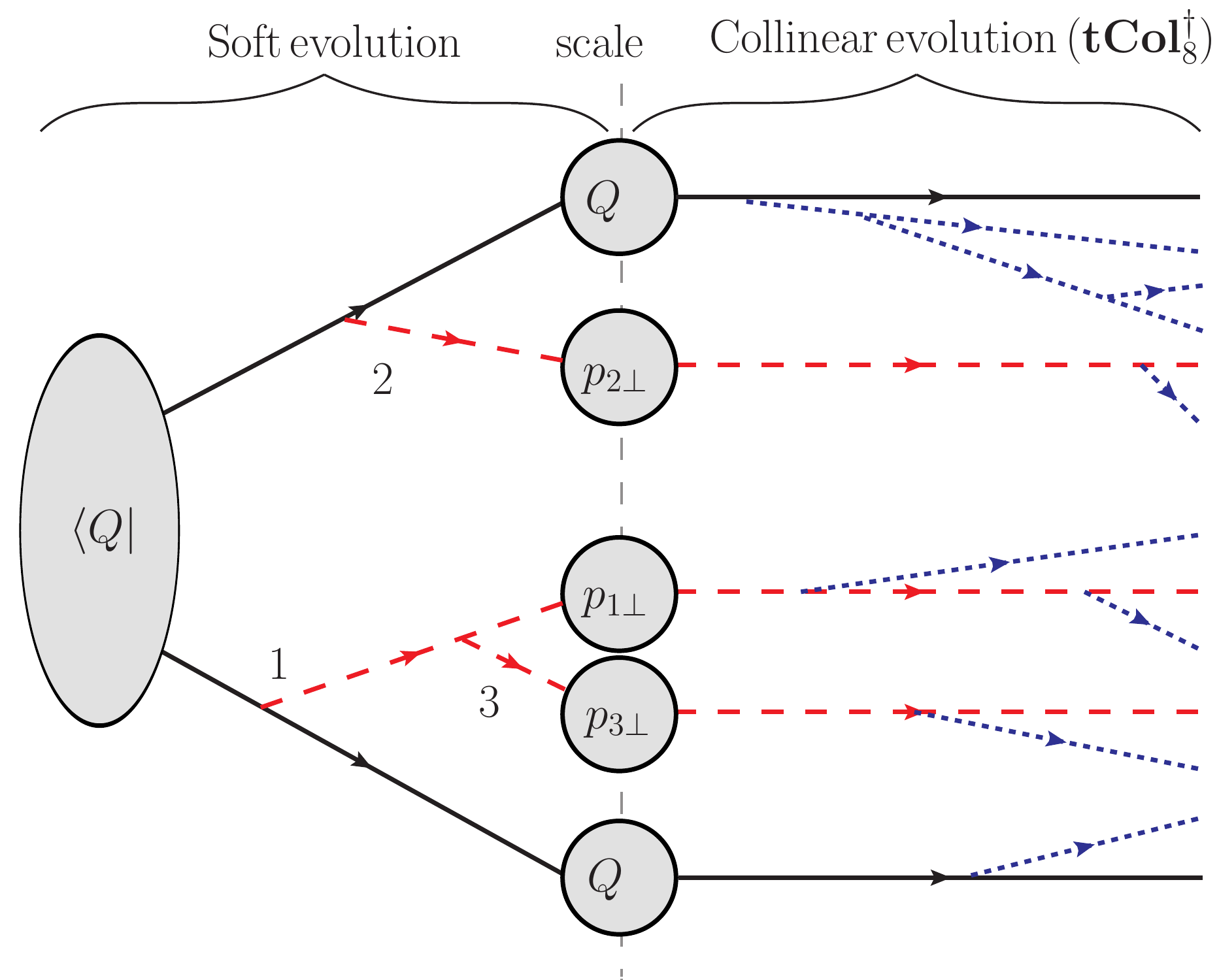}
	\caption{A diagram illustrating factorised parton evolution. Red dashed lines represent the emission of soft gluons and blue dotted lines represent collinear emissions. Circles represent the hard scale from which the subsequent evolution proceeds. Loops (Sudakov factors) have not been drawn.}
	\label{fig:SC1Freals}
\end{figure}

Now we are going to go ahead and factorise the collinear physics completely. Once again, the manipulations are essentially the same for either variant of our algorithm. To be concrete, we will use variant B whenever an exact definition must be given. As before, we will begin by stating the final result:
\begin{eqnarray}
\label{eqn:full_col_factorisation}
\Sigma(\mu) = \int \sum_{n} \left(\prod^{n}_{i = 1} \td \Pi_{i} \right) \sum^{n}_{m=0} \Tr \,\left(\v{tCol}^{\dagger}_{m}(\mu) \circ  \v{tCol}_{m}(\mu) \v{A}^{\TT{soft}}_{n-m}(\mu) \right),
\end{eqnarray}
where have we defined the following operators:
\begin{eqnarray}
\begin{split}
\v{tCol}_{0}(q_{\bot}) &= \v{V}^{\TT{tcol}}_{q_{\bot},Q}, \\
\v{tCol}_{m}(q_{\bot}) &= \v{V}^{\TT{tcol}}_{q_{\bot},q_{m \, \bot}}\tilde{\v{C}}_{m} \, \v{tCol}_{m-1}(q_{m \, \bot}) \, \Theta(q_{\bot} \leq q_{m \, \bot}), \\
\v{V}^{\TT{tcol}}_{a,b} &= \exp \left[- \frac{\alpha_{s}}{\pi} \sum_{j} \int^{b}_{a} \frac{\td k^{(j\vec{n})}_{\bot}}{k^{(j\vec{n})}_{\bot}} \Theta(q^{(j\vec{n})}_{i\,\bot} \leq p_{j \, \bot}) \, \sum_{\upsilon} \mathbb{T}^{\bar{\upsilon} \, 2}_{j} \int \frac{\td z \, \td \phi}{8\pi} \mathcal{P}^{\,\circ}_{\upsilon \upsilon_{j}} \right], \\
\tilde{\v{C}}_{i} &= \sum_{j} \frac{q^{(j\vec{n})}_{i\,\bot}}{2\sqrt{z_{i}}}\Delta_{ij}\, \v{P}_{ij} \, \Theta(q^{(j\vec{n})}_{i\,\bot} \leq p_{j \, \bot}),
\end{split} \label{eqn:full_col_defs}
\end{eqnarray}
and the index $j$ runs over all partons (hard, collinear and soft). We continue to leave aside the functions $\mathfrak{R}^{\TT{soft}}$, $\mathfrak{R}^{\TT{coll}}$, from emission operators and Sudakov factors. These can readily be re-instated as in Sections \ref{sec:SC1f} and \ref{sec:SC2f+}. $\v{A}^{\TT{soft}}_{n}(\mu)$ is as defined in Section \ref{sec:hard_factorisation} and evolves the same as $\v{A}_{n}(\mu)$ except that $\v{C}_{i} \mapsto 0$ and $\v{V}_{a,b} \mapsto \v{V}_{a,b}(\v{V}^{\TT{tcol}}_{a,b})^{-1} \equiv \v{V}^{\TT{soft}}_{a,b}$, i.e. it corresponds to summing over diagrams such as the one in Figure \ref{fig:A_soft}. Ignoring the effects of recoil (and Coulomb exchanges) and using variant A, $\v{A}^{\TT{soft}}_{n}(\mu)$ corresponds to FKS evolution \cite{SoftEvolutionAlgorithm}. One of the possible contributions to $\v{tCol}_4^{\dagger}$ is represented in Figure \ref{fig:SC1F}. In the figure, we have sacrificed the intuitive picture of a parton cascade in lieu of providing more detail on the evolution. To construct Figure \ref{fig:SC1F}, we have employed the Casimir structure of $\v{V}^{\TT{tcol}}_{a,b}$ to split it apart as $$\v{V}^{\TT{tcol}}_{a,b} = \prod_{j}\v{U}^{j}_{a,b}$$ where 
\begin{align}
\v{U}^{j}_{a,b} = \exp \left[- \frac{\alpha_{s}}{\pi} \int^{b}_{a} \frac{\td k^{(j\vec{n})}_{\bot}}{k^{(j\vec{n})}_{\bot}} \Theta(q^{(j\vec{n})}_{i\,\bot} \leq p_{j \, \bot}) \, \sum_{\upsilon} \mathbb{T}^{\bar{\upsilon} \, 2}_{j} \int \frac{\td z \, \td \phi}{8\pi} \mathcal{P}^{\,\circ}_{\upsilon \upsilon_{j}} \right],
\end{align}
and the product over $j$ is over all partons. In the figures we will be explicit with the labelling so that it is clear whether $\v{U}^{j}_{a,b}$ is associated with a hard parton (labelled by $Q$), a soft parton or a collinear parton, i.e. $j \in \{Q, 1^{\TT{soft}}, 2^{\TT{soft}}, ..., (n-m)^{\TT{soft}}, 1^{\TT{coll}}, 2^{\TT{coll}}, ..., m^{\TT{coll}} \}$. Something more in the style of our previous diagrams is illustrated in Figure \ref{fig:SC1Freals}. 

We will now prove \eqref{eqn:full_col_factorisation} by induction. First, we assume that 
\begin{align}
\Tr \, \v{A}_{n}(\mu) = \sum^{n}_{m=0} \Tr \,\left(\v{tCol}^{\dagger}_{m}(\mu) \circ  \v{tCol}_{m}(\mu) \v{A}^{\TT{soft}}_{n-m}(\mu) \right), \label{eqn:assumption}
\end{align}
where $\v{A}_{n}$ is computed as usual from \eqref{eq:Aevo}. We can see that this is true for $n=1$ by expanding out the $\v{tCol}$ operators and $\v{A}^{\TT{soft}}$:
\begin{align}
&\sum^{1}_{m=0} \Tr \,\left(\v{tCol}^{\dagger}_{m}(\mu) \circ  \v{tCol}_{m}(\mu) \v{A}^{\TT{soft}}_{n-m}(\mu) \right) \nonumber \\
&= \Tr \left(\v{V}^{\TT{tcol}}_{\mu,q_{1 \, \bot}} \tilde{\v{C}}_{1} \v{V}^{\TT{tcol}}_{q_{1 \, \bot},Q} \v{V}^{\TT{soft}}_{\mu,Q} \v{H}(Q) \v{V}^{\TT{soft} \, \dagger}_{\mu ,Q} \v{V}^{\TT{tcol} \, \dagger}_{q_{1 \, \bot},Q} \tilde{\v{C}}^{\dagger}_{1}  \v{V}^{\TT{tcol} \,\dagger}_{\mu,q_{1 \, \bot}} \right.\nonumber \\
& \qqqquad \left. + \v{V}^{\TT{tcol}}_{\mu,Q}\v{V}^{\TT{soft}}_{\mu,q_{1 \, \bot}} \v{S}_{1} \v{V}^{\TT{soft}}_{q_{1 \, \bot},Q} \v{H}(Q) \v{V}^{\TT{soft} \, \dagger}_{q_{1 \, \bot},Q} \v{S}^{\dagger}_{1}  \v{V}^{\TT{soft} \, \dagger}_{\mu,q_{1 \, \bot}} \v{V}^{\TT{tcol} \, \dagger}_{\mu,Q}  \right)  \nonumber \\
&= \Tr \left(\v{V}^{\TT{soft}}_{\mu,q_{1 \, \bot}}\v{V}^{\TT{tcol}}_{\mu,q_{1 \, \bot}} \v{C}_{1} \v{V}^{\TT{tcol}}_{q_{1 \, \bot},Q}  \v{V}^{\TT{soft}}_{q_{1 \, \bot},Q} \v{H}(Q) \v{V}^{\TT{soft} \, \dagger}_{\mu ,Q} \v{V}^{\TT{tcol} \, \dagger}_{q_{1 \, \bot},Q} \v{C}^{\dagger}_{1}  \v{V}^{\TT{tcol} \,\dagger}_{\mu,q_{1 \, \bot}} \right.\nonumber \\
& \qqqquad \left. + \v{U}^{1}_{\mu,q_{1 \, \bot}}\v{V}^{\TT{col}}_{\mu,q_{1 \, \bot}}\v{V}^{\TT{soft}}_{\mu,q_{1 \, \bot}} \v{S}_{1} \v{V}^{\TT{col}}_{q_{1 \, \bot},Q}\v{V}^{\TT{soft}}_{q_{1 \, \bot},Q} \v{H}(Q) \v{V}^{\TT{soft} \, \dagger}_{q_{1 \, \bot},Q} \v{S}^{\dagger}_{1}  \v{V}^{\TT{soft} \, \dagger}_{\mu,q_{1 \, \bot}} \v{V}^{\TT{tcol} \, \dagger}_{\mu,Q}  \right)  \nonumber \\
&= \Tr \left(\v{V}_{\mu,q_{1 \, \bot}} \v{D}_{1} \v{V}_{q_{1 \, \bot},Q} \v{H}(Q) \v{V}^{\dagger}_{q_{1 \, \bot},Q} \v{D}^{\dagger}_{1}  \v{V}^{\dagger}_{\mu,q_{1 \, \bot}} \right),
\end{align}
where we have used $\tilde{\v{C}}_{1} \equiv \overline{\v{C}}_{1} \equiv \v{C}_{1}$ as it only acts on hard legs. We have also used the commutators $[\v{V}_{a,b}(\v{V}^{\TT{col}}_{a,b})^{-1}, \v{V}^{\TT{col}}_{c,d} ] \simeq 0$ and $[\v{V}_{a,b} (\v{V}^{\TT{col}}_{a,b})^{-1} , \overline{\v{C}}_{j} ]\simeq 0$, derived in the previous section, and $\v{V}_{c,a} = \v{V}_{c,b} \v{V}_{b,a}$. Notice in the above expressions the theta functions present in $\tilde{\v{C}}_{1}$ and $\v{V}^{\TT{tcol}}_{q_{1 \, \bot},Q}$ are always unity on hard legs as the ordering guarantees their argument is satisfied. We will now show that if \eqref{eqn:assumption} is true for $\v{A}_{n}$, it is also true for $\v{A}_{n+1}$. We begin by noting that from the Markovian way our algorithm evolves, we can write $\v{A}_{n+1} \equiv \hat{\v{A}}_{n}(\mu , q_{1 \, \bot})$ where $\hat{\v{A}}_{n}(\mu , q_{1 \, \bot})$ is computed using our algorithm (as described in \eqref{eq:Aevo}) however with the evolution initiated by $\hat{\v{H}}(q_{1 \, \bot}) = \v{D}_{1} \v{V}_{q_{1 \, \bot},Q} \v{H}(Q) \v{V}^{\dagger}_{q_{1 \, \bot},Q} \v{D}^{\dagger}_{1}$ and with the parton momentum indexed as $2, 3, 4, ... \,$. From this we can use \eqref{eqn:assumption} to write 
\begin{align}
\Tr \, \v{A}_{n+1}(\mu) = \Tr \, \hat{\v{A}}_{n}(\mu , q_{1 \, \bot}) = \sum^{n}_{m=0} \Tr \,\left(\hat{\v{tCol}}^{\dagger}_{m}(\mu, q_{1 \, \bot}) \circ  \hat{\v{tCol}}_{m}(\mu, q_{1 \, \bot}) \hat{\v{A}}^{\TT{soft}}_{n-m}(\mu, q_{1 \, \bot}) \right),
\end{align}
where $\hat{\v{A}}^{\TT{soft}}_{n-m}(\mu, q_{1 \, \bot})$ are generated by the same algorithm as $\v{A}^{\TT{soft}}_{n-m}(\mu)$ however using $\hat{\v{H}}(q_{1 \, \bot})$ as the initial condition. $\hat{\v{tCol}}_{m}(\mu, q_{1 \, \bot})$ are generated using the iterative relation in \eqref{eqn:full_col_defs} but with an initial condition $\hat{\v{tCol}_{0}}(q_{\bot}, q_{1 \, \bot}) = \v{V}^{\TT{tcol}}_{q_{\bot},q_{1 \, \bot}}$. Next we split apart $\hat{\v{H}}(q_{1 \, \bot})$ as 
\begin{align}
\hat{\v{H}}(q_{1 \, \bot}) =&  \v{S}_{1} \v{V}^{\TT{tcol}}_{q_{1 \, \bot},Q} \v{V}^{\TT{soft}}_{q_{1 \, \bot},Q} \v{H}(Q) \v{V}^{\TT{soft} \, \dagger}_{q_{1 \, \bot},Q} \v{V}^{\TT{tcol} \, \dagger}_{q_{1 \, \bot},Q}  \v{S}^{\dagger}_{1} \nonumber \\ 
& + \tilde{\v{C}}_{1} \v{V}^{\TT{tcol}}_{q_{1 \, \bot},Q} \v{V}^{\TT{soft}}_{q_{1 \, \bot},Q} \v{H}(Q) \v{V}^{\TT{soft} \, \dagger}_{q_{1 \, \bot} ,Q} \v{V}^{\TT{tcol} \, \dagger}_{q_{1 \, \bot},Q} \tilde{\v{C}}^{\dagger}_{1}.
\end{align}
Using the commutation relations from Section \ref{sec:hard_factorisation}, we can move the collinear operators in $\hat{\v{H}}(q_{1 \, \bot})$ past the soft operators which construct $\hat{\v{A}}^{\TT{soft}}_{n-m}(\mu, q_{1 \, \bot})$ to arrive at 
\begin{align}
&\Tr \, \v{A}_{n+1}(\mu) = \sum^{n}_{m=0} \Tr \,\left(\v{V}^{\TT{tcol} \, \dagger}_{q_{1 \, \bot},Q}\hat{\v{tCol}}^{\dagger}_{m}(\mu, q_{1 \, \bot}) \circ  \hat{\v{tCol}}_{m}(\mu, q_{1 \, \bot}) \v{V}^{\TT{tcol}}_{q_{1 \, \bot},Q} \v{A}^{\TT{soft}}_{n+1-m}(\mu, Q) \right) \nonumber \\
& + \sum^{n}_{m=0} \Tr \,\left(\v{V}^{\TT{tcol} \, \dagger}_{q_{1 \, \bot},Q} \tilde{\v{C}}^{\dagger}_{1} \hat{\v{tCol}}^{\dagger}_{m}(\mu, q_{1 \, \bot}) \circ  \hat{\v{tCol}}_{m}(\mu, q_{1 \, \bot}) \tilde{\v{C}}_{1}  \v{V}^{\TT{tcol}}_{q_{1 \, \bot},Q} \v{A}^{\TT{soft}}_{n-m}(\mu, Q) \right).
\end{align}
We can now combine the collinear operators using 
$$\hat{\v{tCol}}_{m}(\mu, q_{1 \, \bot}) \tilde{\v{C}}_{1}  \v{V}^{\TT{tcol}}_{q_{1 \, \bot},Q} = \v{tCol}_{m+1}(\mu)\theta(q^{\TT{coll}}_{1 \, \bot} > q^{\TT{soft}}_{1 \, \bot})$$
and 
$$\hat{\v{tCol}}_{m}(\mu, q_{1 \, \bot}) \v{V}^{\TT{tcol}}_{q_{1 \, \bot},Q} = \v{tCol}_{m}(\mu)\theta(q^{\TT{coll}}_{1 \, \bot} < q^{\TT{soft}}_{1 \, \bot}),$$
where in the second equality we need to relabel the momenta of collinear partons again so that they are indexed as $1,2,3,... \,$. We have denoted the momentum of the hardest collinear emission generated by the collinear operators, $\v{tCol}$, as $q^{\TT{coll}}_{1}$ and the hardest soft momentum in $\v{A}^{\TT{soft}}_{n-m}(\mu, Q)$ as $q^{\TT{soft}}_{1}$. Combining the two sums and theta functions, we arrive at 
\begin{align}
\Tr \, \v{A}_{n+1}(\mu) = \sum^{n+1}_{m=0} \Tr \,\left(\v{tCol}^{\dagger}_{m}(\mu) \circ  \v{tCol}_{m}(\mu) \v{A}^{\TT{soft}}_{n+1-m}(\mu) \right).
\end{align}
Thus we have proven that \eqref{eqn:full_col_factorisation} holds for $n \to n+1$. It is important to note the role of the theta functions in the definitions of $\tilde{\v{C}}_{i}$ and $\v{V}^{\TT{tcol}}_{a,b}$. These ensure that the commutation relations from Section \ref{sec:hard_factorisation} can always applied. They do this by squeezing to zero the phase space of any collinear partons generated by $\tilde{\v{C}}_{1}$ and $\v{V}^{\TT{tcol}}_{q_{1 \, \bot},Q}$ from not-hard legs. To illustrate this point, we will consider the relevant Feynman diagrams:
\begin{eqnarray}
\begin{split}
& \left[
\begin{array}{c}
\includegraphics[width=0.2\textwidth]{DC_commutator_1} 
\end{array} + 
\begin{array}{c}
\includegraphics[width=0.2\textwidth]{DC_commutator_2} 
\end{array} \right] \Theta(q_{j \, \bot} - q_{i \, \bot}) \Theta(q_{i \, \bot} - \mu) \Theta(Q - q_{j \, \bot})\\
&+ \left[
\begin{array}{c}
\includegraphics[width=0.2\textwidth]{DC_commutator_3} 
\end{array} + 
\begin{array}{c}
\includegraphics[width=0.2\textwidth]{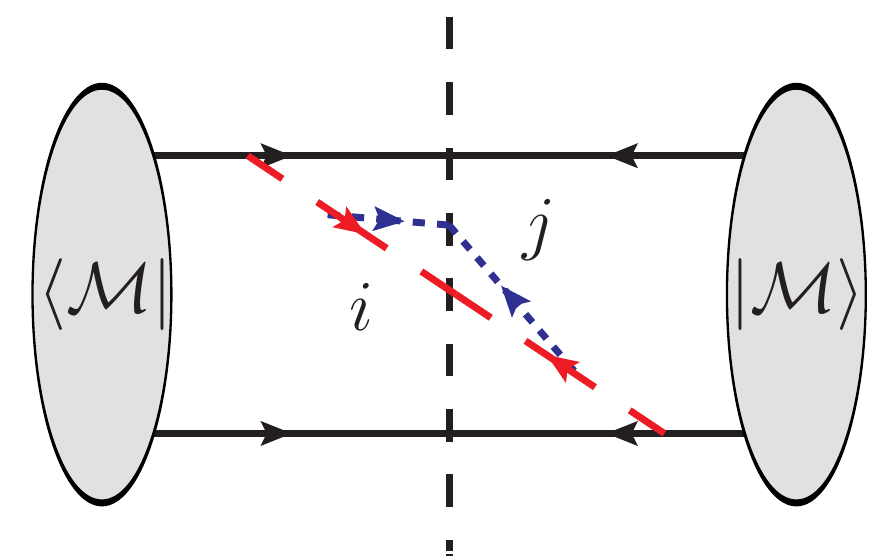} 
\end{array}  \right] \Theta(q_{i \, \bot} - q_{j \, \bot}) \Theta(q_{j \, \bot} - \mu) \Theta(Q - q_{i \, \bot})\\ 
&= 
\begin{array}{c}
\includegraphics[width=0.2\textwidth]{DC_commutator_3} 
\end{array} \Theta(q_{i \, \bot} - \mu) \Theta(Q - q_{i \, \bot}) \Theta(q_{j \, \bot} - \mu) \Theta(Q - q_{j \, \bot})\\
& + 
\begin{array}{c}
\includegraphics[width=0.2\textwidth]{extra_term_1} 
\end{array} \Theta(q_{i \, \bot} - q_{j \, \bot}) \Theta(q_{j \, \bot} - \mu) \Theta(Q - q_{i \, \bot}).
\end{split} \label{eqn:problem_with_factorisation}
\end{eqnarray}
Note that the last term on either side of the equation cannot be manipulated using our commutators and there are no more diagrams we could include which they may cancel against. Nevertheless terms of this form are generated by our algorithm. They represent a collinear parton, emitted from a soft parton, restricted so that its transverse momentum is smaller than the transverse momentum of the soft parton. Using $\tilde{\v{C}}_{i}$, equation \eqref{eqn:problem_with_factorisation} reduces to
\begin{align}
&\< \mathcal{M} \rkl \v{D}^{\dagger}_{1}\v{D}^{\dagger}_{2}\v{D}_{2}\v{D}_{1} \lkl \mathcal{M} \> \Theta(q_{1 \, \bot} - q_{2 \, \bot}) \Theta(q_{2 \, \bot} - \mu) \Theta(Q - q_{1 \, \bot}) \nonumber \\ 
&= \< \mathcal{M} \rkl \v{S}^{\dagger}_{1}\tilde{\v{C}}^{\dagger}_{2}\tilde{\v{C}}_{2}\v{S}_{1} \lkl \mathcal{M} \> \Theta(Q > q_{1 \, \bot} > \mu ) \Theta(Q > q_{2 \, \bot} > \mu ).
\end{align}

\subsection{Partial collinear factorisation with Coulomb interactions}
\label{sec:factorise_with_coloumb}
\begin{figure}[h]
	\centering
	\includegraphics[width=0.98\textwidth]{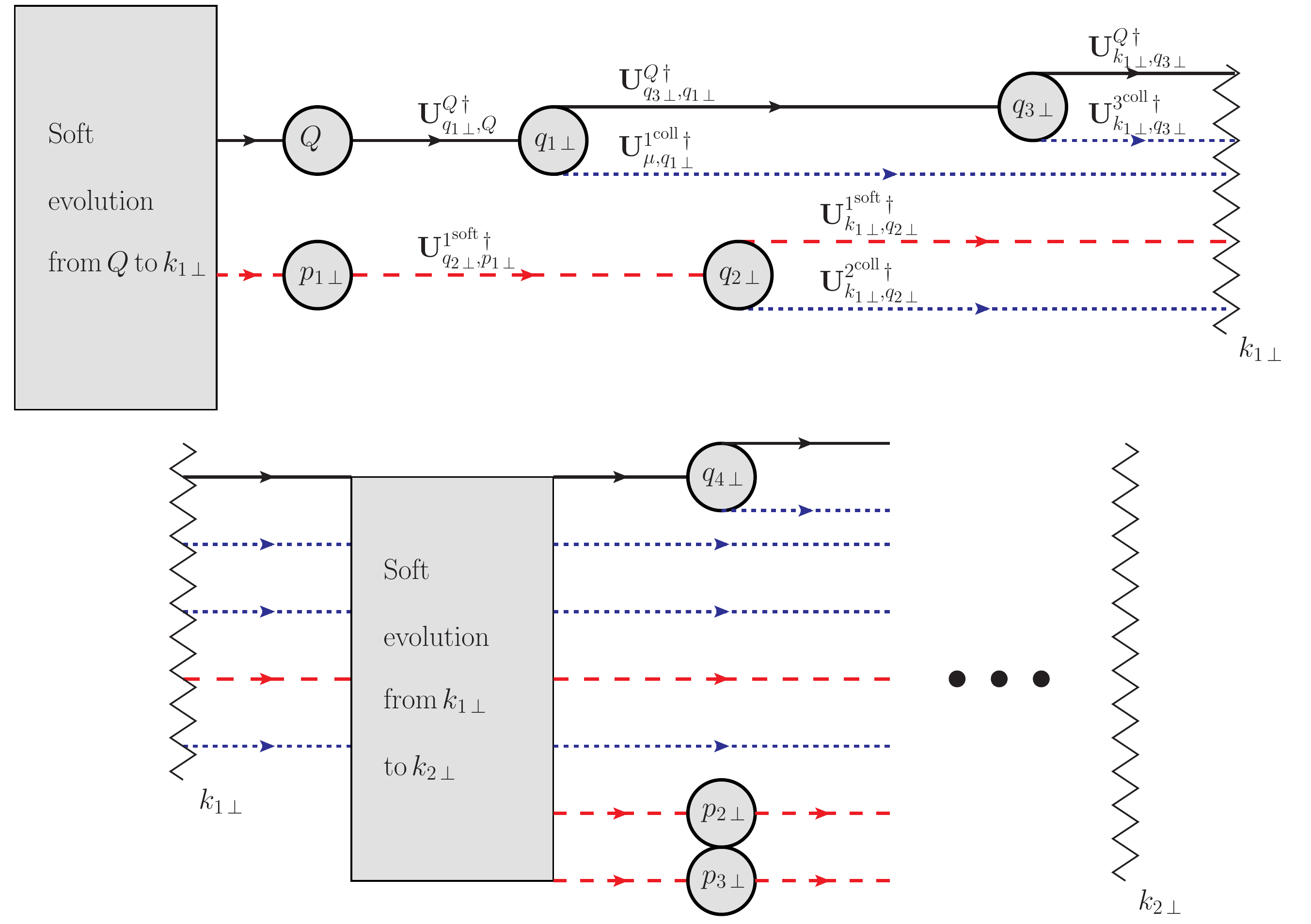}
	\caption{A diagram illustrating factorised parton evolution including Coulomb exchanges.  Red dashed lines represent soft gluons and blue dotted lines represent collinear partons. Each line is associated with a Sudakov factor. Circles represent the scale from which the subsequent evolution proceeds. Circles from which two lines leave represent the action of the operator $\tilde{\v{C}}$. Circles from which one line leaves contain the scale information from the preceeding soft evolution. Coulomb exchanges are indicated by vertical zig-zag lines.  Momenta are ordered from `left to right', as in Figure \ref{fig:SC1F}, including Coulomb exchanges. (The top half of the diagram lies to the `left' of the bottom half.)}
	\label{fig:SC1FCoulomb}
\end{figure}
Though it is not possible to use the identities in \eqref{eqn:commutators} to factorise collinear physics past a Coulomb exchange ($i\pi$ term), it is possible to perform a partial factorisation. Our approach is to expand each Sudakov operator as a series in the number of Coulomb exchanges it resums. Consequently, this enables $\v{A}_{n}$ to be expanded as a series in the number of Coulomb exchanges. We can then factorise soft physics from collinear physics either side of a Coulomb exchange, using our work in the previous section. The partial factorisation is illustrated in Figure \ref{fig:SC1FCoulomb}. 
We begin by expanding the Sudakov operator:
\begin{equation}
\begin{split}
\v{V}_{a,b} =& \hat{\v{V}}_{a,b} - \frac{\alpha_{s}}{\pi} \sum_{i_{1}<j_{1}} \int^{b}_{a} \frac{\td k^{(i_{1}j_{1})}_{1 \, \bot}}{k^{(i_{1}j_{1})}_{1 \, \bot}} \hat{\v{V}}_{a,k_{1\, \bot}} (\mathbb{T}^{g}_{i_{1}} \cdot \mathbb{T}^{g}_{j_{1}}) \, i\pi \, \tilde{\delta}_{i_{1}j_{1}} \hat{\v{V}}_{k_{1 \, \bot},b}\\
& + \left(\frac{\alpha_{s}}{\pi}\right)^{2}\sum_{i_{2}<j_{2}} \int^{b}_{a} \frac{\td k^{(i_{1}j_{1})}_{1 \, \bot}}{k^{(i_{1}j_{1})}_{1 \, \bot}} \sum_{i_{1}<j_{1}} \int_{a}^{k^{(i_{1}j_{1})}_{1 \, \bot}}  \frac{\td k^{(i_{2}j_{2})}_{2 \, \bot}}{k^{(i_{2}j_{2})}_{2 \, \bot}}   \hat{\v{V}}_{a,k_{2 \, \bot}} (\mathbb{T}^{g}_{i_{2}} \cdot \mathbb{T}^{g}_{j_{2}}) \, i\pi \, \tilde{\delta}_{i_{2}j_{2}} \\
& \qqquad \times \hat{\v{V}}_{k_{2 \, \bot},k_{1 \, \bot}} (\mathbb{T}^{g}_{i_{1}} \cdot \mathbb{T}^{g}_{j_{1}}) \, i\pi \tilde{\delta}_{i_{1}j_{1}} \hat{\v{V}}_{k_{1 \, \bot},b} - ...,
\end{split} \label{eqn:expand_around_pi}
\end{equation}
where $\hat{\v{V}}_{a,b}$ is equal to $\v{V}_{a,b}$ with $\tilde{\delta}_{ij} = 0$. Consider using this expanded Sudakov in a parton cascade. The theta functions describing the integral limits on each $i\pi$ term can be used to constrain the limits on the transverse momenta of subsequent emissions (after the $i\pi$). For instance
\begin{align}
&\v{D}_{3}\v{V}_{q_{3 \, \bot},q_{2 \, \bot}}\v{D}_{2}\v{V}_{q_{2 \, \bot},q_{1 \, \bot}}\v{D}_{1} \nonumber \\
&\quad = ... \, + \v{D}_{3}\int^{q_{2 \, \bot}}_{q_{3 \, \bot}}\frac{\td k_{2 \, \bot}}{k_{2 \, \bot}}\hat{\v{V}}_{q_{3 \, \bot},k_{2 \, \bot}}\sum_{i_{2}<j_{2}}(\mathbb{T}^{g}_{i_{2}} \cdot \mathbb{T}^{g}_{j_{2}})  \, i \pi \, \hat{\v{V}}_{k_{2 \, \bot},q_{2 \, \bot}} \nonumber \\
&\qqquad \times \v{D}_{2} \int^{q_{1 \, \bot}}_{q_{2 \, \bot}}\frac{\td k_{1 \, \bot}}{k_{1 \, \bot}} \hat{\v{V}}_{q_{2 \, \bot},k_{1 \, \bot}} \sum_{i_{1}<j_{1}}(\mathbb{T}^{g}_{i_{1}} \cdot \mathbb{T}^{g}_{j_{1}}) \, i \pi \, \hat{\v{V}}_{k_{1 \, \bot},q_{1 \, \bot}} \v{D}_{1} + ... \, , \nonumber
\end{align}
\begin{align}
&\quad \equiv ... \, + \v{D}_{3}\int^{Q}_{\mu}\frac{\td k_{2 \, \bot}}{k_{2 \, \bot}}\hat{\v{V}}_{q_{3 \, \bot},k_{2 \, \bot}}\sum_{i_{2}<j_{2}}(\mathbb{T}^{g}_{i_{2}} \cdot \mathbb{T}^{g}_{j_{2}})  \, i \pi \, \hat{\v{V}}_{k_{2 \, \bot},q_{2 \, \bot}} \nonumber \\
&\qqquad \times \v{D}_{2} \int^{Q}_{\mu}\frac{\td k_{1 \, \bot}}{k_{1 \, \bot}} \hat{\v{V}}_{q_{2 \, \bot},k_{1 \, \bot}} \sum_{i_{1}<j_{1}}(\mathbb{T}^{g}_{i_{1}} \cdot \mathbb{T}^{g}_{j_{1}}) \, i \pi \, \hat{\v{V}}_{k_{1 \, \bot},q_{1 \, \bot}} \v{D}_{1} \nonumber \\
&\qqquad \times \Theta(k_{2 \, \bot} > q_{3 \, \bot}) \Theta(k_{1 \, \bot} > q_{2 \, \bot} > k_{1 \, \bot})\Theta(q_{1 \, \bot} > k_{1 \, \bot}) + .... \, .
\end{align}
Therefore, we can treat each Coulomb scale as hard relative to the emissions that follow it and soft relative to the emission before it. Thus we can perform a factorised evolution on a hard process up to the scale of the first $i\pi$ term ($k_{1 \bot}$). We can take the output from this evolution,
\begin{eqnarray}
\begin{split}
\sum_{n}\v{A}_{n}(k_{1 \, \bot}) =& - \frac{\alpha_{s}}{\pi} \sum_{i_{2}<j_{2}} \frac{\td k^{(i_{1}j_{1})}_{1 \, \bot}}{k^{(i_{1}j_{1})}_{1 \, \bot}} i\pi \tilde{\delta}_{i_{1}j_{1}} \sum_{n} \sum^{n}_{m=0} \\
& \times \Tr \,\left( \v{A}^{\TT{soft}}_{n-m}(k_{1 \, \bot}) \sum_{i_{1}<j_{1}}  \v{Col}^{\dagger}_{m}(k_{1 \, \bot}) \circ (\mathbb{T}^{g}_{i_{1}} \cdot \mathbb{T}^{g}_{j_{1}})  \v{Col}_{m}(k_{1 \, \bot}) \right),
\end{split}
\end{eqnarray}
and use it as a new hard process $\v{H}(k_{1 \, \bot})$ from which a second factorised evolution can be initiated. This process can be iterated for each $i\pi$ term in the expansion, as illustrated in Figure \ref{fig:SC1FCoulomb}. To complete the computation of $\Sigma$, each $k_{i \, \bot}$ must be integrated over the range $[\mu , Q]$. Interestingly, note that any term in the evolution terminating on a $i\pi$ term to the left of the hard process will cancel against an equivalent term terminating with an $i\pi$ term to the right. Hence collinear emissions can always be factorized below the scale of the last Coulomb exchange. This is consistent with the collinear factorisation shown by Collins, Soper and Sterman \cite{Collins:1987pm,Collins:1988ig}.

\subsection{Observations on factorisation}

Before we leave our discussion on factorisation a few comments are in order. Firstly, we have not been able to achieve factorisation of collinear emissions past Coulomb exchanges. This is to be expected and there is already extensive literature exploring this subject \cite{Collins:1988ig,Mangano:1990by,BERENDS1988759,factorisationBreaking,Catani:2011st,SuperleadingLogs,Rogers:2010dm,Rogers:2013zha,Aybat:2008ct}. That said, it should be possible to factorise more completely than we have done, by re-expressing the evolution so that all Coulomb terms are only attached to the initial state partons \cite{factorisationBreaking}, i.e. so we would have complete factorisation on all final state legs.

Secondly, in order to factorise the collinear physics on all legs we had to keep track of intermediate soft scales, from which to initialise the collinear evolution. The number of scales required is equal to the number of soft emissions that occurred prior to factorisation.  This means the fully factorised algorithm is no-longer Markovian. We anticipate that our attempts to factorise the collinear physics should bring us in to contact with exact resummations and soft-collinear effective theory (SCET).

It also should be noted that by factorising collinear emissions from the soft evolution, the soft evolution can be explicitly seen to be independent of spin. This is less evident in the interleaved variants of the algorithm. Soft gluons, and subsequent collinear partons, trapped between Coulomb exchanges might conceivably contribute non-trivial spin correlations. This is because, despite equal probabilities for the probability of emission of positive and negative helicity gluons, a collinear emission originating from a soft gluon may depend on its helicity (specifically  $g\rightarrow qq$ splitting). This has also been explored in the literature, where it has been noted that soft gluons in the presence of Coulomb/Glauber exchanges can generate spin asymmetries \cite{Rogers:2013zha}. Further discussions on the spin evolution of the algorithm after factorisation can be found in Appendix \ref{Appendix_B}.

It is also interesting to consider the consequences of factorisation in the case of variant B with a universal recoil. A universal recoil allows B to be partitioned in terms of colour-diagonal evolution generated by $\v{C}_{i}$ and colour off-diagonal evolution generated by $\v{S}_{i}$. Hence, provided the recoil prescription does not change the commutators in \eqref{eqn:commutators}, the proofs of collinear factorisation we have presented become proofs of the complete factorisation of colour-diagonal physics from colour off-diagonal. This is for observables insensitive to the presence of Coulomb exchanges. Since we know that Coulomb exchanges can be factorised onto the initial state \cite{factorisationBreaking}, this means that there is a complete factorisation of colour-diagonal from colour off-diagonal physics in lepton-lepton, deep-inelastic and Drell-Yan scattering.

Finally, we should remark that it is possible to write down infra-red finite versions of each of the factorised versions of our algorithm, using the procedure in Section \ref{sec:algorithmIR}. 

\section{Phenomenology and resummations}
\label{sec:pheno}

In this section we will first demonstrate how DGLAP evolution emerges. After that, we illustrate the use of the algorithm by calculating thrust, the hemisphere jet mass and gaps-between-jets in $e^+ e^-$. 

\subsection{DGLAP evolution}

We will now show how our algorithm can be used to generate DGLAP evolution, which resums the collinear physics into the running of parton distribution functions. We focus on unpolarised incoming hadrons that collide and produce some high-$p_T$ system of interest. We will neglect threshold effects as sub-leading, which is shown carefully in \cite{Skands:2009tb,Nagy:2009re,Dokshitzer:2008ia}. The methods employed in this section can readily be extended to other processes, including those dependent on fragmentation functions.  

DGLAP evolution \cite{APSplitting,DOKSHITZER1980269} states that
\begin{eqnarray}
\mu \frac{\partial f_{i}(x,\mu)}{\partial \mu} = \frac{\alpha_{s}}{\pi} \sum_{j} \int^{1}_{x} \frac{\td z}{z} P_{ij}(z) \, f_{j}(x/z,\mu),
\end{eqnarray}
where $f_i(x,\mu)$ is the parton distribution function for partons of type $i$. $P_{ij}(z)$ are the regularised splitting functions defined at the end of Appendix \ref{Appendix_A}. Iterative solutions can be found by expanding the parton distributions:
\begin{eqnarray}
f_{i}(x,Q) = f^{(0)}_{i}(x) + \sum_{n=1}^{\infty} \left(\frac{\alpha_{s}}{\pi}\right)^{n} f^{(n)}_{i}(x,Q),
\end{eqnarray}
where $f^{(n)}_{i}(x,\mu) = 0$ for all $n \geq 1$. This gives
\begin{eqnarray}
f^{(n+1)}_{i}(x,q_{m-1 \, \bot}) = \int^{q_{m-1 \, \bot}}_{\mu} \frac{\td q_{m \, \bot}}{q_{m \, \bot}} \sum_{j} \int^{1}_{x} \frac{\td z_{m}}{z_{m}}\, P_{ij} ( z_{m} ) f^{(n)}_{j} (x/z_{m}, q_{m \, \bot} ),
\end{eqnarray}
which has a separable solution of the form $f^{(n)}_{i}(x,Q) = f^{(n)}_{i}(x) \frac{1}{n!} \ln^{n}(Q/\mu)$, where $f^{(n)}_{i}(x)$ satisfies
\begin{eqnarray}
f^{(n+1)}_{i}(x) = \sum_{j} \int^{1}_{x} \frac{\td z_{m}}{z_{m}}\, P_{ij} ( z_{m} ) f^{(n)}_{j} (x/z_{m}).
\label{eqn:DGLAP_seperable_solution}
\end{eqnarray}
We can write this in terms of the unregularised splitting functions (e.g. see \cite{DOKSHITZER1980269})
\begin{eqnarray}
f^{(n+1)}_{i}(x) = \sum_{j} \int^{1}_{0} \frac{\td z_{m}}{z_{m}}\, \left(\mathcal{P}_{ij} ( z_{m} ) f^{(n)}_{j} (x/z_{m}) - z^{2}_{m} \mathcal{P}_{ji} ( z_{m} ) f^{(n)}_{i} (x) \right), \label{eqn:DGLAP}
\end{eqnarray}
where $f^{(n)}_{j} (x) = 0$ for $x>1$ and we have removed factors of $n_{f}$ from $\mathcal{P}_{qg}$.
\begin{figure}[t]
	\centering
\includegraphics[width=\textwidth]{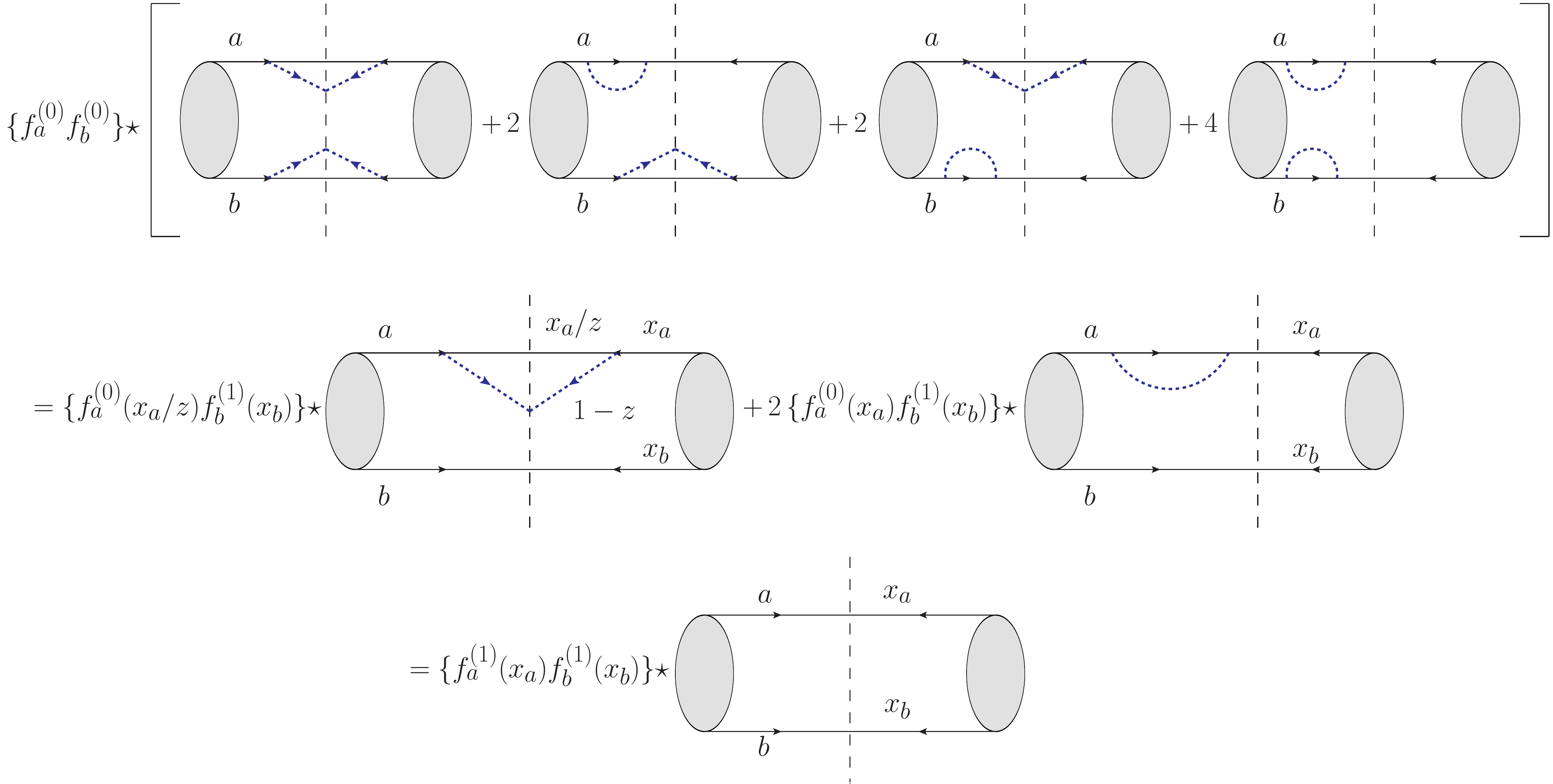}
	\caption{How DGLAP and fragmentation evolution can be constructed from the $\alpha_{s}$ expansion of our algorithm. The vertical dashed lines here correspond to a cut on all external legs (incoming and outgoing) and the grey blobs represent the hard process, i.e. the amplitude evolves from the right grey blob to the vertical dashed line and the conjugate amplitude evolves from the left grey blob to the vertical dashed line. Solid lines indicate hard partons and blue lines collinear partons. }
	\label{fig:DGLAP}
\end{figure}

For hadron-hadron collisions, we label the two incoming partons as $a$ and $b$ and their momentum fractions in the hard process as $x_{a}$ and $x_{b}$. We can take the factorised expression corresponding to variant B of our algorithm \eqref{eqn:hard_col_factorisation} and attach parton distribution functions:
\begin{eqnarray}
\begin{split}
\Sigma = \int \sum_{n} \left(\prod^{n}_{i = 1} \td \Pi_{i} \right)\sum^{n}_{m=0}  \sum^{n-m}_{p=0} \int \td x_{a} \td x_{b} \,& \Tr \bigg( \v{Col}^{\dagger}_{m}(\mu) \circ \v{Col}_{m}(\mu) \\ 
&\times \v{B}^{p}_{n-m-p}(\mu)\bigg) \star \left\{f^{(0)}_{a}(x_{a_{m}})f^{(0)}_{b}(x_{b_{m}}) \right\}.
\end{split}
\label{eqn:col-FKS_hadronic_convo}
\end{eqnarray}
$x_{a_{m}}$ and $x_{b_{m}}$ are the momentum fractions of partons $a$ and $b$ respectively after $m$ collinear emissions generated by $ \v{Col}^{\dagger}_{m}(\mu) \circ \v{Col}_{m}(\mu)$; they can be related back to $x_{a}$ and $x_{b}$ by momentum conservation along the collinear cascade. The $\star$ operator acts to attach parton distributions of the correct flavour/species to partons $a$ and $b$. There is a technicality relating to parton flavour. That is because DGLAP evolution cares about quark flavour whilst we have defined the splitting operators to sum over quark flavours (in the case $g \to q \bar{q}$). We could have avoided this technicality by defining the splitting operators per flavour (i.e. set $n_f=1$ throughout Appendix \ref{Appendix_A}). Then we would have to sum over quark flavours throughout the rest of the paper. Instead, we choose to handle quark flavour by keeping track of flavour along the evolution chain, and whenever a $g \to q \bar{q}$ splitting occurs we label the subsequent parton flavour generically, i.e. for two-flavours the relevant set of parton flavours would be $\{ u, \bar{u}, d, \bar{d}, q, g\}$. Note that since we evolve {\it away from} the hard scattering, a $g \to q \bar{q}$ branching from an incoming $g$ actually corresponds to a $q \to q g$ (or $\bar{q} \to \bar{q} g$) splitting in the usual DGLAP sense. This can be seen in \eqref{eq:appsplit}, where the terms involving $\delta^\text{initial}_j$ and $s_j = \pm 1$ involve the $\mathcal{P}_{gq}$ splitting function (the $\mathcal{P}_{qg}$ splitting function appears in the corresponding $\delta^\text{initial}$ terms). With this in mind we have that
\begin{align}
\td \sigma \, \star \left\{ f^A \, f^B  \right\} = \sum_{\alpha,\beta} \td \sigma_{\alpha,\beta} \, f^A_\alpha \, f^B_\beta , 
\end{align}   
where $\alpha$ and $\beta$ label parton type. For $n_f = 2$, we would have $\alpha,\beta \in \{ u, \bar{u}, d, \bar{d}, q, g \}$, and $2 n_f f_q$ is the singlet distribution function, i.e.  $f_q = (u + \bar{u} + d + \bar{d})/(2 n_f)$ for $n_f = 2$. For completeness, we here also attach labels $A$ and $B$ to indicate the type of hadron (we will drop that label elsewhere in this section).  

After expanding $\v{Col}^{\dagger}_{m}(\mu) \circ \v{Col}_{m}(\mu)$ in powers of $\alpha_{s}$, then spin averaging at every vertex, substituting for \eqref{eqn:DGLAP_seperable_solution} and evaluating the transverse momentum integrals, we find
\begin{eqnarray}
\begin{split}
\Sigma =& \int \sum_{n} \sum^{n}_{m=0}  \sum^{n-m}_{p=0} \left(\prod^{n-m}_{i = 1} \td \Pi_{i} \right) \\
&\times \int \td x_{a} \td x_{b}\, \Tr\left( \v{B}^{p}_{n-m-p}(\mu)\right)  \star \left\{f_{a}(x_{a},Q)f_{b}(x_{b},Q) + \mathcal{O}(\alpha^{m+1}_{s})\right\}.
\end{split} \label{eqn:SC_DGLAP}
\end{eqnarray}
Figure \ref{fig:DGLAP} illustrates how terms in our algorithm should be grouped in order to generate the iterative relation in \eqref{eqn:DGLAP} and so arrive at \eqref{eqn:SC_DGLAP}. Hence we see that variant B iteratively generates DGLAP evolution up to the hard scale. The derivation of fragmentation function evolution of final-state partons proceeds similarly.

For processes where Coulomb exchanges are relevant, DGLAP evolution is generated up to the scale of the last Coulomb exchange.  Also note that, in the infra-red finite reformulations of A and B, DGLAP can be found in the $\v{B}_{n}$ for $n \geq 1$. 
\subsection{Some familiar resummations}

In this subsection we show how to resum a number of observables in $e^{+}e^{-}\rightarrow \TT{hadrons}$. The idea is to use well-known results to illustrate the use of the algorithm. The simplicity of the hard process means we can use $\v{H}(Q) = N^{-1}_{c}\sigma_{\TT{H}} \mathbbm{1}$  for the hard-scattering matrix. We perform all calculations using the LL recoil from Section \ref{sec:LLrecoil}.

\subsubsection{Thrust}

The resummed thrust distribution was initially computed at LL accuracy in \cite{Binetruy:1980hd}, then at NLL in \cite{Resum_large_logs_ee}. The current state-of-the-art computation is at $\TT{N}^{3}$LL \cite{Becher:2008cf}. Thrust is defined as
\begin{eqnarray}
T = \max_{\v{n}}  \frac{\sum_{\forall \v{p}}|\v{p}\cdot\v{n}|}{\sum_{\forall \v{p}}|\v{p}|},
\end{eqnarray}
where the thrust axis $\v{n}$ points along the initial hard parton axis at leading-order in the soft and collinear limits; see Section 3.1 in \cite{Resum_large_logs_ee}. We will only need to define thrust for 3 partons; of which 2 are hard ($p_{1}$ and $p_{2}$) and one is soft ($k$). We can work in the dipole zero-momentum frame using $p_{q}=E_{q}(1,0,0,1)$, $p_{\bar{q}}=E_{\bar{q}}(1,0,0,-1)$, $k=(k_{\bot}\cosh y, \vec{k}_{\bot}, k_{\bot}\sinh y)$, and we can fix $2E_{q} = 2E_{\bar{q}}=Q$. Thrust is evaluated as
\begin{eqnarray}
T = 1 - \frac{k_{\bot}\cosh |y| - k_{\bot}\sinh |y|}{Q} + \mathcal{O}\left(\frac{k^{2}_{\bot}}{Q^{2}}\right).
\end{eqnarray}
When calculated in the hard-collinear limit, we can let $T=1 + \mathcal{O}(k^{2}_{\bot})$ as all partons lie on the thrust axis up to sub-leading contributions.
The thrust distribution $R_{T}$ is defined as
\begin{eqnarray}
R_{T} =  \int  \td T' \frac{1}{\Tr(\v{H}(Q))} \tdf{\Sigma}{T'}.
\end{eqnarray}
As thrust is global, the calculation is most readily performed using the manifestly infra-red finite version of variant A (see Section \ref{sec:algorithmIR}). Using this, all terms with one or more emissions cancel exactly. The measurement function $u(k, \{ \emptyset \})$ is 
\begin{eqnarray}
u(k, \{ \emptyset \})= \Theta\left( 1 - \frac{k^{(q\bar{q})}_{\bot}\cosh |y| - k^{(q\bar{q})}_{\bot}\sinh |y|}{Q} - T \right).
\end{eqnarray}
This is unity for a hard-collinear emission, since $|y| \rightarrow \infty$ at LL accuracy. This kills the hard-collinear terms, since they contain a factor $(1-u(k, \{ \emptyset \}))= 0$, which is as expected since they contribute no double logarithms. Thus we can immediately write
\begin{eqnarray}
\begin{split}
\Sigma(T) &= \Tr(\overline{\v{V}}_{0,Q}\overline{\v{V}}^{\dagger}_{0,Q})\sigma_{\TT{H}}\\
&= \Tr \left( \exp \left[ \frac{2\alpha_{s}}{\pi} \mathbb{T}^{g}_{q} \cdot \mathbb{T}^{g}_{\bar{q}} \int^{Q}_{0} \frac{\td k^{(q\bar{q})}_{\bot}}{k^{(q\bar{q})}_{\bot}}\int \frac{\td S_{2} }{4\pi}\theta_{ij}(k)\right. \right. \\ 
& \qquad \left. \left. \times \Theta\left(T- 1 + \frac{k^{(q\bar{q})}_{\bot}\cosh |y| - k^{(q\bar{q})}_{\bot}\sinh |y|}{Q} \right)   k_{\bot}^{2}\frac{p_{q}\cdot p_{\bar{q}}}{(p_{q}\cdot k)(p_{\bar{q}} \cdot k)} \right] \right)N^{-1}_{c}\sigma_{\TT{H}}.
\end{split}
\end{eqnarray}
After integrating
\begin{eqnarray}
\begin{split}
\Sigma(T) &=  \Tr \left( \exp \left[ \frac{2\alpha_{s}}{\pi} \mathbb{T}^{g}_{q} \cdot \mathbb{T}^{g}_{\bar{q}} \int^{Q}_{0} \frac{\td k^{(q\bar{q})}_{\bot}}{k^{(q\bar{q})}_{\bot}} 2 \int^{\infty}_{0} \td y \, \theta_{ij}(k)\Theta\left(\tfrac{k_{\bot}}{Q}e^{-y} - (1-T) \right) \right] \right)N^{-1}_{c}\sigma_{\TT{H}}, \\
&= \Tr \left( \exp \left[ \frac{\alpha_{s}}{\pi} \mathbb{T}^{g}_{q} \cdot \mathbb{T}^{g}_{\bar{q}} \ln^{2}\left( \frac{1}{1-T}\right)  \right] \right)N^{-1}_{c}\sigma_{\TT{H}}.
\end{split}
\end{eqnarray}
Here we used the fact that $\theta_{ij}(k)$ restricts the integration so that $k_{0} < Q$. The colour trace can be evaluated to give
\begin{eqnarray}
\begin{split}
\frac{\Sigma(T)}{N^{-1}_{c}\sigma_{\TT{H}}} &= \Tr(\mathbbm{1}) + \frac{\alpha_{s}}{\pi}\ln^{2}\left( \frac{1}{1-T}\right) \Tr \left( \mathbb{T}^{g}_{q} \cdot \mathbb{T}^{g}_{\bar{q}} \right) \\ 
& \qquad + \frac{1}{2!} \left[\frac{\alpha_{s}}{\pi}\ln^{2}\left( \frac{1}{1-T}\right) \right]^{2}\Tr \left( \mathbb{T}^{g}_{q} \cdot \mathbb{T}^{g}_{\bar{q}}\mathbb{T}^{g}_{q} \cdot \mathbb{T}^{g}_{\bar{q}} \right) + ... \\
&= \Tr(\mathbbm{1})  - \frac{\alpha_{s}}{\pi}\ln^{2}\left( \frac{1}{1-T}\right) \Tr \left( \mathbb{T}^{g}_{q} \cdot \mathbb{T}^{g}_{q} \right) \\ 
& \qquad + \frac{1}{2!}\left[\frac{\alpha_{s}}{\pi}\ln^{2}\left( \frac{1}{1-T}\right) \right]^{2}\Tr \left( \mathbb{T}^{g}_{q} \cdot \mathbb{T}^{g}_{q}\mathbb{T}^{g}_{q} \cdot \mathbb{T}^{g}_{q} \right) + ... \\
&= \Tr(\mathbbm{1})  - \frac{\alpha_{s}}{\pi}\mathcal{C}_{\TT{F}}\ln^{2}\left( \frac{1}{1-T}\right) \Tr(\mathbbm{1}) +\left[\frac{\alpha_{s}}{\pi}\mathcal{C}^{2}_{\TT{F}}\ln^{2}\left( \frac{1}{1-T}\right) \right]^{2}\Tr(\mathbbm{1}) + ... \\
&= N_\TT{c} \exp \left[ - \frac{\alpha_{s}}{\pi} \mathcal{C}_{\TT{F}} \ln^{2}\left( \frac{1}{1-T}\right)  \right].
\end{split}
\end{eqnarray}
And so we obtain the familiar result:
\begin{eqnarray}
\begin{split}
R_{T} &=  \int  \td T' \frac{1}{\Tr(\v{H}(Q))} \tdf{\Sigma}{T'} \\
&=- \frac{\alpha_{s}}{\pi} \mathcal{C}_{\TT{F}} \int \td T' \, \frac{\ln\left(1-T'\right)}{1-T'} \exp \left[ - \frac{\alpha_{s}}{\pi} \mathcal{C}_{\TT{F}} \ln^{2}\left( \frac{1}{1-T'}\right)  \right].
\end{split}
\end{eqnarray}

\subsubsection{Hemisphere jet mass}

The hemisphere jet mass is subject to non-global logarithms, which greatly increase the challenge of resummation. It was first resummed at LL and LC in \cite{Dasgupta:2001sh}. The current state-of-the-art is split between fixed-order computation ($\alpha^{5}_{s}$ with leading colour \cite{Schwartz:2014wha} and $\alpha^{4}_{s}$ with full colour \cite{Khelifa-Kerfa:2015mma}) and resummation using numerical techniques to introduce full colour dependence with sub-leading logarithms \cite{Hagiwara:2015bia}. The measurement function corresponding to the hemisphere jet mass in $e^{+}e^{-}\rightarrow \TT{hadrons}$ is
\begin{eqnarray}
u_{\pm}(\{ q_{i} \})= \prod_{q \in \{ q_{i} \}}\left(\Theta(q \in S^{\mp}_{2}) + \Theta(q \in S^{\pm}_{2})\Theta(\rho - m_{\pm})\right),
\end{eqnarray}
where $S^{+}_{2}$ and $S^{-}_{2}$ are the hemispheres centred on the two primary jets. $m_{\pm}$ is the total invariant mass in the $S^{\pm}_{2}$ hemisphere. The measurement function can be simplified by considering $m_{\pm}$
\begin{eqnarray}
m_{\pm}^2 = \sum_{q_{i},q_{j} \in S^{\pm}_{2}} 2 q_{i} \cdot q_{j} = \sum_{q_{i},q_{j} \in S^{\pm}_{2}} 2E_{i}E_{j}\left(1-\sqrt{1-\frac{q^{2}_{i \, \bot}}{E^{2}_{i}}} \, \right).
\end{eqnarray}
At the order we will perform the calculation we only need to consider one emission, hence
\begin{eqnarray}
\begin{split}
u_{\pm}(q_{1})= \Theta(q_{1} \in S^{\mp}_{2}) + \Theta(q_{1} \in S^{\pm}_{2})\Theta\left(\rho^2 - 2E_{1}E_{\pm}+2E_{1}E_{\pm}\sqrt{1-q^{2}_{1 \, \bot}/E^{2}_{1}}\right), \\
= \left\{\begin{array}{ll}
\Theta(q_{1} \in S^{\mp}_{2}) + \Theta(q_{1} \in S^{\pm}_{2})\Theta\left(\rho^2 - Q(q_{1 \,  \bot}\cosh y - q_{1 \, \bot}\sinh y)\right) & \; \TT{for} \, E_{1} \ll E_{\pm},\\
\Theta(q_{1} \in S^{\mp}_{2}) + \Theta(q_{1} \in S^{\pm}_{2})\Theta\left(\rho^2 - \frac{E_{\pm}}{E_{1}}q^{2}_{1 \, \bot})\right) & \; \TT{for} \, q_{1 \, \bot} \ll E_{1},
\end{array} \right.
\end{split}
\end{eqnarray}
where $E_{\pm}$ is the energy of the quark/anti-quark defining the $S^{\pm}_{2}$ hemisphere and $Q=2E_{\pm}$. Note the similarity between this and the measurement function for thrust, which is expected since it is well known that, at lowest-order, thrust can be expressed as the sum over the two hemisphere jet masses defined by the thrust axis.

Again, we can use the manifestly infra-red finite version of A to find $\Sigma(\rho)$:
\begin{eqnarray}
\begin{split}
\Sigma(\rho) =& \Tr(\overline{\v{V}}_{0,Q}\overline{\v{V}}^{\dagger}_{0,Q})N^{-1}_{c}\sigma_{\TT{H}} + \int \td \Pi_{1} \Tr \left[\overline{\v{V}}_{0,q_{1 \, \bot}}\v{D}_{1}\overline{\v{V}}_{q_{1 \, \bot},Q}\overline{\v{V}}^{\dagger}_{q_{1 \, \bot},Q} \v{D}^{\dagger}_{1} \overline{\v{V}}^{\dagger}_{0,q_{1 \, \bot}}u(q_{1})  \right. \\
& \left. \quad\overline{\v{V}}_{0,q_{1 \, \bot}}\left\{ \overline{\v{V}}_{q_{1 \, \bot},Q}\overline{\v{V}}^{\dagger}_{q_{1 \, \bot},Q}, \frac{1}{2}\v{D}^{2}_{1} \right\}\overline{\v{V}}^{\dagger}_{0,q_{1 \, \bot}} u(q_{1}, \{\emptyset\}) \right]\Theta(Q - q_{1 \, \bot})N^{-1}_{c}\sigma_{\TT{H}} + ...
\end{split}
\end{eqnarray}
 From the calculation in the previous section, we can immediately write
\begin{eqnarray}
\begin{split}
\Tr(\overline{\v{V}}_{0,Q}\overline{\v{V}}^{\dagger}_{0,Q}) = N_\TT{c}\exp \left[ - \frac{2\alpha_{s}}{\pi} \mathcal{C}_{\TT{F}} \ln^{2}\left( \frac{Q}{\rho}\right)  \right].
\end{split}
\end{eqnarray}
This gives the global contribution. The non-global contributions are found by evaluating the remaining terms (corresponding to summing over real emissions in \eqref{eq:nglsum}). This calculation can be found in \cite{SoftEvolutionAlgorithm}, where the non-global terms are evaluated using the FKS algorithm, which is entirely sufficient in this case. Hence we find
\begin{eqnarray}
\begin{split}
\Sigma(\rho)  =& \sigma_{\TT{H}} \exp\left[- \frac{2\alpha_{s}}{\pi} \mathcal{C}_{\TT{F}} \ln^{2}\left( Q/\rho \right)\right] \left(1 - \mathcal{C}_{\TT{A}}\mathcal{C}_{\TT{F}}\zeta(2)\left(\frac{\alpha_{s}}{\pi}\right)^{2}\frac{\ln(Q/\rho)^{2}}{2} \right. \\
& \left. - \mathcal{C}^{2}_{\TT{A}}\mathcal{C}_{\TT{F}}\zeta(3)\left(\frac{\alpha_{s}}{\pi}\right)^{3}\frac{\ln(Q/\rho)^{3}}{3!} + ...\right).
\end{split}
\end{eqnarray}

\subsubsection{Gaps-between-jets}

The LL measurement function in the case of gaps-between-jets is
\begin{eqnarray}
u_{n} (q_{1},..,q_{n}) = \prod_{m=1}^{n}(\Theta_{\TT{out}}(q_{m})+\Theta_{\TT{in}}(q_{m})\Theta(Q_{0}-q_{m, \bot})),
\end{eqnarray} 
where the `in' region corresponds to two cones centred on the two leading jets and the `out' region is the region between these cones. The observable vetoes emissions in the out region that have transverse momentum greater than $Q_0$. 
At order $\alpha^{5}_{s}$ this observable is sensitive to super-leading logarithms \cite{Forshaw:2006fk,SuperleadingLogs}. These will be correctly calculated using variants A, B and their manifestly infra-red finite versions, but not their factorised form unless Coulomb exchanges are interleaved as in Section \ref{sec:factorise_with_coloumb}). Using the manifestly infra-red finite version of variant A we correctly find that
\begin{eqnarray}
\begin{split}
\Sigma(\mu) &= \sigma_{\TT{H}}\exp\left[- \frac{2 \alpha_{s} \mathcal{C}_{\TT{F}}}{\pi} Y \ln(Q/Q_{0}) \right]\left(1 + \mathcal{O}(\alpha^{2}_{s})\right),
\end{split}
\end{eqnarray}
where $Y$ is the rapidity range of the out region and the $\left(1 + \mathcal{O}(\alpha^{2}_{s})\right)$ factor is the stack of non-global logarithms, which can be computed by considering real gluon emission into the out region, as encoded in \eqref{eq:nglsum}. These were calculated up to $\mathcal{O}(\alpha^{5}_{s})$ in \cite{SuperleadingLogs,Forshaw:2006fk,Keates:2009dn}. We note a kinematic maximum on the rapidity of an emitted gluon, i.e. $2 |y| < Y_{\TT{max}} = \ln\left(\frac{Q}{2Q_{0}}+\sqrt{\frac{Q}{2Q_{0}}-1}\right)$. This means that as $Y \to Y_{\TT{max}}$ all soft radiation goes into the in region.  At leading-order in the soft approximation $Y_{\TT{max}} = \ln(Q/Q_{0})$, i.e. for $Y \geq Y_\text{max}$ the observable becomes doubly logarithmic.

\section{Conclusions}
Our primary goal in writing this paper is to provide the theoretical basis for
the future development of a computer code that is able systematically to resum
enhanced logarithms due to soft and/or collinear partons including quantum
mechanical interference effects. The algorithm we present, and its variants,
are (mostly) Markovian and their recursive nature makes them well suited for
the task. First steps towards this goal are under development, using the
\verb=CVolver= code to perform the colour evolution
\cite{Platzer:2013fha,QCDTalks,HARPSTalks}.

The algorithms in this paper correctly account for the leading soft and/or
collinear logarithms, though we have been careful to try and present them in
such a way as to make the extension beyond LL possible. For example, we have
taken account of the momentum re-mappings that are necessary in order to
account for energy-momentum conservation and we have included $g \rightarrow q
\bar{q}$ transitions which are strictly single logarithmic. 

\section{Acknowledgments}

This work has received funding from the UK Science and Technology Facilities
Council (grant no. ST/P000800/1), the European Union’s Horizon 2020 research
and innovation programme as part of the Marie Skłodowska-Curie Innovative
Training Network MCnetITN3 (grant agreement no. 722104). JRF thanks the
Institute for Particle Physics Phenomenology in Durham for the award of an
Associateship and the Particle Physics Group of the University of Vienna for
financial support and kind hospitality. JH thanks the UK Science and
Technology Facilities Council for the award of a studentship. SP acknowledges
partial support by the COST actions CA16201 PARTICLEFACE and CA16108
VBSCAN. We would like to thank Matthew De Angelis for extensive
discussions. We would also like to thank Mrinal Dasgupta, Jack Helliwell and
Mike Seymour for helpful discussions and comments on the manuscript. Figures
have been prepared using JaxoDraw \cite{JaxoDraw}.

\newpage
\appendix
\section{Splitting functions}
\label{Appendix_A}

The splitting operator $\v{P}_{ij}$ (see Figure \ref{fig:momenta_definitions}), which is explicitly used in variant B of our algorithm, is built from the spin dependent DGLAP splitting functions \cite{HelicitySplitting,HelicitySplitting2}. It is an operator in colour and helicity spaces and is defined using the spinor-helicity formalism \cite{HelicityTechniques}. We use the convention $v(p,\lambda)=C\bar{u}^{\TT{T}}(p,\lambda)$ where $C=i\gamma^{2}\gamma^{0}$, which defines our crossing symmetry to have no global minus sign. Using rotational symmetry and parity invariance one generally can write $\mathcal{M}(\{\lambda_{i}\})=\mathcal{M}^{*}(\{-\lambda_{i}\})$ where $\mathcal{M}$ is a matrix element and $\{\lambda_{i}\}$ the set of helicity states on which $\mathcal{M}$ depends. Together these define the correct treatment for antiparticles, which should evolve as if they are particles with the opposite helicity. Thus $\v{P}_{ij}$ is 
\begin{eqnarray}
\label{eq:appsplit}
\begin{split}
	\v{P}_{ij} & =  \delta_{s_{j}, \frac{1}{2}} \delta^{\TT{final}}_{j} \left( \sqrt{\frac{\mathcal{P}_{qq}}{2\mathcal{C}_{\TT{F}}(1+z_{i}^{2})}} \frac{1}{\<q_{i} \tilde{p}_{j}\>} (\mathbb{T}^{g}_{j} \otimes \mathbb{S}^{+1_{i}}) + \sqrt{\frac{z_{i}^{2}\mathcal{P}_{qq}}{2\mathcal{C}_{\TT{F}}(1+z_{i}^{2})}} \frac{1}{ [\tilde{p}_{j} q_{i}]} (\mathbb{T}^{g}_{j} \otimes \mathbb{S}^{-1_{i}})\right. \\ 
	& \qquad  + \sqrt{\frac{\mathcal{P}_{gq}}{2\mathcal{C}_{\TT{F}}(2-2z_{i}+z_{i}^{2})}} \frac{1}{\<\tilde{p}_{j} q_{i}\>} \mathbb{W}^{ij} (\mathbb{T}^{g}_{j} \otimes \mathbb{S}^{+1_{i}}) 
	\left. + \sqrt{\frac{(1-z_{i})^{2}\mathcal{P}_{gq}}{2\mathcal{C}_{\TT{F}}(2-2z_{i}+z_{i}^{2})}} \frac{1}{ [q_{i} \tilde{p}_{j}]} \mathbb{W}^{ij} (\mathbb{T}^{g}_{j} \otimes \mathbb{S}^{-1_{i}}) \right) \\
	& + \delta_{s_{j}, -\frac{1}{2}} \delta^{\TT{final}}_{j} \left( \sqrt{\frac{\mathcal{P}_{qq}}{2\mathcal{C}_{\TT{F}}(1+z_{i}^{2})}} \frac{1}{ [\tilde{p}_{j} q_{i}] } (\mathbb{T}^{g}_{j} \otimes \mathbb{S}^{-1_{i}}) + \sqrt{\frac{z_{i}^{2}\mathcal{P}_{qq}}{2\mathcal{C}_{\TT{F}}(1+z_{i}^{2})}} \frac{1}{ \< q_{i} \tilde{p}_{j}\>} (\mathbb{T}^{g}_{j} \otimes \mathbb{S}^{+1_{i}}) \right.\\ 
	& \qquad  + \sqrt{\frac{\mathcal{P}_{gq}}{2\mathcal{C}_{\TT{F}}(2-2z_{i}+z_{i}^{2})}} \frac{1}{ [q_{i} \tilde{p}_{j}] } \mathbb{W}^{ij} (\mathbb{T}^{g}_{j} \otimes \mathbb{S}^{-1_{i}}) \left. + \sqrt{\frac{(1-z_{i})^{2}\mathcal{P}_{gq}}{2\mathcal{C}_{\TT{F}}(2-2z_{i}+z_{i}^{2})}} \frac{1}{ \< \tilde{p}_{j} q_{i} \> }  \mathbb{W}^{ij} (\mathbb{T}^{g}_{j} \otimes \mathbb{S}^{+1_{i}}) \right) \\
	& + \delta_{s_{j}, 1} \delta^{\TT{final}}_{j} \left( \sqrt{\frac{(1-z_{i})^{2}\mathcal{P}_{qg}}{2T_{\TT{R}}(1-2z_{i}(1-z_{i}))}} \frac{1}{ [\tilde{p}_{j} q_{i}] } (\mathbb{W}^{ij} - \mathbbm{1})(\mathbb{T}^{q}_{j} \otimes \mathbb{P}^{1}_{j} \mathbb{P}^{2}_{j}\mathbb{S}^{+\frac{1}{2}_{i}}) \right. \\ 
	& \qquad + \sqrt{\frac{z_{i}^{2}\mathcal{P}_{qg}}{2T_{\TT{R}}(1-2z_{i}(1-z_{i}))}} \frac{1}{ [\tilde{p}_{j} q_{i}]} (\mathbb{W}^{ij} - \mathbbm{1})(\mathbb{T}^{q}_{j} \otimes \mathbb{P}^{2}_{j}\mathbb{S}^{-\frac{1}{2}_{i}}) \\
	& \qquad  + \sqrt{\frac{\mathcal{P}_{gg}}{2\mathcal{C}_{\TT{A}}(1-z_{i}+z_{i}^{2})^2}}\frac{1}{ \< q_{i} \tilde{p}_{j}\> } (\mathbb{T}^{g}_{j} \otimes \mathbb{S}^{+1_{i}}) \\
	& \qquad + \sqrt{\frac{z_{i}^{4}\mathcal{P}_{gg}}{2\mathcal{C}_{\TT{A}}(1-z_{i}+z_{i}^{2})^2}} \frac{1}{ [ q_{i} \tilde{p}_{j}] } (\mathbb{T}^{g}_{j} \otimes \mathbb{S}^{-1_{i}}) \left. + \sqrt{\frac{\mathcal{P}_{gg}(1-z_{i})^{4}}{2\mathcal{C}_{\TT{A}}(1-z_{i}+z_{i}^{2})^2}} \frac{1}{ [ \tilde{p}_{j} q_{i} ] } (\mathbb{T}^{g}_{j} \otimes \mathbb{P}^{1}_{j} \mathbb{S}^{+1_{i}}) \right)  \\
	& + \delta_{s_{j}, -1} \delta^{\TT{final}}_{j} \left( \sqrt{\frac{(1-z_{i})^{2}\mathcal{P}_{qg}}{2T_{\TT{R}}(1-2z_{i}(1-z_{i}))}} \frac{1}{ \< q_{i} \tilde{p}_{j}\> } (\mathbb{W}^{ij} - \mathbbm{1})(\mathbb{T}^{q}_{j} \otimes \mathbb{P}^{1}_{j}\mathbb{P}^{2}_{j}\mathbb{S}^{-\frac{1}{2}_{i}}) \right. \\
	& \qquad + \sqrt{\frac{ z_{i}^{2}\mathcal{P}_{qg}}{2T_{\TT{R}}(1-2z_{i}(1-z_{i}))}} \frac{1}{  \< q_{i} \tilde{p}_{j}\> } (\mathbb{W}^{ij} - \mathbbm{1})(\mathbb{T}^{q}_{j} \otimes \mathbb{P}^{2}_{j}\mathbb{S}^{+\frac{1}{2}_{i}}) \\ 
	& \qquad + \sqrt{\frac{\mathcal{P}_{gg}}{2\mathcal{C}_{\TT{A}}(1-z_{i}+z_{i}^{2})^2}}\frac{1}{ [\tilde{p}_{j} q_{i}] } (\mathbb{T}^{g}_{j} \otimes \mathbb{S}^{-1_{i}}) \\
	& \qquad + \sqrt{\frac{z_{i}^{4}\mathcal{P}_{gg}}{2\mathcal{C}_{\TT{A}}(1-z_{i}+z_{i}^{2})^2}} \frac{1}{ \< \tilde{p}_{j} q_{i} \> } (\mathbb{T}^{g}_{j} \otimes \mathbb{S}^{+1_{i}}) \left. + \sqrt{\frac{\mathcal{P}_{gg}(1-z_{i})^{4}}{2\mathcal{C}_{\TT{A}}(1-z_{i}+z_{i}^{2})^2}} \frac{1}{ \< q_{i} \tilde{p}_{j}\> } (\mathbb{T}^{g}_{j} \otimes \mathbb{P}^{1}_{j} \mathbb{S}^{-1_{i}}) \right) \\
\end{split} \nonumber
\end{eqnarray}
\begin{eqnarray}
\begin{split}
	 \quad \; & +  \delta_{s_{j}, \frac{1}{2}} \delta^{\TT{initial}}_{j} \sqrt{\frac{1}{z_{i}}} \left( \sqrt{\frac{\mathcal{P}_{qq}}{\mathcal{C}_{\TT{F}}(1+z_{i}^{2})}} \frac{1}{\<q_{i} p_{j}\>} (\mathbb{T}^{g}_{j} \otimes \mathbb{S}^{+1_{i}}) + \sqrt{\frac{z_{i}^{2}\mathcal{P}_{qq}}{\mathcal{C}_{\TT{F}}(1+z_{i}^{2})}} \frac{1}{ [p_{j} q_{i}]} (\mathbb{T}^{g}_{j} \otimes \mathbb{S}^{-1_{i}}) \right. \\ 
	& \qquad + \sqrt{\frac{(1-z_{i})^{2}\mathcal{P}_{qg}}{n_{f}\mathcal{C}_{F}(1-2z_{i}(1-z_{i}))}} \frac{1}{ [p_{j} q_{i}] } \mathbb{W}^{ij} (\mathbb{T}^{g}_{j} \otimes \mathbb{S}^{+1_{i}}) \\ 
	& \qquad \left. + \sqrt{\frac{z_{i}^{2}\mathcal{P}_{qg}}{n_{f}\mathcal{C}_{F}(1-2z_{i}(1-z_{i}))}} \frac{1}{  \< q_{i} p_{j}\> } \mathbb{W}^{ij} (\mathbb{T}^{g}_{j} \otimes \mathbb{S}^{-1_{i}}) \right) \\
	 \quad \; & + \delta_{s_{j}, -\frac{1}{2}} \delta^{\TT{initial}}_{j} \sqrt{\frac{1}{z_{i}}} \left( \sqrt{\frac{\mathcal{P}_{qq}}{\mathcal{C}_{\TT{F}}(1+z_{i}^{2})}} \frac{1}{ [p_{j} q_{i}] } (\mathbb{T}^{g}_{j} \otimes \mathbb{S}^{-1_{i}}) + \sqrt{\frac{z_{i}^{2}\mathcal{P}_{qq}}{\mathcal{C}_{\TT{F}}(1+z_{i}^{2})}} \frac{1}{ \< q_{i} p_{j}\>} (\mathbb{T}^{g}_{j} \otimes \mathbb{S}^{+1_{i}}) \right.\\ 
	& \qquad + \sqrt{\frac{(1-z_{i})^{2}\mathcal{P}_{qg}}{n_{f}\mathcal{C}_{F}(1-2z_{i}(1-z_{i}))}} \frac{1}{ \< q_{i} p_{j}\> } \mathbb{W}^{ij} (\mathbb{T}^{g}_{j} \otimes \mathbb{S}^{-1_{i}}) \\
	& \left. \qquad + \sqrt{\frac{z_{i}^{2}\mathcal{P}_{qg}}{n_{f}\mathcal{C}_{F}(1-2z_{i}(1-z_{i}))}} \frac{1}{ [p_{j} q_{i}]} \mathbb{W}^{ij} (\mathbb{T}^{g}_{j} \otimes \mathbb{S}^{+1_{i}}) \right) \\
	& + \delta_{s_{j}, 1} \delta^{\TT{initial}}_{j} \sqrt{\frac{1}{z_{i}}} \left(  \sqrt{\frac{2n_{f} \mathcal{P}_{gq}}{T_{\TT{R}}(2-2z_{i}+z_{i}^{2})}} \frac{1}{\<p_{j} q_{i}\>} (\mathbb{T}^{q}_{j} \otimes \mathbb{P}^{2}_{j}\mathbb{S}^{+\frac{1}{2}_{i}}) \right. \\ 
	& \qquad + \sqrt{\frac{2n_{f} (1-z_{i})^{2}\mathcal{P}_{gq}}{ T_{\TT{R}}(2-2z_{i}+z_{i}^{2})}} \frac{1}{ [q_{i} p_{j}]} (\mathbb{T}^{q}_{j} \otimes \mathbb{P}^{1}_{j}\mathbb{P}^{2}_{j}\mathbb{S}^{-\frac{1}{2}_{i}}) + \sqrt{\frac{\mathcal{P}_{gg}}{\mathcal{C}_{\TT{A}}(1-z_{i}+z_{i}^{2})^2}}\frac{1}{ \< q_{i} p_{j}\> } (\mathbb{T}^{g}_{j} \otimes \mathbb{S}^{+1_{i}}) \\
	& \qquad + \sqrt{\frac{z_{i}^{4}\mathcal{P}_{gg}}{\mathcal{C}_{\TT{A}}(1-z_{i}+z_{i}^{2})^2}} \frac{1}{ [ q_{i} p_{j}] } (\mathbb{T}^{g}_{j} \otimes \mathbb{S}^{-1_{i}}) \left. + \sqrt{\frac{\mathcal{P}_{gg}(1-z_{i})^{4}}{\mathcal{C}_{\TT{A}}(1-z_{i}+z_{i}^{2})^2}} \frac{1}{ \< q_{i} p_{j}\> } (\mathbb{T}^{g}_{j} \otimes \mathbb{P}^{1}_{j} \mathbb{S}^{-1_{i}}) \right)  \\
	& + \delta_{s_{j}, -1} \delta^{\TT{initial}}_{j} \sqrt{\frac{1}{z_{i}}} \left( \sqrt{\frac{2n_{f} \mathcal{P}_{gq}}{T_{\TT{R}}(2-2z_{i}+z_{i}^{2})}} \frac{1}{ [q_{i} p_{j}] } (\mathbb{T}^{q}_{j} \otimes \mathbb{P}^{2}_{j}\mathbb{S}^{-\frac{1}{2}_{i}}) \right. \\
	& \qquad + \sqrt{\frac{2n_{f} (1-z_{i})^{2}\mathcal{P}_{gq}}{T_{\TT{R}}(2-2z_{i}+z_{i}^{2})}} \frac{1}{ \< p_{j} q_{i} \> } (\mathbb{T}^{q}_{j} \otimes \mathbb{P}^{1}_{j}\mathbb{P}^{2}_{j}\mathbb{S}^{+\frac{1}{2}_{i}}) + \sqrt{\frac{\mathcal{P}_{gg}}{\mathcal{C}_{\TT{A}}(1-z_{i}+z_{i}^{2})^2}}\frac{1}{ [p_{j} q_{i}] } (\mathbb{T}^{g}_{j} \otimes \mathbb{S}^{-1_{i}}) \\
	& \qquad + \sqrt{\frac{z_{i}^{4}\mathcal{P}_{gg}}{\mathcal{C}_{\TT{A}}(1-z_{i}+z_{i}^{2})^2}} \frac{1}{ \< p_{j} q_{i} \> } (\mathbb{T}^{g}_{j} \otimes \mathbb{S}^{+1_{i}}) \left. + \sqrt{\frac{\mathcal{P}_{gg}(1-z_{i})^{4}}{\mathcal{C}_{\TT{A}}(1-z_{i}+z_{i}^{2})^2}} \frac{1}{ [ p_{j} q_{i} ] } (\mathbb{T}^{g}_{j} \otimes \mathbb{P}^{1}_{j} \mathbb{S}^{+1_{i}}) \right) . 
\end{split} 
\end{eqnarray}
Here $s_{j}$ is the spin/helicity of parton $j$ and $z_{i}$ is the momentum fraction between parton $i$ and its parent parton, $j$ (as in \eqref{eqn:momentum_fraction}). 
$\mathbb{T}^{g}_{j}$ are the basis-independent colour-charge operators for the emission of a gluon \cite{SoftEvolutionAlgorithm,QCD_colour_flow}.  $\mathbb{T}^{q}_{j}$ is the colour charge operator for the emission of a $q\bar{q}$ pair from a gluon. In the colour flow basis it is 
\begin{eqnarray}
\mathbb{T}^{q}_{j} = \sqrt{T_{\TT{R}}}\mathbbm{1} - \frac{\sqrt{T_{\TT{R}}}}{N} \mathbb{\tau}_{j},
\end{eqnarray}
where $\mathbb{\tau}_{j}$ exchanges the anti-colour lines associated with the colour line of parton $j$.  For example, let parton $j$ have colour line $c_{2}$ and anti-colour $\bar{c}_{5}$, $\mathbb{\tau}_{j}$ would exchange anti-colour lines $\bar{c}_{2}$ and $\bar{c}_{5}$. A full definition of $\mathbb{\tau}_{j}$, and other colour flow operators, can be found in \cite{SoftEvolutionAlgorithm}, where $\mathbb{\tau}_{j}$ is written $\v{s}_{\alpha,\beta}$. Note $\mathbb{T}^{q}_{j} \cdot \mathbb{T}^{q}_{j} = T_\text{R} \mathbbm{1}$.\footnote{Strictly speaking this is only valid when acting on a physical matrix element.}
We have defined $\mathbb{S}^{s}$ as the operator that adds a parton with helicity $s$. Just as $\mathbb{T}^{g}_{i} \cdot \mathbb{T}^{g}_{i} = \mathcal{C}_{i} \mathbbm{1} $, it can be shown that $\mathbb{S}^{s} \cdot \mathbb{S}^{s} = \mathbbm{1}$. We have also defined a `swap' operator, $\mathbb{W}^{ij}$, which swaps the colour and helicity of particles $i$ and $j$. Finally, we defined $\mathbb{P}^{1}_{i}$ as the operator that flips the helicity of parton $i$ and $\mathbb{P}^{2}_{i}$ that halves the helicity of $i$. There is some freedom in how we introduce operators to keep track of the evolving helicity state (for example one could have instead made use of $(\mathbb{S}^{s})^{\dag}$, which deletes a parton of helicity $s$). 
Examples to illustrate the use of these helicity operators can be found in Appendix \ref{Appendix_B}.
The (unregularised) collinear splitting functions are 
\begin{eqnarray}
\begin{split}
\mathcal{P}_{qq} &= \mathcal{C}_{\TT{F}} \frac{1+z^{2}}{1-z}, \\
\mathcal{P}_{gq} &= \mathcal{C}_{\TT{F}} \frac{1+(1-z)^{2}}{z}, \\
\mathcal{P}_{qg} &= n_{f} T_{\TT{R}}(1-2z(1-z)), \\
\mathcal{P}_{gg} &= 2 \mathcal{C}_{\TT{A}} \left( z(1-z) + \frac{z}{1-z} + \frac{1-z}{z} \right).
\end{split} 
\end{eqnarray}
It should be understood that $\v{P}_{ij}$ always acts as
\begin{align}
\v{P}_{ij} \mathcal{O} \v{P}^{\dagger}_{ij} = \sum_{\upsilon, \, s_{i}, \, s_{i}'}S^{(\upsilon_{j} \rightarrow \upsilon)}_{s_{i},s_{i}'}  \mathbb{T}^{\bar{\upsilon}}_{j} \otimes \mathbb{S}^{(\upsilon_{j} \rightarrow \upsilon)}_{s_{i}} \, \mathcal{O} \, \mathbb{T}^{\bar{\upsilon} \, \dagger}_{j} \otimes \mathbb{S}^{(\upsilon_{j} \rightarrow \upsilon) \, \dagger}_{s_{i}'} \, ,
\end{align}
where $\mathbb{S}^{(\upsilon_{j} \rightarrow \upsilon)}_{s_{i}}$ is a generalised spin operator and $S^{(\upsilon_{j} \rightarrow \upsilon)}_{s_{i},s_{i}'}$ is a c-number coefficient corresponding to a $\upsilon_{j} \rightarrow \upsilon$ splitting. For example, when $j$ is a quark
\begin{align}
\v{P}_{ij} \mathcal{O} \v{P}^{\dagger}_{ij} &= S^{(q \rightarrow q)}_{1 , 1}  \mathbb{T}^{g}_{j} \otimes \mathbb{S}^{1_{i}} \, \mathcal{O} \, \mathbb{T}^{g \, \dagger}_{j} \otimes \mathbb{S}^{1_{i} \, \dagger} + S^{(q \rightarrow q)}_{1 , -1}  \mathbb{T}^{g}_{j} \otimes \mathbb{S}^{1_{i}} \, \mathcal{O} \, \mathbb{T}^{g \, \dagger}_{j} \otimes \mathbb{S}^{- 1_{i} \, \dagger} \nonumber \\
& + S^{(q \rightarrow q)}_{-1 , 1}  \mathbb{T}^{g}_{j} \otimes \mathbb{S}^{- 1_{i}} \, \mathcal{O} \, \mathbb{T}^{g \, \dagger}_{j} \otimes \mathbb{S}^{1_{i} \, \dagger}  + S^{(q \rightarrow q)}_{-1 , -1}  \mathbb{T}^{g}_{j} \otimes \mathbb{S}^{-1_{i}} \, \mathcal{O} \, \mathbb{T}^{g \, \dagger}_{j} \otimes \mathbb{S}^{-1_{i} \, \dagger} \nonumber \\
& + S^{(q \rightarrow g)}_{1 , 1}  \mathbb{T}^{g}_{j} \otimes \mathbb{W}^{ij}\mathbb{S}^{1_{i}} \, \mathcal{O} \, \mathbb{T}^{g \, \dagger}_{j} \otimes \mathbb{S}^{1_{i} \, \dagger}\mathbb{W}^{ij} + S^{(q \rightarrow g)}_{1 , -1}  \mathbb{T}^{g}_{j} \otimes \mathbb{W}^{ij} \mathbb{S}^{1_{i}} \, \mathcal{O} \, \mathbb{T}^{g \, \dagger}_{j} \otimes \mathbb{S}^{- 1_{i} \, \dagger}\mathbb{W}^{ij} \nonumber \\
& + S^{(q \rightarrow g)}_{-1 , 1}  \mathbb{T}^{g}_{j} \otimes \mathbb{W}^{ij}\mathbb{S}^{- 1_{i}} \, \mathcal{O} \, \mathbb{T}^{g \, \dagger}_{j} \otimes \mathbb{S}^{1_{i} \, \dagger} \mathbb{W}^{ij} + S^{(q \rightarrow g)}_{-1 , -1}  \mathbb{T}^{g}_{j} \otimes \mathbb{W}^{ij}\mathbb{S}^{-1_{i}} \, \mathcal{O} \, \mathbb{T}^{g \, \dagger}_{j} \otimes \mathbb{S}^{-1_{i} \, \dagger}\mathbb{W}^{ij} ,
\end{align}
where
\begin{align}
\sum_{\upsilon, \, s_{i}}S^{(q \rightarrow \upsilon)}_{s_{i},s_{i}} =  \mathcal{P}_{q q} \, \mathcal{C}^{-1}_{\TT{F}} \left( \frac{\delta^{\TT{final}}_{j}}{4 \, q_{i} \cdot \tilde{p}_{j}} + \frac{\delta^{\TT{initial}}_{j}}{2 \, z_{i} \, q_{i} \cdot p_{j}} \right) + \mathcal{P}_{g q} \,\mathcal{C}^{-1}_{\TT{F}} \left( \frac{\delta^{\TT{final}}_{j}}{4 \, q_{i} \cdot \tilde{p}_{j}} + \frac{\delta^{\TT{initial}}_{j}}{2 \, z_{i} \, q_{i} \cdot p_{j}} \right).
\end{align}
The Sudakov factors in variant B can be written in a variety of ways using $$\int^{1 - x}_{x}\td z \, \mathcal{P}_{gg} = 2\int^{1 - x}_{x}\td z \, z \mathcal{P}_{gg} \quad \TT{and} \quad \int^{1 - x}_{x}\td z \, (\mathcal{P}_{qq} + \mathcal{P}_{gq}) = 2\int^{1 - x}_{x}\td z \, z \mathcal{P}_{qq}.$$ 
Note also that there is a subtle factor of two difference between initial state and final state splittings in $\v{P}_{ij}$. The factor arises as partons in the initial state must be convoluted with PDFs which changes the pole structure of splittings and increases the number of diagrams that must be summed over relative to splittings in the final state. In the final state (without fragmentation function dependence), soft poles from real emissions can be found at both $z=1$ and $z=0$. These poles cancel the poles from loop diagrams. For real emissions in the initial state, the $z=0$ poles are absent due to kinematics whilst the $z=1$ poles cancel the poles from loops. The factor of 2 ensures the correct pattern of cancellations.

Finally, we also note the factors of $n_f$, the number of quark flavours, in $\v{P}_{ij}$. They are present since we sum democratically over flavours whenever there is a $g \to q \bar{q}$ branching. Note that since we always evolve away from the hard process this means that we sum over quark flavours in the case of an initial-state $q \to g q$ branching. Care must be taken however, since if the branching cascade terminates with an initial-state quark (or anti-quark) then it is necessary to divide by a factor of $n_f$ before convoluting with the corresponding parton distribution function. The same holds in the case where fragmentation functions are needed. In Section \ref{sec:pheno}, we introduced the $\star$ notation to handle this. Of course, one could set $n_f=1$ in the above splitting operators, after which it would be necessary to sum over flavours as appropriate.

For variant A, we need the hard-collinear emission operator $\overline{\v{P}}_{ij}$. This operator is defined at cross-section level through the relation
\begin{align}
\overline{\v{P}}_{ij}\mathcal{O}\overline{\v{P}}^{\dagger}_{ij} = \v{P}_{ij}\mathcal{O}\v{P}^{\dagger}_{ij}& - \delta^{\TT{final}}_{j}\mathbb{T}^{g}_{j}\otimes\left( \frac{\mathbb{S}^{1_{i}}}{\sqrt{1-z_{i}}\< q_{i} \tilde{p}_{j} \>} + \frac{\mathbb{S}^{-1_{i}}}{\sqrt{1-z_{i}}[\tilde{p}_{j} q_{i}]} + \frac{\mathbb{W}^{ij}\mathbb{S}^{1_{i}}}{\sqrt{z_{i}}\< \tilde{p}_{j} q_{i} \>} + \frac{\mathbb{W}^{ij}\mathbb{S}^{-1_{i}}}{\sqrt{z_{i}}[q_{i} \tilde{p}_{j}]} \right)  \nonumber \\
&~\times \, \mathcal{O} \, \mathbb{T}^{g \, \dagger}_{j}\otimes\left( \frac{\mathbb{S}^{1_{i}}}{\sqrt{1-z_{i}}\< q_{i} \tilde{p}_{j} \>} + \frac{\mathbb{S}^{-1_{i}}}{\sqrt{1-z_{i}}[\tilde{p}_{j} q_{i}]} + \frac{\mathbb{W}^{ij}\mathbb{S}^{1_{i}}}{\sqrt{z_{i}}\< \tilde{p}_{j} q_{i} \>} + \frac{\mathbb{W}^{ij}\mathbb{S}^{-1_{i}}}{\sqrt{z_{i}}[q_{i} \tilde{p}_{j}]} \right)^{\dagger} \nonumber \\
& - 2 \delta^{\TT{initial}}_{j} \mathbb{T}^{g}_{j}\otimes\left( \frac{\mathbb{S}^{1_{i}}}{\sqrt{1-z_{i}}\< q_{i} p_{j} \>} + \frac{\mathbb{S}^{-1_{i}}}{\sqrt{1-z_{i}}[p_{j} q_{i}]} \right) \nonumber \\
&~ \times \, \mathcal{O} \, \mathbb{T}^{g \, \dagger}_{j}\otimes\left( \frac{\mathbb{S}^{1_{i}}}{\sqrt{1-z_{i}}\< q_{i} p_{j} \>} + \frac{\mathbb{S}^{-1_{i}}}{\sqrt{1-z_{i}}[p_{j} q_{i}]} \right)^{\dagger},
 \label{eqn:Pstrict}
\end{align}
where $\mathcal{O}$ is a generalised operator. Note that $...\v{P}_{ij} \mathcal{O} \v{P}^{\dagger}_{ij}...$ is not necessarily Casimir in colour. However, as we observed in section \ref{sec:Coll}, ignoring Coulomb contributions, the collinear physics can be factorised and becomes colour-diagonal after taking the trace. Therefore, for processes where Coulomb terms do not contribute (e.g. $e^{+}e^{-}$ and DIS) we could use the emergent colour-diagonal structure to greatly simplify the $\v{P}_{ij}$ and $\overline{\v{P}}_{ij}$ operators. For example, we could redefine $\overline{\v{P}}_{ij}$ with a simpler amplitude-level statement. To this end, we can introduce the hard-collinear splitting functions;
\begin{eqnarray}
\begin{split}
	\overline{\mathcal{P}}_{qq} &= \mathcal{P}_{qq} - 2 \mathcal{C}_{\TT{F}} \frac{1}{1-z} = - \mathcal{C}_{\TT{F}} (1+z), \\
	\overline{\mathcal{P}}^{\TT{initial}}_{gg} &= \mathcal{P}_{gg} - 2 \mathcal{C}_{\TT{A}} \frac{1}{1-z}  = 2 \mathcal{C}_{\TT{A}} \left(\frac{1}{z} + z(1-z) - 2 \right) \\
	\overline{\mathcal{P}}^{\TT{final}}_{gg} &= \mathcal{P}_{gg} - 2 \mathcal{C}_{\TT{A}} \frac{1}{1-z} - 2 \mathcal{C}_{\TT{A}} \frac{1}{z}  = 2 \mathcal{C}_{\TT{A}} \left(z(1-z) - 2 \right) \\
	\overline{\mathcal{P}}^{\TT{final}}_{gq} &= \mathcal{P}_{gq} - 2 \mathcal{C}_{\TT{F}} \frac{1}{z} = \mathcal{C}_{\TT{F}}\left(\frac{1+(1-z)^{2}}{z} - \frac{2}{z}\right), \qquad \overline{\mathcal{P}}^{\TT{initial}}_{gq} = \mathcal{P}_{gq}, \\
	\overline{\mathcal{P}}_{qg} &= \mathcal{P}_{qg}.
\end{split} 
\end{eqnarray}
The newly simplified $\overline{\v{P}}_{ij}$ is equal to the operator found by substituting $\mathcal{P} \mapsto \overline{\mathcal{P}}$ inside $\v{P}_{ij}$, i.e. $$\delta^{\TT{final}}_{j} \, \mathcal{P}_{gq} \mapsto \delta^{\TT{final}}_{j} \, \overline{\mathcal{P}}^{\TT{final}}_{gq} \quad \TT{and} \quad \delta^{\TT{initial}}_{j} \, \mathcal{P}_{gq} \mapsto \delta^{\TT{initial}}_{j} \, \overline{\mathcal{P}}^{\TT{initial}}_{gq}.$$ This new $\overline{\v{P}}_{ij}$ operator is constructed so that when used in the LHS of \eqref{eqn:Pstrict} the expression becomes exact after a trace is taken. Additionally, it correctly computes spin correlations after collinear factorisation. Simplifying the collinear emission operators would be very pertinent to an efficient computational implementation of our algorithm.

$\overline{\mathcal{P}}^{\,\circ}_{\upsilon_{i}\upsilon_{j}}$ and $\mathcal{P}^{\,\circ}_{\upsilon_{i}\upsilon_{j}}$ are splitting functions used exclusively in our Sudakov factors and they are defined with all colour factors removed:
\begin{eqnarray}
\begin{split}
	\overline{\mathcal{P}}^{\,\circ}_{qq} &= \mathcal{P}^{\,\circ}_{qq} -  \frac{1}{1-z} = - \frac{1}{2}(1+z), \\
	\overline{\mathcal{P}}^{\,\circ}_{gq} &= \mathcal{P}^{\,\circ}_{gq}- \frac{1}{z} = \frac{1+(1-z)^{2}}{2z} - \frac{1}{z}, \\
	\overline{\mathcal{P}}^{\,\circ}_{qg} &= \mathcal{P}^{\,\circ}_{qg} = n_{f}(1-2z(1-z)), \\
	\overline{\mathcal{P}}^{\,\circ}_{gg} &= \mathcal{P}^{\,\circ}_{gg} - \frac{1}{1-z} - \frac{1}{z} = \left( z(1-z) - 2 \right).
\end{split} 
\end{eqnarray}

In Section \ref{sec:SC2f+} and Section \ref{sec:pheno} we make use of the plus prescription (see \eqref{eq:plusP}). Applying the plus prescription means
\begin{align}
\int_0^1 \td x  \, f(x)_{+} \, g(x) = \int_0^1 \td x  \, [f(x)g(x) - f(x)g(1)]. \label{eqn:plus_prescription}
\end{align}
The plus prescription is, in our case, is defined by
\begin{align}
\int \td x  \, \v{P}(x)_{+} \mathcal{O} \, \v{P}^{\dagger}(x)_{+} \, u(x) = \int \td x  \, \bigg[&\v{P}(x) \mathcal{O} \, \v{P}^{\dagger}(x) \, u(x) \nonumber \\ 
&- \v{P}^{\dagger}_\text{V}(x) \v{P}_\text{V}(x) \, \mathcal{O} \, \frac{u(1)}{2} - \mathcal{O} \, \v{P}^{\dagger}_\text{V}(x) \v{P}_\text{V}(x) \, \frac{u(1)}{2}\bigg],
\end{align}
where the structure of the subtraction terms is determined by the corresponding structure of the virtual corrections and this simply means that $\mathbf{P}_\text{V}$ is determined using \eqref{eq:appsplit} but with parton $j$ always treated as if it is final state.
The two splitting functions affected by the plus prescription are
\begin{eqnarray}
\begin{split}
P_{qq} &= \mathcal{C}_{\TT{F}} \left(\frac{1+z^{2}}{1-z}\right)_{+} \equiv \mathcal{C}_{\TT{F}} \left[\frac{1+z^{2}}{(1-z)_{+}} + \frac{3}{2} \delta(1-z) \right], \\
P_{gg} &= 2 \mathcal{C}_{\TT{A}} \left(\frac{z}{(1-z)_{+}} + z(1-z) + \frac{1-z}{z} \right) + \frac{1}{6}(11\mathcal{C}_{\TT{A}} - 4n_{f}T_{\TT{R}})\delta(1-z) .
\end{split} 
\end{eqnarray}
For the other parton branchings, $P_{ij} = \mathcal{P}_{ij}$ where $i,j$ labels parton type.

\section{Connecting to other work on spin}
\label{Appendix_B}

Our goal in this appendix is to show how our treatment of spin connects with the work of others, specifically that of Collins \cite{Collins:1987cp} and Knowles \cite{KNOWLES1990271}. We will begin by re-capping the calculation of the tree-level $q\rightarrow qg$ collinear splitting using the standard notation. The matrix element is
\begin{eqnarray}
\begin{split}
	\mathcal{M}^{n+1}_{s_{1}...s_{i}\lambda_{j}}(...,p_{i}, p_{j})= ig T^{a}_{f} \epsilon^{*a}_{\lambda_{j} \, \mu}(p_{j})\bar{u}_{s_{i}}(p_{i}) \gamma^{\mu}\dfrac{i \slashed{p}_{ij}}{p_{ij}^{2}+i\epsilon} \hat{\mathcal{M}}^{n}_{s_{1}...}(...,p_{ij}),
\end{split} 
\end{eqnarray}
where $p_{ij} = p_{i} + p_{j}$. $\mathcal{M}^{n}_{s_{1}...s_{n}}$ is the spin-dependent $n$-particle matrix element, carrying $n$ spin indices. $\hat{\mathcal{M}}^{n}$ is defined so that $\bar{u}_{s_{ij}}(p_{ij})\hat{\mathcal{M}}^{n}_{s_{1}...}(...,p_{ij})=\mathcal{M}^{n}_{s_{1}...s_{ij}}(...,p_{ij})$. In the collinear limit, $\slashed{p}_{ij}$ is on shell and so we can express it as a product of on-shell spinors, i.e. $\slashed{p}_{ij} = \sum_{s_{ij}} u_{s_{ij}}(p_{ij})\bar{u}_{s_{ij}}(p_{ij})$. We can then further simplify by replacing Dirac spinors with massless Weyl spinors, defined in the chiral basis as $u_{s}=(x_{s \, \alpha},y^{\dagger \dot{\alpha}}_{s})^{\TT{T}}$. To avoid clutter, we will temporarily drop colour factors, factors of $g$ and the denominator of the propagator. We find
\begin{eqnarray}
\begin{split} \textstyle
	\mathcal{M}^{n+1}_{s_{1}...s_{i}\lambda_{j}}(...,p_{i}, p_{j}) \propto \, &\epsilon^{*}_{\lambda_{j} \, \mu}(p_{j}) y^{\alpha}_{\frac{1}{2}}(p_{i}) \sigma^{\mu}_{\alpha\dot{\beta}} y^{\dagger\dot{\beta}}_{\frac{1}{2}}(p_{ij}) \mathcal{M}^{n}_{s_{1}...\frac{1}{2}}(...,p_{ij}) \delta_{s_{i} \frac{1}{2}} \\
	&\textstyle + \epsilon^{*}_{\lambda_{j} \, \mu}(p_{j})x^{\dagger}_{-\frac{1}{2}\dot{\alpha}}(p_{i}) \bar{\sigma}^{\mu\,\dot{\alpha}\beta} x_{-\frac{1}{2}\beta}(p_{ij}) \mathcal{M}^{n}_{s_{1}...-\frac{1}{2}}(...,p_{ij}) \delta_{s_{i} -\frac{1}{2}}.
\end{split} 
\end{eqnarray}
We can now employ the spinor-helicity formalism \cite{HelicityTechniques}. Also applying a Sudakov decomposition, as defined in Section \ref{sec:SC2}, the matrix element becomes
\begin{eqnarray}
\begin{split}
\mathcal{M}^{n+1}_{s_{1}...s_{i}\lambda_{j}}(...,p_{i}, p_{j}) =& g T_{f} \sqrt{\frac{\mathcal{P}_{qq}}{\mathcal{C}_{\TT{F}}(1+z^{2})}} \frac{1}{\<p_{j}p_{i}\>} \mathcal{M}^{n}_{s_{1}...\frac{1}{2}}(...,p_{ij})\delta_{s_{i},\frac{1}{2}}\delta_{\lambda_{j}, 1} \\
&+g T_{f} \sqrt{\frac{z^{2}\mathcal{P}_{qq}}{\mathcal{C}_{\TT{F}}(1+z^{2})}} \frac{1}{\<p_{j}p_{i}\>}  \mathcal{M}^{n}_{s_{1}...-\frac{1}{2}}(...,p_{ij})\delta_{s_{i}, -\frac{1}{2}}\delta_{\lambda_{j}, 1} \\
&+g T_{f} \sqrt{\frac{z^{2}\mathcal{P}_{qq}}{\mathcal{C}_{\TT{F}}(1+z^{2})}} \frac{1}{[p_{i}p_{j}]} \mathcal{M}^{n}_{s_{1}...\frac{1}{2}}(...,p_{ij})\delta_{s_{i}, \frac{1}{2}}\delta_{\lambda_{j}, -1} \\
&+g T_{f} \sqrt{\frac{\mathcal{P}_{qq}}{\mathcal{C}_{\TT{F}}(1+z^{2})}} \frac{1}{[p_{i}p_{j}]} \mathcal{M}^{n}_{s_{1}...-\frac{1}{2}}(...,p_{ij})\delta_{s_{i}, -\frac{1}{2}}\delta_{\lambda_{j}, -1}.
\end{split} \label{eqn:usual_spin}
\end{eqnarray}
Therefore, for each fixed value of $\lambda_{j}$ there is an amplitude level decay matrix $\mathcal{D}^{(\lambda_{j})}_{s_{i}s_{ij}}$ describing the transition of a quark with spin $s_{ij}$ to two partons with spin $s_{i}$ and $\lambda_{s}$ so that $\mathcal{M}^{n+1}_{s_{1}...s_{i}\lambda_{j}} = \mathcal{D}^{(\lambda_{j})}_{s_{i}s_{ij}}\mathcal{M}^{n}_{s_{1}...s_{ij}}$, which can be determined from the above expression. Equivalent calculations lead to decay matrices for each possible collinear splitting. When computed for initial-state collinear splittings, these matrices are amplitude-level spin-density matrices and we denote them with an $\mathcal{S}$ instead of a $\mathcal{D}$.

Current parton showers deal with spin by algorithmically evaluating cross-section level spin density matrices. Consider a $2 \rightarrow 2$ scattering, where each hard parton is coloured. Then
\begin{eqnarray}
\begin{split}
\td \sigma \propto \rho^{(1)}_{s_{1}s'_{1}} \rho^{(2)}_{s_{2}s'_{2}} \mathcal{M}_{s_{1}s_{2}s_{3}s_{4}} \mathcal{M}^{*}_{s'_{1}s'_{2}s'_{3}s'_{4}} D^{(3)}_{s_{3}s'_{3}} D^{(4)}_{s_{4}s'_{4}},
\end{split} 
\end{eqnarray}
where $\mathcal{M}$ is the full spin-dependent hard matrix element. Summation over spin indices is implicit in this expression. $\rho^{(1)}_{s_{1}s'_{1}}$ and $\rho^{(2)}_{s_{2}s'_{2}}$ are cross-section level spin-density matrices. $D^{(3)}_{s_{3}s'_{3}}$ and $D^{(4)}_{s_{4}s'_{4}}$ are cross-section level decay matrices. $D$ and $\rho$ are calculated from products of amplitude level matrices, $\mathcal{D}$ and $\mathcal{S}$ respectively. For instance, after $n$ emissions from parton $1$:
$$
\rho^{(1)}_{s_{1}s'_{1}} = \sum_{\{\lambda\}} [\mathcal{S}^{\lambda_{1}}\mathcal{S}^{\lambda_{2}}\mathcal{S}^{\lambda_{3}} ... \mathcal{S}^{\lambda_{n}}  \mathcal{S}^{\lambda_{n} \, \dagger}...\mathcal{S}^{\lambda_{3} \, \dagger}\mathcal{S}^{\lambda_{2} \, \dagger}\mathcal{S}^{\lambda_{1} \, \dagger}]_{s_{1}s'_{1}},
$$ 
where usual matrix multiplication is implied. The algorithm of Collins and Knowles is able to determine the spin density and decay matrices such that computational time only grows linearly with the number of partons \cite{Collins:1987cp,KNOWLES1990271,Richardson_2001}.

Now let us turn to the calculation of splitting functions in our notation.  We write $\mathcal{M}^{n}_{s_{1}...s_{n}} = \<s_{1}...s_{n}\rkl \left. n \>$, which ignores colour since it is not our focus here, i.e. more correctly we should write $\mathcal{M}^{n}_{c_{1}...c_{n},s_{1}...s_{n}} = (\<c_{1}...c_{n}\rkl\otimes \<s_{1}...s_{n}\rkl) \lkl n \>$. We wish to define an operator $\v{P}_{k\rightarrow ij}$ that adds a new (collinear) particle ($j$) to $\lkl n \>$ that is emitted off leg $k$, i.e. $\lkl n + 1_{\TT{col}} \> = \sum_{k \in \{n\}}\v{P}_{k\rightarrow ij}\lkl n \>$. 

As before, we will focus on the $q \rightarrow qg$ collinear splitting. Note that $\mathcal{M}^{n+1}_{s_{1}...s_{i},\lambda_{j}} = \<s_{1}...s_{i}, \lambda_{j} \rkl \v{P}_{k\rightarrow ij} \lkl n \>$. Inserting the identity gives 
\begin{eqnarray}
\begin{split}
\mathcal{M}^{n+1}_{s_{1}...s_{i},\lambda_{j}} &= \<s_{1}...s_{i} , \lambda_{j} \rkl \v{P}_{k\rightarrow ij} \sum_{s'_{1}...s'_{k}}\lkl s'_{1}...s'_{k} \> \< s'_{1}...s'_{k} \right. \lkl n \> \\ 
&= \sum_{s'_{k}} \<s_{1}...s_{i}, \lambda_{j} \rkl \v{P}_{k\rightarrow ij} \lkl s_{1}...s'_{k} \> \mathcal{M}^{n}_{s_{1}...s'_{k}}.
\end{split} 
\end{eqnarray}
Comparing to \eqref{eqn:usual_spin}, it follows that $\v{P}_{k\rightarrow ij}$ is (for $q \to qg$)
\begin{eqnarray}
\begin{split}
\<s_{1}...s_{i}, \lambda_{j} \rkl \v{P}_{k\rightarrow ij} \lkl s_{1}...s'_{k} \> =& g T_{f} \sqrt{\frac{\mathcal{P}_{qq}}{\mathcal{C}_{\TT{F}}(1+z^{2})}} \frac{1}{\<p_{j}p_{i}\>} \<s_{1}...s_{i}, 1_{j} \rkl \mathbb{S}^{1_{j}} \lkl s_{1}...s'_{k} \> \delta_{s_{i}, \frac{1}{2}} \delta_{\lambda_{j}, 1} \\
+&g T_{f} \sqrt{\frac{z^{2}\mathcal{P}_{qq}}{\mathcal{C}_{\TT{F}}(1+z^{2})}} \frac{1}{\<p_{j}p_{i}\>} \<s_{1}...s_{i}, 1_{j} \rkl \mathbb{S}^{1_{j}} \lkl s_{1}...s'_{k} \> \delta_{s_{i}, -\frac{1}{2}} \delta_{\lambda_{j}, 1} \\
+&g T_{f} \sqrt{\frac{z^{2}\mathcal{P}_{qq}}{\mathcal{C}_{\TT{F}}(1+z^{2})}} \frac{1}{[p_{i}p_{j}]} \<s_{1}...s_{i}, -1_{j} \rkl \mathbb{S}^{-1_{j}} \lkl s_{1}...s'_{k} \> \delta_{s_{i}, \frac{1}{2}} \delta_{\lambda_{j}, -1} \\
+&g T_{f} \sqrt{\frac{\mathcal{P}_{qq}}{\mathcal{C}_{\TT{F}}(1+z^{2})}} \frac{1}{[p_{i}p_{j}]} \<s_{1}...s_{i}, -1_{j} \rkl \mathbb{S}^{-1_{j}} \lkl s_{1}...s'_{k} \> \delta_{s_{i}, - \frac{1}{2}} \delta_{\lambda_{j}, -1}.
\end{split} 
\end{eqnarray}
$\mathbb{S}^{s_{j}}$ must satisfy $\<s_{1}...s_{i}, s_{j} \rkl \mathbb{S}^{s_{j}} \lkl s_{1}...s'_{k} \> = \delta_{s_{i}, s'_{k}} $. More generally, we require
$$\<s_{1}...s_{i}, s_{j} \rkl \mathbb{S}^{s'_{j}} \lkl s'_{1}...s'_{k} \> = \delta_{s_{1}, s'_{1}}...\delta_{s_{i}, s'_{k}} \delta_{s_{j}, s'_{j}}.$$ This is the definition for $\mathbb{S}^{s}$ presented in Appendix \ref{Appendix_A}\footnote{Repeating this procedure for the other splitting operators leads us to introduce the operators $\mathbb{W}^{ij},\mathbb{P}^{1}_{i}$ and $\mathbb{P}^{2}_{i}$ used in Appendix A.}.

We will now construct a decay matrix $D^{(j)}_{s_{j}s'_{j}}$, for a final-state hard parton $j$ using the spin operators we have just introduced. Let us first consider the situation where there are no soft interactions and only include emissions from the initial primary leg, $j$: 
\begin{eqnarray}
\begin{split}
&\td \sigma \propto \\
& \; \sum_{n}\sum_{\{i\}} \< n ; j \rkl \v{V}^{\dagger}_{1,Q} \v{P}^{\dagger}_{i_{1}j}\v{V}^{\dagger}_{2,1}...\v{V}^{\dagger}_{n,n-1}\v{P}^{\dagger}_{i_{n}j}\v{V}^{\dagger}_{0,n}\v{V}_{0,n}\v{P}_{i_{n}j} \v{V}_{2,1} ... \v{V}_{2,1} \v{P}_{i_{1}j} \v{V}_{1,Q} \lkl n ; j \>, \label{eqn:spin_chain}
\end{split} 
\end{eqnarray}
where the partons in the set $\{i\}$ are transverse momentum ordered. $\v{V}_{a,b}$ is a Sudakov factor:
\begin{eqnarray}
\begin{split}
\v{V}^{\dagger}_{a,b} = \exp &\left[- \frac{\alpha_{s}}{\pi} \int^{q_{b \, \bot}}_{q_{a \, \bot}} \frac{\td k_{\bot}}{k_{\bot}} \sum_{\upsilon} \mathbb{T}^{\bar{\upsilon} \, 2}_{j} \int \frac{\td z \, \td \phi}{8\pi} \mathcal{P}^{\,\circ}_{\upsilon \upsilon_{j}} \right].
\end{split} 
\end{eqnarray}
We can evaluate \eqref{eqn:spin_chain} by inserting identity operators and extracting Sudakov factors, which are proportional to identity operators, into a single numerical factor. Hence
\begin{eqnarray}
\begin{split}
&\td \sigma \propto \, \sum_{n}\sum_{\{i\}}\sum_{s_{j}s'_{j}} \#^{s_{j}s'_{j}}_{i_{1}...i_{n}i'_{1}...i'_{n}} \< n ; j \rkl \left. s_{j} \>\< s_{j} \rkl \v{P}^{\dagger}_{i_{1}}...\v{P}^{\dagger}_{i_{n}}\v{P}_{i'_{n}}... \v{P}_{i'_{1}}\lkl s'_{j} \> \< s'_{j} \right. \lkl n ; j \>,
\end{split} 
\end{eqnarray}
where each $\v{P}^{\dagger}_{i}$ is a `pure' colour-helicity operator with no scalar pre-factor. For instance $\v{P}_{i}=\mathbb{T}^{g}_{j}\otimes\mathbb{S}^{1_{i}}$ in the case of a $q^{+\frac{1}{2}}\rightarrow q^{+\frac{1}{2}}g^{+1}$ splitting or $\v{P}_{i} = \mathbb{T}^{q}_{j} \otimes \mathbb{P}^{1}_{j} \mathbb{P}^{2}_{j} \mathbb{S}^{-1_{i}}$ in the case of a $g^{-1}\rightarrow q^{+\frac{1}{2}}q^{-\frac{1}{2}}$ splitting. $\#^{s_{j}s'_{j}}_{i_{1}...i_{n}i'_{1}...i'_{n}}$ is a c-number coefficient built from helicity dependent splitting functions and expanded Sudakov factors. We can now make the link with the previous approach, i.e. 
\begin{eqnarray}
\begin{split}
D^{(j)}_{s_{j}s'_{j}} = \sum_{n}\sum_{\{i\}} \#^{s_{j}s'_{j}}_{i_{1}...i_{n}i'_{1}...i'_{n}} \< s_{j} \rkl \v{P}^{\dagger}_{i_{1}}...\v{P}^{\dagger}_{i_{n}}\v{P}_{i'_{n}}... \v{P}_{i'_{1}}\lkl s'_{j} \>.
\end{split} 
\end{eqnarray}
The expectation value is calculable and equals a product of $n$ Casimir co-coefficients, e.g. if parton $j$ is a gluon and each operator $\v{P}_{i}$ corresponds to a $g \rightarrow gg$ splitting then the expectation value equals $\mathcal{C}^{n}_{\TT{A}}$ (up to a normalisation for the colour evolution). 

Following this procedure, spin-density and decay matrices can be derived using the algorithm presented in this paper. Let's see this explicitly. Knowles' algorithm calculates spin-density and decay matrices using other intermediate matrices $\rho '$, $\rho ''$, $D '$ and $D ''$ \cite{KNOWLES1990271}. We will calculate $\rho '$ using the factorised form of variant B (which we refer to as B-f), with LL recoil. $\rho'_{ss'}$ describes the distribution of spin states for a give parton after a single collinear emission. It is normalised by the trace of itself so that it maintains a probabilistic interpretation. Knowles begins by defining
\begin{eqnarray}
\begin{split}
\rho'_{ss'} = \frac{\sum_{s_{1},s'_{1},s_{2},s'_{2}}\rho_{s_{1}s'_{1}}V_{s_{1}s_{2}s} V^{*}_{s'_{1}s'_{2}s'} \delta_{s_{2}s'_{2}}}{\sum_{s,s_{1},s'_{1},s_{2},s'_{2}}\rho_{s_{1}s'_{1}}V_{s_{1}s_{2}s} V^{*}_{s'_{1}s'_{2}s}\delta_{s_{2}s'_{2}}},
\end{split} 
\end{eqnarray}
where $\rho$ is a spin density matrix for a parton in the hard process that is to be inherited by a forwardly evolving shower. In the language of this paper $\rho_{s_{1}s'_{1}}\propto \<s_{1}\rkl \v{H}(Q) \lkl s'_{1} \>$\footnote{For simplicity we suppose $\v{H}(Q)$ to contain a single propagating particle. If we were to introduce more particles we would have more indices/states to keep track of. This is because collinear emissions do not involve interference terms.}. $V_{s_{1}s_{2}s}$ is the spin-dependent collinear splitting function for the transition $s_{1}\rightarrow s s_{2}$ with parton type indices suppressed. Importantly, parton type indices are not summed over in $V_{s_{1}s_{2}s}$. When using Knowles' algorithm, it is assumed that the structure of a cascade has already been fully decided; all except the spin that is. 

Consider a term from B-f corresponding to one collinear emission from a final-state hard parton. Labelling this term $P'$, we have
\begin{eqnarray}
\begin{split}
P'_{ss'} &= \frac{\sum_{s_{2}s'_{2}} \< s, s_{2} \lkl \overline{\v{V}}_{\mu,q_{n+1 \, \bot}} \overline{\v{D}}_{n+1} \overline{\v{V}}_{q_{n+1 \, \bot},Q} \, \v{d} \boldsymbol{\sigma}(n, 0) \overline{\v{V}}^{\dagger}_{q_{n+1 \, \bot},Q} \overline{\v{D}}^{\dagger}_{n+1}  \overline{\v{V}}^{\dagger}_{\mu,q_{n+1 \, \bot}} \rkl s', s'_{2} \> \delta_{s_{2}s'_{2}} }{\Tr(\overline{\v{V}}_{\mu,q_{n+1 \, \bot}} \overline{\v{D}}_{n+1} \overline{\v{V}}_{q_{n+1 \, \bot},Q} \, \v{d} \boldsymbol{\sigma}(n, 0) \overline{\v{V}}^{\dagger}_{q_{n+1 \, \bot},Q} \overline{\v{D}}^{\dagger}_{n+1}  \overline{\v{V}}^{\dagger}_{\mu,q_{n+1 \, \bot}})}, \\
&= \frac{\sum_{s_{2}s'_{2}} \< s, s_{2} \lkl \v{P}_{2 \, 1} \v{H}(Q) \v{P}^{\dagger}_{2 \, 1} \rkl s', s'_{2} \> \delta_{s_{2}s'_{2}} }{\Tr(\v{P}_{2 \, 1} \v{H}(Q) \v{P}^{\dagger}_{2 \, 1})}.
\end{split} 
\end{eqnarray}
We have used the LL recoil with variant B and so integrals over the recoil functions were trivial. In the second line, we have labelled the collinear parton as parton 2 and the hard parton as parton 1. We can insert identity operators and evaluate the trace explicitly to find
\begin{eqnarray}
\begin{split}
P'_{ss'} = \frac{\sum_{s_{1},s'_{1},s_{2},s'_{2}} \< s, s_{2} \rkl \v{P}_{2 \, 1} \lkl s_{1} \> \< s'_{1} \rkl \v{P}^{\dagger}_{2 \, 1} \lkl s', s'_{2} \> \rho_{s_{1}s'_{1}}  \delta_{s_{2}s'_{2}} }{\sum_{s,s_{1},s'_{1},s_{2},s'_{2}} \< s, s_{2} \rkl \v{P}_{2 \, 1} \lkl s_{1} \> \< s'_{1} \rkl \v{P}^{\dagger}_{2 \, 1} \lkl s, s'_{2} \> \rho_{s_{1}s'_{1}} \delta_{s_{2}s'_{2}}}. \label{eqn:knowles_matched}
\end{split} 
\end{eqnarray}
Now note that $\< s s_{2} \rkl \v{P}_{2 \, 1} \lkl s_{1} \> = V_{s_{1}s_{2}s}$, with the possibilities of parton 2 being a gluon or quark summed over. Hence
\begin{eqnarray}
\begin{split}
P'_{ss'} = \frac{\sum_{2 \in \{q,g\}}\sum_{s_{1},s'_{1},s_{2},s'_{2}}\rho_{s_{1}s'_{1}}V_{s_{1}s_{2}s} V^{*}_{s'_{1}s'_{2}s'}\delta_{s_{2}s'_{2}}}{\sum_{2 \in \{q,g\}}\sum_{s,s_{1},s'_{1},s_{2},s'_{2}}\rho_{s_{1}s'_{1}}V_{s_{1}s_{2}s} V^{*}_{s'_{1}s'_{2}s}\delta_{s_{2}s'_{2}}}. 
\end{split} 
\end{eqnarray}
Thus, we have made the link to Collins and Knowles. If we pick either the quark or gluon term in the numerator and set it 0,  then renormalise $P'_{ss'}$ against the trace of itself,  we find 
$$
P^{' (1\rightarrow 23)}_{ss'} = \frac{\sum_{s_{1},s'_{1},s_{2},s'_{2}}\rho_{s_{1}s'_{1}}V_{s_{1}s_{2}s} V^{*}_{s'_{1}s'_{2}s'}\delta_{s_{2}s'_{2}}}{\sum_{s,s_{1},s'_{1},s_{2},s'_{2}}\rho_{s_{1}s'_{1}}V_{s_{1}s_{2}s} V^{*}_{s'_{1}s'_{2}s}\delta_{s_{2}s'_{2}}} 
 = \rho'_{ss'}.
$$
When comparing with Collins and Knowles it was necessary for us to pick a species for parton 2 as their algorithm is defined for pre-determined decay chains. This is why $\rho'_{ss'}$ is typically used without a label specifying the species of the partons involved, as their species is always provided by context.

We will finish off by calculating $\rho'_{ss'}$ for a $q\rightarrow qg$ splitting. $\rho_{s_{1}s'_{1}}$ is hermitian and so can be expressed as $\rho_{s_{1}s'_{1}} = \mathbbm{1} + \rho_{i}\sigma_{i}$ where $\sigma_{i}$ are the Pauli matrices. Using \eqref{eqn:knowles_matched} and normalising correctly gives
\begin{eqnarray}
\begin{split}
P'^{\, (q\rightarrow qg)}_{++} & = \frac{2q.\tilde{p}_{i}\sqrt{\frac{\mathcal{P}_{qq}}{1+z^{2}}}^{2}\rho_{++}+2q.\tilde{p}_{i}\sqrt{\frac{z^{2}\mathcal{P}_{qq}}{1+z^{2}}}^{2}\rho_{++}}{\left(2q.\tilde{p}_{i}\sqrt{\frac{\mathcal{P}_{qq}}{1+z^{2}}}^{2}+2q.\tilde{p}_{i}\sqrt{\frac{z^{2}\mathcal{P}_{qq}}{1+z^{2}}}^{2}\right)(\rho_{++}+\rho_{--})}=\frac{1}{2}(1+\rho_{3}), \\
P'^{\, (q\rightarrow qg)}_{+-}  & = \frac{2q.\tilde{p}_{i}\sqrt{\frac{\mathcal{P}_{qq}}{1+z^{2}}}\sqrt{\frac{z^{2}\mathcal{P}_{qq}}{1+z^{2}}}\rho_{++}+2q.\tilde{p}_{i}\sqrt{\frac{\mathcal{P}_{qq}}{1+z^{2}}}\sqrt{\frac{z^{2}\mathcal{P}_{qq}}{1+z^{2}}}\rho_{++}}{\left(2q.\tilde{p}_{i}\sqrt{\frac{\mathcal{P}_{qq}}{1+z^{2}}}^{2}+2q.\tilde{p}_{i}\sqrt{\frac{z^{2}\mathcal{P}_{qq}}{1+z^{2}}}^{2}\right)(\rho_{++}+\rho_{--})} = \frac{z}{1+z^{2}}(\rho_{1}-i\rho_{2}),\\
\end{split} 
\end{eqnarray}
where $q$ is the momentum of the gluon and $1-z$ is its momentum fraction. We also used $\< q \tilde{p}_{i} \> \< q \tilde{p}_{i} \>^{*} = [ q \tilde{p}_{i} ] [ q \tilde{p}_{i} ]^{*} = 2 q \cdot \tilde{p}_{i}$ . It follows that
\begin{eqnarray}
\begin{split}
P'^{\, (q\rightarrow qg)} &= \frac{1}{2}\left(\begin{array}{ll}
1+\rho_{3} & \frac{2z}{1+z^{2}}(\rho_{1}-i\rho_{2}) \\
\frac{2z}{1+z^{2}}(\rho_{1}+i\rho_{2}) & 1-\rho_{3} \\
\end{array} \right)
\end{split} = \rho'^{\, (q\rightarrow qg)}.
\end{eqnarray}
Similarly, matrices for the other collinear splittings can be found. The most algebraically complex is the $g\rightarrow gg$ splitting (as usual). In that case
\begin{eqnarray}
P'^{\, (g\rightarrow gg)}_{++}= 1+\frac{\left(\frac{z}{1-z} + 2(1-z)\right)\rho_{3}}{\frac{1}{2}P_{gg}+z(1-z)(\cos(2\phi)\rho_{1}+\sin(2\phi)\rho_{2})} = \rho'^{\, (g\rightarrow gg)}_{++},
\end{eqnarray}
where $\phi$ is the azimuthal angle to the plane of the splitting. The exact angular dependence depends on the definition of the Weyl spinor products. We have chosen the definition so as to match with the matrices defined in \cite{KNOWLES1990271}, where a factor $e^{i(s_{1}-s_{2}-s)\phi}$ has been pulled out from the definitions of $V_{s_{1}s_{2}s}$.  

\bibliographystyle{JHEP}
\bibliography{An_amplitude_lev_parton_shower}

\providecommand{\href}[2]{#2}\begingroup\raggedright\begin{thebibliography}{10}

\bibitem{Bauer:2000ew}
C.~W. Bauer, S.~Fleming and M.~E. Luke, \emph{{Summing Sudakov logarithms in $B
  \rightarrow X_s + \gamma$ in effective field theory}},
  \href{https://doi.org/10.1103/PhysRevD.63.014006}{\emph{Phys. Rev.}
  {\bfseries D63} (2000) 014006},
  [\href{https://arxiv.org/abs/hep-ph/0005275}{{\ttfamily hep-ph/0005275}}].

\bibitem{Bauer:2000yr}
C.~W. Bauer, S.~Fleming, D.~Pirjol and I.~W. Stewart, \emph{{An Effective field
  theory for collinear and soft gluons: Heavy to light decays}},
  \href{https://doi.org/10.1103/PhysRevD.63.114020}{\emph{Phys. Rev.}
  {\bfseries D63} (2001) 114020},
  [\href{https://arxiv.org/abs/hep-ph/0011336}{{\ttfamily hep-ph/0011336}}].

\bibitem{Bauer:2001yt}
C.~W. Bauer, D.~Pirjol and I.~W. Stewart, \emph{{Soft collinear factorization
  in effective field theory}},
  \href{https://doi.org/10.1103/PhysRevD.65.054022}{\emph{Phys. Rev.}
  {\bfseries D65} (2002) 054022},
  [\href{https://arxiv.org/abs/hep-ph/0109045}{{\ttfamily hep-ph/0109045}}].

\bibitem{SCET}
T.~Becher, A.~Broggio and A.~Ferroglia, \emph{{Introduction to Soft-Collinear
  Effective Theory}},
  \href{https://doi.org/10.1007/978-3-319-14848-9}{\emph{Lect. Notes Phys.}
  {\bfseries 896} (2015) pp.1--206},
  [\href{https://arxiv.org/abs/1410.1892}{{\ttfamily 1410.1892}}].

\bibitem{colour_dipole_model}
L.~{L\"onnblad}, \emph{{ARIADNE version 4: A Program for simulation of QCD
  cascades implementing the color dipole model}},
  \href{https://doi.org/10.1016/0010-4655(92)90068-A}{\emph{Comput. Phys.
  Commun.} {\bfseries 71} (1992) 15--31}.

\bibitem{Herwig_dipole_shower}
S.~{Pl\"atzer} and S.~Gieseke, \emph{{Dipole Showers and Automated NLO Matching
  in Herwig++}},
  \href{https://doi.org/10.1140/epjc/s10052-012-2187-7}{\emph{Eur. Phys. J.}
  {\bfseries C72} (2012) 2187},
  [\href{https://arxiv.org/abs/1109.6256}{{\ttfamily 1109.6256}}].

\bibitem{Herwig_shower}
S.~Gieseke, P.~Stephens and B.~Webber, \emph{{New formalism for QCD parton
  showers}}, \href{https://doi.org/10.1088/1126-6708/2003/12/045}{\emph{JHEP}
  {\bfseries 12} (2003) 045},
  [\href{https://arxiv.org/abs/hep-ph/0310083}{{\ttfamily hep-ph/0310083}}].

\bibitem{DIRE}
S.~Höche and S.~Prestel, \emph{{The midpoint between dipole and parton
  showers}}, \href{https://doi.org/10.1140/epjc/s10052-015-3684-2}{\emph{Eur.
  Phys. J.} {\bfseries C75} (2015) 461},
  [\href{https://arxiv.org/abs/1506.05057}{{\ttfamily 1506.05057}}].

\bibitem{Banfi:2004yd}
A.~Banfi, G.~P. Salam and G.~Zanderighi, \emph{{Principles of general
  final-state resummation and automated implementation}},
  \href{https://doi.org/10.1088/1126-6708/2005/03/073}{\emph{JHEP} {\bfseries
  03} (2005) 073}, [\href{https://arxiv.org/abs/hep-ph/0407286}{{\ttfamily
  hep-ph/0407286}}].

\bibitem{Pythia}
T.~{Sj\"ostrand} and P.~Z. Skands, \emph{{Transverse-momentum-ordered showers
  and interleaved multiple interactions}},
  \href{https://doi.org/10.1140/epjc/s2004-02084-y}{\emph{Eur. Phys. J.}
  {\bfseries C39} (2005) 129--154},
  [\href{https://arxiv.org/abs/hep-ph/0408302}{{\ttfamily hep-ph/0408302}}].

\bibitem{Pythia8}
T.~Sjöstrand, S.~Ask, J.~R. Christiansen, R.~Corke, N.~Desai, P.~Ilten et~al.,
  \emph{{An Introduction to PYTHIA 8.2}},
  \href{https://doi.org/10.1016/j.cpc.2015.01.024}{\emph{Comput. Phys. Commun.}
  {\bfseries 191} (2015) 159--177},
  [\href{https://arxiv.org/abs/1410.3012}{{\ttfamily 1410.3012}}].

\bibitem{Dasgupta:2001sh}
M.~Dasgupta and G.~P. Salam, \emph{{Resummation of nonglobal QCD observables}},
  \href{https://doi.org/10.1016/S0370-2693(01)00725-0}{\emph{Phys. Lett.}
  {\bfseries B512} (2001) 323--330},
  [\href{https://arxiv.org/abs/hep-ph/0104277}{{\ttfamily hep-ph/0104277}}].

\bibitem{Forshaw:2006fk}
J.~R. Forshaw, A.~Kyrieleis and M.~H. Seymour, \emph{{Super-leading logarithms
  in non-global observables in QCD}},
  \href{https://doi.org/10.1088/1126-6708/2006/08/059}{\emph{JHEP} {\bfseries
  08} (2006) 059}, [\href{https://arxiv.org/abs/hep-ph/0604094}{{\ttfamily
  hep-ph/0604094}}].

\bibitem{SuperleadingLogs}
J.~R. Forshaw, A.~Kyrieleis and M.~H. Seymour, \emph{{Super-leading logarithms
  in non-global observables in QCD: Colour basis independent calculation}},
  \href{https://doi.org/10.1088/1126-6708/2008/09/128}{\emph{JHEP} {\bfseries
  09} (2008) 128}, [\href{https://arxiv.org/abs/0808.1269}{{\ttfamily
  0808.1269}}].

\bibitem{Banfi:2010xy}
A.~Banfi, G.~P. Salam and G.~Zanderighi, \emph{{Phenomenology of event shapes
  at hadron colliders}},
  \href{https://doi.org/10.1007/JHEP06(2010)038}{\emph{JHEP} {\bfseries 06}
  (2010) 038}, [\href{https://arxiv.org/abs/1001.4082}{{\ttfamily 1001.4082}}].

\bibitem{Catani:2011st}
S.~Catani, D.~de~Florian and G.~Rodrigo, \emph{{Space-like (versus time-like)
  collinear limits in QCD: Is factorization violated?}},
  \href{https://doi.org/10.1007/JHEP07(2012)026}{\emph{JHEP} {\bfseries 07}
  (2012) 026}, [\href{https://arxiv.org/abs/1112.4405}{{\ttfamily 1112.4405}}].

\bibitem{factorisationBreaking}
J.~R. Forshaw, M.~H. Seymour and A.~Si{\'o}dmok, \emph{{On the Breaking of
  Collinear Factorization in QCD}},
  \href{https://doi.org/10.1007/JHEP11(2012)066}{\emph{JHEP} {\bfseries 11}
  (2012) 066}, [\href{https://arxiv.org/abs/1206.6363}{{\ttfamily 1206.6363}}].

\bibitem{Dasgupta:2018nvj}
M.~Dasgupta, F.~A. Dreyer, K.~Hamilton, P.~F. Monni and G.~P. Salam,
  \emph{{Logarithmic accuracy of parton showers: a fixed-order study}},
  \href{https://doi.org/10.1007/JHEP09(2018)033}{\emph{JHEP} {\bfseries 09}
  (2018) 033}, [\href{https://arxiv.org/abs/1805.09327}{{\ttfamily
  1805.09327}}].

\bibitem{Platzer:2018pmd}
S.~Plätzer, M.~Sj{\"o}dahl and J.~Thorén, \emph{{Color matrix element
  corrections for parton showers}},
  \href{https://doi.org/10.1007/JHEP11(2018)009}{\emph{JHEP} {\bfseries 11}
  (2018) 009}, [\href{https://arxiv.org/abs/1808.00332}{{\ttfamily
  1808.00332}}].

\bibitem{Platzer:2012np}
S.~{Pl\"atzer} and M.~Sj{\"o}dahl, \emph{{Subleading $N_c$ improved Parton
  Showers}}, \href{https://doi.org/10.1007/JHEP07(2012)042}{\emph{JHEP}
  {\bfseries 07} (2012) 042},
  [\href{https://arxiv.org/abs/1201.0260}{{\ttfamily 1201.0260}}].

\bibitem{Nagy:2017ggp}
Z.~Nagy and D.~E. Soper, \emph{{What is a parton shower?}},
  \href{https://doi.org/10.1103/PhysRevD.98.014034}{\emph{Phys. Rev.}
  {\bfseries D98} (2018) 014034},
  [\href{https://arxiv.org/abs/1705.08093}{{\ttfamily 1705.08093}}].

\bibitem{Nagy:2008eq}
Z.~Nagy and D.~E. Soper, \emph{{Parton showers with quantum interference:
  Leading color, with spin}},
  \href{https://doi.org/10.1088/1126-6708/2008/07/025}{\emph{JHEP} {\bfseries
  07} (2008) 025}, [\href{https://arxiv.org/abs/0805.0216}{{\ttfamily
  0805.0216}}].

\bibitem{Nagy:2012bt}
Z.~Nagy and D.~E. Soper, \emph{{Parton shower evolution with subleading
  color}}, \href{https://doi.org/10.1007/JHEP06(2012)044}{\emph{JHEP}
  {\bfseries 06} (2012) 044},
  [\href{https://arxiv.org/abs/1202.4496}{{\ttfamily 1202.4496}}].

\bibitem{Nagy:2015hwa}
Z.~Nagy and D.~E. Soper, \emph{{Effects of subleading color in a parton
  shower}}, \href{https://doi.org/10.1007/JHEP07(2015)119}{\emph{JHEP}
  {\bfseries 07} (2015) 119},
  [\href{https://arxiv.org/abs/1501.00778}{{\ttfamily 1501.00778}}].

\bibitem{SoftEvolutionAlgorithm}
R.~Ángeles Martínez, M.~De~Angelis, J.~R. Forshaw, S.~Plätzer and M.~H.
  Seymour, \emph{{Soft gluon evolution and non-global logarithms}},
  \href{https://doi.org/10.1007/JHEP05(2018)044}{\emph{JHEP} {\bfseries 05}
  (2018) 044}, [\href{https://arxiv.org/abs/1802.08531}{{\ttfamily
  1802.08531}}].

\bibitem{Nagy:2019pjp}
Z.~Nagy and D.~E. Soper, \emph{{Parton showers with more exact color
  evolution}},  \href{https://arxiv.org/abs/1902.02105}{{\ttfamily
  1902.02105}}.

\bibitem{ColoumbGluonsOrdering}
R.~Ángeles Martínez, J.~R. Forshaw and M.~H. Seymour, \emph{{Coulomb gluons
  and the ordering variable}},
  \href{https://doi.org/10.1007/JHEP12(2015)091}{\emph{JHEP} {\bfseries 12}
  (2015) 091}, [\href{https://arxiv.org/abs/1510.07998}{{\ttfamily
  1510.07998}}].

\bibitem{ColoumbGluonsOrderingLetter}
R.~Ángeles Martínez, J.~R. Forshaw and M.~H. Seymour, \emph{{Ordering
  multiple soft gluon emissions}},
  \href{https://doi.org/10.1103/PhysRevLett.116.212003}{\emph{Phys. Rev. Lett.}
  {\bfseries 116} (2016) 212003},
  [\href{https://arxiv.org/abs/1602.00623}{{\ttfamily 1602.00623}}].

\bibitem{BMSEquation}
A.~Banfi, G.~Marchesini and G.~Smye, \emph{{Away from jet energy flow}},
  \href{https://doi.org/10.1088/1126-6708/2002/08/006}{\emph{JHEP} {\bfseries
  08} (2002) 006}, [\href{https://arxiv.org/abs/hep-ph/0206076}{{\ttfamily
  hep-ph/0206076}}].

\bibitem{Collins:1987pm}
J.~C. Collins and D.~E. Soper, \emph{{The Theorems of Perturbative QCD}},
  \href{https://doi.org/10.1146/annurev.ns.37.120187.002123}{\emph{Ann. Rev.
  Nucl. Part. Sci.} {\bfseries 37} (1987) 383--409}.

\bibitem{Collins:1988ig}
J.~C. Collins, D.~E. Soper and G.~F. Sterman, \emph{{Soft Gluons and
  Factorization}},
  \href{https://doi.org/10.1016/0550-3213(88)90130-7}{\emph{Nucl. Phys.}
  {\bfseries B308} (1988) 833--856}.

\bibitem{Dokshitzer:1977sg}
Y.~L. Dokshitzer, \emph{{Calculation of the Structure Functions for Deep
  Inelastic Scattering and $e^+ e^-$ Annihilation by Perturbation Theory in
  Quantum Chromodynamics.}}, {\emph{Sov. Phys. JETP} {\bfseries 46} (1977)
  641--653}.

\bibitem{Gribov:1972ri}
V.~N. Gribov and L.~N. Lipatov, \emph{{Deep inelastic e p scattering in
  perturbation theory}}, {\emph{Sov. J. Nucl. Phys.} {\bfseries 15} (1972)
  438--450}.

\bibitem{APSplitting}
G.~Altarelli and G.~Parisi, \emph{Asymptotic freedom in parton language},
  \href{https://doi.org/https://doi.org/10.1016/0550-3213(77)90384-4}{\emph{Nuclear
  Physics B} {\bfseries 126} (1977) 298 -- 318}.

\bibitem{Dokshitzer:1991wu}
Y.~L. Dokshitzer, V.~A. Khoze, A.~H. Mueller and S.~I. Troian, \emph{{Basics of
  perturbative QCD}}.
\newblock 1991.

\bibitem{DOKSHITZER1980269}
Y.~Dokshitzer, D.~Dyakonov and S.~Troyan, \emph{Hard processes in quantum
  chromodynamics},
  \href{https://doi.org/https://doi.org/10.1016/0370-1573(80)90043-5}{\emph{Physics
  Reports} {\bfseries 58} (1980) 269 -- 395}.

\bibitem{Bassetto:1984ik}
A.~Bassetto, M.~Ciafaloni and G.~Marchesini, \emph{{Jet Structure and Infrared
  Sensitive Quantities in Perturbative QCD}},
  \href{https://doi.org/10.1016/0370-1573(83)90083-2}{\emph{Phys. Rept.}
  {\bfseries 100} (1983) 201--272}.

\bibitem{HelicitySplitting}
Z.~Bern, V.~Del~Duca, W.~B. Kilgore and C.~R. Schmidt, \emph{{The infrared
  behavior of one loop QCD amplitudes at next-to-next-to leading order}},
  \href{https://doi.org/10.1103/PhysRevD.60.116001}{\emph{Phys. Rev.}
  {\bfseries D60} (1999) 116001},
  [\href{https://arxiv.org/abs/hep-ph/9903516}{{\ttfamily hep-ph/9903516}}].

\bibitem{HelicityTechniques}
H.~K. Dreiner, H.~E. Haber and S.~P. Martin, \emph{{Two-component spinor
  techniques and Feynman rules for quantum field theory and supersymmetry}},
  \href{https://doi.org/10.1016/j.physrep.2010.05.002}{\emph{Phys. Rept.}
  {\bfseries 494} (2010) 1--196},
  [\href{https://arxiv.org/abs/0812.1594}{{\ttfamily 0812.1594}}].

\bibitem{Bewick:2019rbu}
G.~Bewick, S.~Ferrario~Ravasio, P.~Richardson and M.~H. Seymour,
  \emph{{Logarithmic Accuracy of Angular-Ordered Parton Showers}},
  \href{https://arxiv.org/abs/1904.11866}{{\ttfamily 1904.11866}}.

\bibitem{Dokshitzer:2005ek}
{\relax Yu}.~L. Dokshitzer and G.~Marchesini, \emph{{Hadron collisions and the
  fifth form-factor}},
  \href{https://doi.org/10.1016/j.physletb.2005.10.009}{\emph{Phys. Lett.}
  {\bfseries B631} (2005) 118--125},
  [\href{https://arxiv.org/abs/hep-ph/0508130}{{\ttfamily hep-ph/0508130}}].

\bibitem{Schumann:2007mg}
S.~Schumann and F.~Krauss, \emph{{A Parton shower algorithm based on
  Catani-Seymour dipole factorisation}},
  \href{https://doi.org/10.1088/1126-6708/2008/03/038}{\emph{JHEP} {\bfseries
  03} (2008) 038}, [\href{https://arxiv.org/abs/0709.1027}{{\ttfamily
  0709.1027}}].

\bibitem{Platzer:recoil}
S.~{Pl\"atzer} and S.~Gieseke, \emph{{Coherent Parton Showers with Local
  Recoils}}, \href{https://doi.org/10.1007/JHEP01(2011)024}{\emph{JHEP}
  {\bfseries 01} (2011) 024},
  [\href{https://arxiv.org/abs/0909.5593}{{\ttfamily 0909.5593}}].

\bibitem{Catani:1996vz}
S.~Catani and M.~H. Seymour, \emph{{A General algorithm for calculating jet
  cross-sections in NLO QCD}},
  \href{https://doi.org/10.1016/S0550-3213(96)00589-5,
  10.1016/S0550-3213(98)81022-5}{\emph{Nucl. Phys.} {\bfseries B485} (1997)
  291--419}, [\href{https://arxiv.org/abs/hep-ph/9605323}{{\ttfamily
  hep-ph/9605323}}].

\bibitem{Mangano:1990by}
M.~L. Mangano and S.~J. Parke, \emph{{Multiparton amplitudes in gauge
  theories}}, \href{https://doi.org/10.1016/0370-1573(91)90091-Y}{\emph{Phys.
  Rept.} {\bfseries 200} (1991) 301--367},
  [\href{https://arxiv.org/abs/hep-th/0509223}{{\ttfamily hep-th/0509223}}].

\bibitem{BERENDS1988759}
F.~Berends and W.~Giele, \emph{Recursive calculations for processes with n
  gluons},
  \href{https://doi.org/https://doi.org/10.1016/0550-3213(88)90442-7}{\emph{Nuclear
  Physics B} {\bfseries 306} (1988) 759 -- 808}.

\bibitem{Rogers:2010dm}
T.~C. Rogers and P.~J. Mulders, \emph{{No Generalized TMD-Factorization in
  Hadro-Production of High Transverse Momentum Hadrons}},
  \href{https://doi.org/10.1103/PhysRevD.81.094006}{\emph{Phys. Rev.}
  {\bfseries D81} (2010) 094006},
  [\href{https://arxiv.org/abs/1001.2977}{{\ttfamily 1001.2977}}].

\bibitem{Rogers:2013zha}
T.~C. Rogers, \emph{{Extra spin asymmetries from the breakdown of
  transverse-momentum-dependent factorization in hadron-hadron collisions}},
  \href{https://doi.org/10.1103/PhysRevD.88.014002}{\emph{Phys. Rev.}
  {\bfseries D88} (2013) 014002},
  [\href{https://arxiv.org/abs/1304.4251}{{\ttfamily 1304.4251}}].

\bibitem{Aybat:2008ct}
S.~M. Aybat and G.~F. Sterman, \emph{{Soft-Gluon Cancellation, Phases and
  Factorization with Initial-State Partons}},
  \href{https://doi.org/10.1016/j.physletb.2008.11.050}{\emph{Phys. Lett.}
  {\bfseries B671} (2009) 46--50},
  [\href{https://arxiv.org/abs/0811.0246}{{\ttfamily 0811.0246}}].

\bibitem{Skands:2009tb}
P.~Z. Skands and S.~Weinzierl, \emph{{Some remarks on dipole showers and the
  DGLAP equation}},
  \href{https://doi.org/10.1103/PhysRevD.79.074021}{\emph{Phys. Rev.}
  {\bfseries D79} (2009) 074021},
  [\href{https://arxiv.org/abs/0903.2150}{{\ttfamily 0903.2150}}].

\bibitem{Nagy:2009re}
Z.~Nagy and D.~E. Soper, \emph{{Final state dipole showers and the DGLAP
  equation}}, \href{https://doi.org/10.1088/1126-6708/2009/05/088}{\emph{JHEP}
  {\bfseries 05} (2009) 088},
  [\href{https://arxiv.org/abs/0901.3587}{{\ttfamily 0901.3587}}].

\bibitem{Dokshitzer:2008ia}
{\relax Yu}.~L. Dokshitzer and G.~Marchesini, \emph{{Monte Carlo and large
  angle gluon radiation}},
  \href{https://doi.org/10.1088/1126-6708/2009/03/117}{\emph{JHEP} {\bfseries
  03} (2009) 117}, [\href{https://arxiv.org/abs/0809.1749}{{\ttfamily
  0809.1749}}].

\bibitem{Binetruy:1980hd}
P.~Binetruy, \emph{{Summing Leading Logs in Thrust Distributions}},
  \href{https://doi.org/10.1016/0370-2693(80)90442-6}{\emph{Phys. Lett.}
  {\bfseries 91B} (1980) 245--248}.

\bibitem{Resum_large_logs_ee}
S.~Catani, L.~Trentadue, G.~Turnock and B.~Webber, \emph{{Resummation of large
  logarithms in $e^+ e^-$ event shape distributions}}, {\emph{Nucl. Phys.}
  {\bfseries B407} (1993) 3}.

\bibitem{Becher:2008cf}
T.~Becher and M.~D. Schwartz, \emph{{A precise determination of $\alpha_s$ from
  LEP thrust data using effective field theory}},
  \href{https://doi.org/10.1088/1126-6708/2008/07/034}{\emph{JHEP} {\bfseries
  07} (2008) 034}, [\href{https://arxiv.org/abs/0803.0342}{{\ttfamily
  0803.0342}}].

\bibitem{Schwartz:2014wha}
M.~D. Schwartz and H.~X. Zhu, \emph{{Nonglobal logarithms at three loops, four
  loops, five loops, and beyond}},
  \href{https://doi.org/10.1103/PhysRevD.90.065004}{\emph{Phys. Rev.}
  {\bfseries D90} (2014) 065004},
  [\href{https://arxiv.org/abs/1403.4949}{{\ttfamily 1403.4949}}].

\bibitem{Khelifa-Kerfa:2015mma}
K.~Khelifa-Kerfa and Y.~Delenda, \emph{{Non-global logarithms at finite N$_{c}$
  beyond leading order}},
  \href{https://doi.org/10.1007/JHEP03(2015)094}{\emph{JHEP} {\bfseries 03}
  (2015) 094}, [\href{https://arxiv.org/abs/1501.00475}{{\ttfamily
  1501.00475}}].

\bibitem{Hagiwara:2015bia}
Y.~Hagiwara, Y.~Hatta and T.~Ueda, \emph{{Hemisphere jet mass distribution at
  finite $N_c$}},
  \href{https://doi.org/10.1016/j.physletb.2016.03.028}{\emph{Phys. Lett.}
  {\bfseries B756} (2016) 254--258},
  [\href{https://arxiv.org/abs/1507.07641}{{\ttfamily 1507.07641}}].

\bibitem{Keates:2009dn}
J.~Keates and M.~H. Seymour, \emph{{Super-leading logarithms in non-global
  observables in QCD: Fixed order calculation}},
  \href{https://doi.org/10.1088/1126-6708/2009/04/040}{\emph{JHEP} {\bfseries
  04} (2009) 040}, [\href{https://arxiv.org/abs/0902.0477}{{\ttfamily
  0902.0477}}].

\bibitem{Platzer:2013fha}
S.~Pl{\"a}tzer, \emph{{Summing Large-$N$ Towers in Colour Flow Evolution}},
  \href{https://doi.org/10.1140/epjc/s10052-014-2907-2}{\emph{Eur. Phys. J.}
  {\bfseries C74} (2014) 2907},
  [\href{https://arxiv.org/abs/1312.2448}{{\ttfamily 1312.2448}}].

\bibitem{QCDTalks}
M.~De~Angelis, \emph{{Non-global Logarithms beyond Leading Colour, talk at
  \url{https://indico.cern.ch/event/662485/}}},  2018.

\bibitem{HARPSTalks}
J.~R. Forshaw and S.~Plätzer, \emph{{Soft Gluon Evolution beyond Leading
  Colour, talks at \url{https://indico.cern.ch/event/729453/}}},  2018.

\bibitem{JaxoDraw}
D.~Binosi and L.~Theussl, \emph{{JaxoDraw: A Graphical user interface for
  drawing Feynman diagrams}},
  \href{https://doi.org/10.1016/j.cpc.2004.05.001}{\emph{Comput. Phys. Commun.}
  {\bfseries 161} (2004) 76--86},
  [\href{https://arxiv.org/abs/hep-ph/0309015}{{\ttfamily hep-ph/0309015}}].

\bibitem{HelicitySplitting2}
L.~J. Dixon, \emph{{Calculating scattering amplitudes efficiently}},  in
  \emph{{QCD and beyond. Proceedings, Theoretical Advanced Study Institute in
  Elementary Particle Physics, TASI-95, Boulder, USA, June 4-30, 1995}},
  pp.~539--584, 1996, \href{https://arxiv.org/abs/hep-ph/9601359}{{\ttfamily
  hep-ph/9601359}},
  \href{http://www-public.slac.stanford.edu/sciDoc/docMeta.aspx?slacPubNumber=SLAC-PUB-7106}{http://www-public.slac.stanford.edu/sciDoc/docMeta.aspx?slacPubNumber=SLAC-PUB-7106}.

\bibitem{QCD_colour_flow}
W.~Kilian, T.~Ohl, J.~Reuter and C.~Speckner, \emph{{QCD in the Color-Flow
  Representation}}, \href{https://doi.org/10.1007/JHEP10(2012)022}{\emph{JHEP}
  {\bfseries 10} (2012) 022},
  [\href{https://arxiv.org/abs/1206.3700}{{\ttfamily 1206.3700}}].

\bibitem{Collins:1987cp}
J.~C. Collins, \emph{{Spin Correlations in Monte Carlo Event Generators}},
  \href{https://doi.org/10.1016/0550-3213(88)90654-2}{\emph{Nucl. Phys.}
  {\bfseries B304} (1988) 794--804}.

\bibitem{KNOWLES1990271}
I.~Knowles, \emph{A linear algorithm for calculating spin correlations in
  hadronic collisions},
  \href{https://doi.org/https://doi.org/10.1016/0010-4655(90)90063-7}{\emph{Computer
  Physics Communications} {\bfseries 58} (1990) 271 -- 284}.

\bibitem{Richardson_2001}
P.~Richardson, \emph{{Spin correlations in Monte Carlo simulations}},
  \href{https://doi.org/10.1088/1126-6708/2001/11/029}{\emph{Journal of High
  Energy Physics} {\bfseries 2001} (Nov., 2001) 029--029}.

\end{thebibliography}\endgroup

\end{document}